\documentclass[journal=jpclcd,manuscript=letter]{achemso}

\usepackage{amsfonts}
\usepackage{bm}
\usepackage{setspace}
\usepackage{graphicx}
\usepackage{color}
\usepackage{verbatim}
\usepackage{float}
\usepackage{array}
\usepackage{booktabs}
\usepackage{amsmath}
\usepackage{cleveref}
\usepackage{amssymb}
\usepackage{graphicx}
\usepackage[version=3]{mhchem}
\usepackage[T1]{fontenc}
\usepackage{subfigure}
\usepackage{subfig}
\usepackage{xr}
\usepackage{threeparttablex}
\usepackage{xcolor}
\usepackage{enumitem}
\usepackage{soul}

\newcommand{\dd}{\mathop{}\!\text{d}}

\title{
    The Exact Second Order Corrections and Accurate Quasiparticle
    Energy Calculations in Density Functional Theory
}

\author{Yuncai Mei}
\author{Zehua Chen}
\affiliation{Department of Chemistry, Duke University, Durham,
    North Carolina 27708, USA}
\footnotetext[1]{Y.M. and Z.C. contributed equally to this work and share
    the first authorship.}

\author{Weitao Yang}
\email{weitao.yang@duke.edu}
\affiliation{Department of Chemistry, Duke University, Durham,
    North Carolina 27708, USA}
\alsoaffiliation{Department of Physics, Duke University, Durham,
    North Carolina 27708, USA}
\date{\today}

\begin{document}
\begin{tocentry}
    \includegraphics[width=1\linewidth]{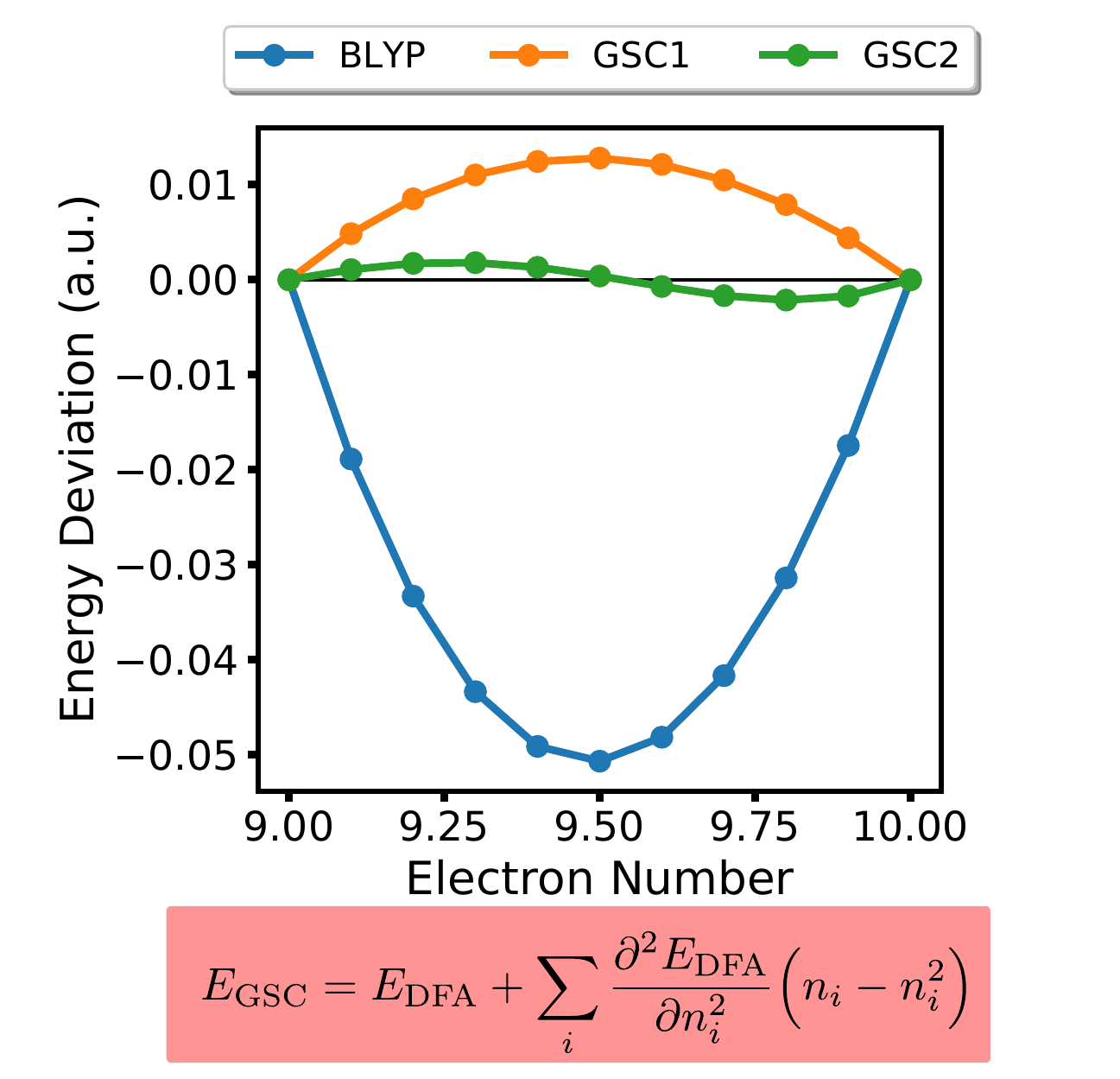}
\end{tocentry}

\begin{abstract}
    We develop a second order correction to commonly used density
    functional approximations (DFA) to eliminate the systematic
    delocalization error. The method, based on the previously developed
    global scaling correction (GSC), is an exact quadratic correction
    to the DFA for the fractional charge behavior and uses the
    analytical second derivatives of the total energy with respect to
    fractional occupation numbers of the canonical molecular orbitals.
    For small and medium-size molecules, this correction leads to
    ground-state orbital energies that are highly accurate
    approximation to the corresponding quasiparticle energies. It
    provides excellent predictions of ionization potentials,
    electron affinities, photoemission spectrum and photoexcitation
    energies beyond previous approximate second order approaches,
    thus showing potential for broad applications in computational
    spectroscopy.
\end{abstract}

\maketitle

The Kohn-Sham density functional theory (DFT)
\cite{1964hohenbergB871,1965kohnA1138,1989parr}
has gained much success in
modern chemistry, materials science and physics.
Most density functional approximations (DFAs)
to the exchange correlation energy $E_{\rm{xc}}$
usually produce reasonable total energies for small and
medium-size molecules,
however, they have major deficiencies in orbital energies.
As known for a long time, for finite systems,
the eigenvalue of the highest occupied molecular orbital
(HOMO) for the exact Kohn-Sham potential is equal to
the negative of the first ionization potential (IP)
for finite systems, based on the asymptotic decay behavior
of the exact electron density,
and the requirement that the Kohn-Sham effective
potential be zero at infinity \cite{1989parr}.
However, in a Kohn-Sham calculation, the local
Kohn-Sham potential can have any additive constant
and still have the same total
energy and density, but different orbital energies.
Thus the argument based on the long range behavior of
density and potential
hinges on a particular choice of the additive constant
of the potential.

The general physical meaning of orbital energies,
for both the HOMO and the lowest
unoccupied molecular orbital (LUMO),
has been established through three key results.
It is based on the property of the total energy
functional (approximate or exact).
First, the Janak theorem
links Kohn-Sham orbital energies to the derivatives of the
total energy with respect to the orbital occupation numbers,
which were not related to any physical observables \cite{1978janak7168}.
Second, the derivatives of the total energy with respective to
the total electron number, which are the chemical potentials,
are respectively the negative of the first ionization potential (IP)
and the first electron affinity (EA) for the exact functional
based on the linear condition on the behavior of energy for fractional number
of electrons. \cite{1982perdew1694}
Finally, the chemical potentials were established to be equal to
the derivatives of the total energy with respect to the
HOMO/LUMO orbital occupation numbers in the Kohn-Sham
calculation with exchange-correlation energy being
functionals of the density, or the
generalized Kohn-Sham calculation with exchange-correlation
energy being functionals of the non-interacting
one-electron density matrix. \cite{2008cohen115123}
Combining these theoretical results, the HOMO and LUMO energies
are the negative of the first IP and the first EA for electron removal and addition
respectively, as approximated by the corresponding DFA used \cite{2008cohen115123}.
This physical interpretation of the HOMO and LUMO energies holds true
for general molecules and bulk systems
with commonly used DFAs.
Exceptions occur in some strongly correlated systems when the DFA used
has the desired explicit derivative discontinuity.
\cite{2009mori-sanchez066403}

Thus the HOMO and LUMO energies are fully established for the theoretical
prediction of the IP and the EA in a DFA calculation.
However, conventional DFAs have the systematic error in
significantly underestimating the IP and
overestimating the EA, and thus the fundamental gap.
\cite{2008mori-sanchez146401,2008cohen115123,2008cohen794,2012cohen320}

To describe the systematic error of DFAs, the concept of the delocalization error
\cite{2008mori-sanchez146401,2008cohen115123,2008cohen794,2012cohen320}
has been developed, and it can be understood from the
perspective of fractional charges.
\cite{2008mori-sanchez146401,2008cohen115123}
For systems of small or moderate physical sizes,
conventional DFAs usually have good accuracy in total energies
when there are integer number of electrons.
When there are fractional number of electrons,
conventional DFAs, however, violate the
Perdew-Parr-Levy-Balduz (PPLB) linearity condition,
\cite{1982perdew1694,2000yang5175,2001zhang348}
which states the exact ground state energy $E_{\text{gs}}(N)$
is a linear function of the fractional electron numbers connecting adjacent integer points.
Inconsistent with the requirement of the PPLB linearity condition,
$E_{\text{gs}}(N)$ curves from conventional DFAs are usually convex,
with drastic underestimation to the ground state energies
of fractional systems.
The convex deviation of conventional DFAs decreases when the systems
are larger, and vanishes at the bulk limit.
\cite{2008mori-sanchez146401}
However, the delocalization
error exhibits in another way, in which the error manifests as too
low relative ground state energies of ionized systems and incorrect
linear $E_{\text{gs}}(N)$ curves with wrong slopes at the bulk limit.

To reduce the error, enormous efforts have been
devoted in the development of new exchange-correlation functionals
during last decades.
These developments,
including global hybrid
\cite{1999adamo6170,1988lee789},
local hybrid,
\cite{2003jaramillo1073,2007arbuznikov168}
double hybrid
\cite{2006grimme034108a,2009zhang4968a,2014su9211}
and range-separated functionals
\cite{1995savin332,1996savin357,2001iikura3544,2004yanai57,
2006vydrov234109a,2008chai6620},
mainly incorporate a certain amount
of Hartree Fock (HF) exchange in the $E_{\rm{xc}}$.
The HF exchange exhibits a concave deviation to the linear condition,
which is opposite to conventional DFAs and called the localization error
\cite{2008mori-sanchez146401,2008cohen115123}.
Some hybrid
and range-separated functionals mostly rely on
system-dependent tuned parameters.\cite{2010baer109}

There are many other efforts to eliminate the systematical error of DFAs,
based on the perspective of the delocalization error or from different
understandings. Self-interaction error (SIE)
\cite{1981perdew5079,2006mori-sanchez201102a}
was the first concept to describe the systematic error of DFAs.
SIE associates the error to the incomplete cancellation
between the electron self-Coulomb and self-exchange energies,
which is a different interpretation of the error source compared
to the concept of delocalization error.
Many approaches have been designed on the basis
of correcting SIE,
\cite{1981perdew5079,2006mori-sanchez201102a,
2006mori-sanchez091102,
2008perdew052513,2014schmidt14367,
2014pederson121103,
2016schmidt165120,
2017yang052505}
including the latest development with
Fermi localized orbitals.
\cite{2014pederson121103,2017yang052505}
Beside SIE, there are other approaches
developed with the focus on specific properties, such as
the Koopmans-compliant functionals
\cite{2014borghi075135,2019colonna1914},
the generalized transition state \cite{2005anisimov075125}
and related methods
\cite{2016ma24924}
extending the straight line condition
with Wannier functions for bad gap predictions of solid.

Following the understanding of the delocalization error,
researchers at the Yang laboratory developed a set of correction methods
to conventional DFAs to systematically reduce the delocalization error,
\cite{2011zheng026403a,2015li053001a,
2018li215,2020su1535,2020mei10277}
in which the PPLB linearity condition is imposed
by applying explicit treatments to systems of fractional charges
to restore the correct behavior of $E_{\text{gs}}(N)$. Specifically,
the global scaling correction (GSC)
\cite{2011zheng026403a}
method imposes the
PPLB linearity condition globally on the (delocalized) canonical
molecular orbital occupation numbers.
On the basis of the accurate description of
ground state energies from conventional DFAs for small integer systems,
the GSC was designed to preserve the energy of integer systems
and correct the convex $E_{\text{gs}}(N)$ curve to be linear for fractional systems.
It should be pointed out that, the GSC is only applicable for small
and moderate size systems,
because the convex deviation of conventional DFAs to the linear line
for fractional charges
decreases with increasing system size, and the delocalization
error manifests as underestimated ground state energies for
integer systems and incorrect linear $E_{\rm{gs}}(N)$ curves
with wrong slops at the bulk limits.
\cite{2008mori-sanchez146401,2008cohen115123,2012cohen320}
To reduce the delocalization error for large systems,
the local scaling correction (LSC) method
\cite{2015li053001a}
was developed with focus on the local regions of molecular
systems to apply the energy correction locally.
Combing the ideas of the GSC and the LSC, the Yang laboratory recently
developed the localized orbital scaling correction (LOSC)
\cite{2018li215,2020su1535,2020mei10277}
to achieve size-consistent and systematic improvement, in
which orbitalets,
orbitals localized in space and in energy and linearly
combining both occupied and virtual orbitals,
were developed to apply the global
or local corrections adaptively.
All these scaling correction methods have shown major improvements
to describe challenging properties for conventional DFAs, including
IPs, EAs, photoemission spectrum and polarizabilities.
\cite{2018li215,2019mei,2019mei673,2020su1535,2020mei10277,
2021mei054302}

Though the GSC method has its limitations for large and bulk systems,
the method is very useful for spectrum properties and
excitation energies of small and moderate size molecules.
Therefore improving the accuracy of GSC can have
a significant impact for large areas of DFT applications.
In addition, because the GSC and the LOSC become the same for small and
moderate size molecules when the orbital localization does not take place,
insight on the GSC can lead to improvements for the LOSC, which is applicable to
general systems.
To improve the accuracy of the GSC method,
Xiao and coworkers developed a correction form that utilizes higher order density expansions
to go beyond the frozen orbital approximation in the original GSC work.
\cite{2015zhang154113,2020yangc} Up to third-order orbital derivatives were
calculated, and remarkable improvements had been achieved.
However, the exchange-correlation component was still treated
approximately as the LDA exchange energy, regardless of the DFA used.
All works \cite{2011zheng026403a,2015zhang154113,2020yangc} up to now are
approximate second-order corrections to the DFA total energy for
fractional-electron systems.

In this work, we present an \emph{exact} formula for
the second order correction to DFA energies with
respect to occupation numbers of
canonical molecular orbitals.
This constitutes an exact global scaling correction if the convex deviation
is quadratic. We will also show that, to be accurate to the second order in the density expansion,
only the first order orbital derivatives are required.
The result of this work
is an energy correction to DFAs under the framework of GSC
but with the exact second-order expansions and much improved accuracy.

To start, we briefly review the methodology of the GSC method. \cite{2011zheng026403a}
The aim of GSC is to correct originally non-straight
\(E(N)\) curves produced by DFAs to straight lines, as required by the PPLB
linearity condition:
\begin{align}
    \label{eq:gsc_deltaE_1}
    E_{\text{GSC}}(N+n) &= E(N+n) + \Delta_{\text{GSC}}(N+n)
    = (1 - n)E(N) + n E(N+1),
\end{align}
where $n$ is a fraction, $0 \leq n \leq 1$.
Therefore, the GSC energy correction can be easily seen as
\begin{align}
    \label{eq:gsc_deltaE_2}
    \Delta_{\text{GSC}} (N+n) &= \left[(1 - n)E(N) + n E(N+1)\right] - E(N+n).
\end{align}
According to Eq.~\ref{eq:gsc_deltaE_1} and Eq.~\ref{eq:gsc_deltaE_2},
the GSC preserves the total energy of integer systems,
while it produces corrections to total energies of fractional systems.
In the original work of the GSC\cite{2011zheng026403a},
$\Delta_{\text{GSC}}$ is evaluated based on the expansion of the density matrix
for $(N+n)$ electrons:
\begin{align}
    \label{eq:rho_expand}
    \rho_{\text{s}}^{N+n}(\mathbf{r}, \mathbf{r}')
    = \rho_{\text{s}}^{N}(\mathbf{r}, \mathbf{r}')
    + n f(\mathbf{r}, \mathbf{r}') + n^2 \gamma(\mathbf{r}, \mathbf{r}')
    + \cdots,
\end{align}
To the first order in \(n\), and with the frozen
orbital approximation in the Fukui function \cite{1984parr4050},
it can be shown that
\begin{align}
    \label{eq:rho_expand_approx}
    \rho_{\text{s}}^{N+n}(\mathbf{r}, \mathbf{r}') \approx
        \rho_{\text{s}}^{N}(\mathbf{r}, \mathbf{r}')
        + n \psi_f(\mathbf{r})\psi_f^*(\mathbf{r}'),
\end{align}
where \(\psi_f\) represents the frontier orbital that has
fractional occupation \(n_f = n\).
By inserting the density matrix expansion (Eq.~\ref{eq:rho_expand_approx})
into Eq.~\ref{eq:gsc_deltaE_2},
and keeping terms up to the second order in \(n\), the GSC
corrected total energy reads
\begin{align}
    \label{eq:gsc_E}
    E_{\mathrm{GSC}}
    = E_{\mathrm{DFA}} +
        \frac12 \kappa \left( n - n^2 \right).
\end{align}
The approximate expression of the coefficients $\kappa$ that was derived based the LDA is
\begin{align}
    \label{eq:kappa_approx}
    \kappa = \int \frac{\rho_f(\mathbf{r}) \rho_f(\mathbf{r'})}
        {|\mathbf{r} - \mathbf{r'}|} \dd \mathbf{r} \dd \mathbf{r'}
        - \frac{2\tau C_{\text{x}}}{3} \int
        \big[\rho_f(\mathbf{r})\big]^{\frac{4}{3}} \dd \mathbf{r},
\end{align}
in which $\rho_f(\mathbf{r}) = |\psi_f(\mathbf{r})|^2$, and
$C_{\text{x}} = \frac{3}{4}\big(\frac{6}{\pi}\big)^{1/3}$. \cite{2011zheng026403a}
The first term corresponds to the Coulomb part,
the second term corresponds to the exchange part,
and the parameter $\tau$ is used to balance the contribution
of two parts, which is set to 1 in the original work.

The outcome of recovering a linear \(E(N)\) curve is
the correction to the orbital energies for integer systems with the
GSC method. Based on the GSC energy expression (Eq.~\ref{eq:gsc_E}),
the chemical potential associated with the frontier orbital is
\begin{align}
    \label{eq:gsc_chem_potential}
    \frac{\partial E_{\text{GSC}}}{\partial n_f} &=
        \epsilon_f^{\text{DFA}}
        + \frac12 \kappa (1 - 2 n_f)
        + \frac12 \frac{\partial \kappa}{\partial n_f}
        (n_f - n_f^2),
\end{align}
where $\epsilon_f^{\rm{DFA}}$ is the frontier orbital energy
from the parent DFA, according to the Janak theorem. \cite{1978janak7168}
Although \(\kappa\) is dependent on \(\psi_f\), which further
depends on the density matrix and the set of occupation numbers,
the contribution from the derivative of \(\frac{\partial \kappa}{\partial n_f}\)
vanishes for integer systems, as $(n_f - n_f^2) = 0$ when $n_f$ is either 1 or 0.
Thus, Eq.~\ref{eq:gsc_chem_potential} leads to
simple corrections for HOMO/LUMO energies of integer systems with
\(
    \epsilon_{\text{HOMO}}^{\rm{GSC}} = \epsilon_{\text{HOMO}}^{\text{DFA}}
        - \kappa/2
\) and
\(
    \epsilon_{\text{LUMO}}^{\rm{GSC}} = \epsilon_{\text{LUMO}}^{\text{DFA}}
        + \kappa/2
\),
which are equal to the negative of IP and EA respectively,
as predicted by the GSC, based on the theoretical developments.
\cite{1982perdew1694,2008cohen115123}

One limitation in the original GSC work is, since the PPLB linearity
condition holds only for ground states, the orbital energy correction
applies to the HOMO and the LUMO only. In most cases, the HOMO and the LUMO
energies are of particular interest, since they are related to IPs, EAs
and thus the fundamental gaps. However, precise descriptions of orbitals
other than HOMO/LUMO can also play important roles in certain cases.
For example, in calculations of the photoemission spectra and excitation energies
from orbital energies. \cite{2019mei673,2019mei2545}
Therefore, a natural extension to excited states is necessary,
and similar constructions for the correction to all orbitals
have already been established in LOSC
\cite{2018li215}
and recent work from Xiao and coworkers on GSC \cite{2020yangc}.
To facilitate the discussion to orbitals
above LUMO and under the HOMO,
we need to extend the ground state energy $E(N)$,
as a function of $N$, the total number of electron,
to $E(\{n_{p\sigma}\})$, the total energy as a function
of the canonical orbital occupation numbers
$\{n_{p\sigma}\}$, which can correspond to some
excited states. This energy function is what was used
in Janak's work \cite{1978janak7168,1989parr},
and it is given by the following minimum
\begin{equation}
    \label{eqn:frac_energy}
    E(\{n_{p\sigma}\}) = \min_{\{\psi_{p\sigma}\}}
    E_v[\{\rho_{\text{s}}^{\sigma}\}],
\end{equation}
where the canonical orbitals $\{\psi_{p\sigma}\}$ are
constrained to be orthogonormal and the density matrix
is $\hat{\rho_{\text{s}}}^\sigma = \sum_{p\sigma} n_{p\sigma} |\psi_{p\sigma}\rangle \langle \psi_{p\sigma}|$.
Following the same procedure for the cases of HOMO/LUMO and generalizing
to all orbitals, the extended definition of the GSC energy correction becomes
\begin{align}
    E_{\mathrm{GSC}}
    = E_{\mathrm{DFA}} +
        \frac12 \sum_{p\sigma} \kappa_{p\sigma}
        \left( n_{p\sigma} - n_{p\sigma}^2 \right).
\end{align}
The second-order correction term depends on the
corresponding orbital density \(\rho_{p\sigma}=|\psi_{p\sigma}|^2\),
and the coefficients $\kappa_{p\sigma}$ has the exact
same expression as shown in Eq.~\ref{eq:kappa_approx},
except with the $\rho_{p\sigma}$ instead of $\rho_f(\mathbf{r})$.
As a result, the orbital energy corrections are
\(
    \epsilon_{i\sigma}^{\rm{GSC}} = \epsilon_{i\sigma}^{\text{DFA}}
        - \kappa_{i\sigma}/2
\)
for occupied orbitals and
\(
    \epsilon_{a\sigma}^{\rm{GSC}} = \epsilon_{a\sigma}^{\text{DFA}}
        + \kappa_{a\sigma}/2
\) for virtual orbitals.

Although the original GSC has shown great improvements to
conventional DFAs for the description of many challenging properties,
it is still possible to achieve better accuracy within the GSC framework.
Recently, it has been reported that involving higher
order terms to evaluate the response of electron
density to the electron number (Eq.~\ref{eq:rho_expand})
can provide more accurate GSC orbital energies, giving
excellent performance for IPs, EAs and other quasihole energetics.
\cite{2015zhang154113,2020yangc}
However, the correlation energy contribution is still missing
in the existing approximate expression of \(\kappa_{p\sigma}\).
Furthermore, the exchange part was derived from the LDA only,
and directly applying it to GGAs and hybrid functionals
may not produce ideal results.
It is therefore the goal of this paper to develop a more accurate form
for the GSC, namely the exact functional expression for
$\{\kappa_{p\sigma}\}$.

In this work, instead of considering the density relaxation
as in Ref. \citenum{2015zhang154113}, we directly
deal with the function $E(\{n_{m\tau}\})$ and
find the correction to the second order.
Consider $E(\{n_{m\tau} + \delta_{m\tau}\})$,
where $0 \leq n_{m\tau} + \delta_{m\tau} \leq 1$,
and expand the energy function in a Taylor
series to
the second order in $\{\delta_{m\tau}\}$.
We have the following relation
\begin{align}
    E(\{n_{m\tau} + \delta_{m\tau}\}) =
    E(\{n_{m\tau}\}) + \sum_{p\sigma}
    \frac{\partial E(\{n_{m\tau}\})}{\partial n_{p\sigma}} \delta_{p\sigma}
    + \frac{1}{2} \sum_{p\sigma}
    \frac{\partial^2 E(\{n_{m\tau}\})}{\partial n_{p\sigma}^2} \delta_{p\sigma}^2
    + \mathcal{O}(\delta_{p\sigma}^3),
    \label{eqn:E_N_expand}
\end{align}
where the partial derivatives are evaluated at
$\delta_{p\sigma}=0$, and $\mathcal{O}(\delta_{m\tau}^3)$
contains all terms with order $\delta_{m\tau}^3$ or higher.

In Eq.~\ref{eqn:E_N_expand}, we omit the cross terms in
the second order, which will be discussed in the Supporting Information.
This reason for the omission here is that we only need to consider changing one specific
occupation number at a time.
Particularly, consider only one specific orbital, $\psi_{p\sigma}$,
with a fractional occupation at a time, while
the rest of orbitals are fully occupied with
$\{n_{i\tau}=1\}, i\neq p$ or unoccupied with
$\{n_{a\tau}=0\}, a\neq p$. Denote this set of occupations
as $[n_{p\tau}]$ and its energy as $E([n_{p\tau}])$.
We now apply Eq.~\ref{eqn:E_N_expand} to the system
with an integer number of electrons,
with $n_{p\tau} + \delta_{p\tau}=1$ for the $N+1$-electron
system and with $n_{p\tau}+\delta_{p\tau}=0$ for the $N$-electron
system, and obtain the total energies of the two integer systems
up to the second order with the error in
$\mathcal{O}(\delta_{p\sigma}^3)$.
Substituting
the results energies of integer systems with
the truncation at the second order into Eq.~\ref{eq:gsc_deltaE_2},
we obtain the energy correction
from the GSC associated with the orbital $\psi_{p\sigma}$ as
\begin{align}
    \Delta_{\rm{GSC}}([n_{p\tau}]) = \frac{1}{2}
        \frac{\partial^2 E(\{n_{m\tau}\})}{\partial n_{p\sigma}^2}
        \left ( n_{p\tau} - n_{p\tau}^2\right),
\end{align}
which is equally valid for the entire range of occupation,
$0 \leq n_{p\tau} \leq 1$, with the
the partial derivative evaluated \emph{locally} at the
corresponding occupation number $n_{p\tau}$. The detailed derivation
is given in the Supporting Information.

Summing up contributions from all the orbitals, we have
\begin{align}
    \label{eq:exact_kappa_ni}
    \Delta_{\text{GSC}}(\{n_{p\sigma}\})
    & = \sum_{p\sigma} \frac12
        \frac{\partial^2 E(\{n_{m\tau}\})}{\partial n_{p\sigma}^2}
    (n_{p\sigma} - n_{p\sigma}^2).
\end{align}
Eq.~\ref{eq:exact_kappa_ni} gives the exact second-order correction for the GSC
method with the coefficients evaluated as the second
order derivative of the energy with respect to occupation numbers.

Compared to the original GSC, the new formalism in this work, Eq.~\ref{eq:exact_kappa_ni},
naturally involves both
the exchange and correlation contribution in $E_{\rm{xc}}$
in the corrections, applicable to all commonly used DFAs,
which was not achieved in previous works.
Another difference is that the energy function expansion in this work
is directly expanded on occupation numbers,
while the original GSC indirectly use the relaxation
of the density matrix.
As shown in the Supporting Information,
both approaches produce identical results as long as everything is
dealt precisely up to the second order.
Using the occupation numbers as the direct variables has the advantage of clearer
definition and cleaner equations obtained.
Using the density matrix as the variable will require the second-order
density matrix relaxation, which is also shown
to require first-order orbital derivatives only (in contrast to
requiring second-order orbital derivatives in previous developments
\cite{2015zhang154113,2020yangc}).
The reason is that the second-order density matrix relaxation part
cancels with other parts to give zero net contributions.
(See the Supporting Information for details.)

In addition, we emphasize that the only approximation for the
new formalism shown in Eq.~\ref{eq:exact_kappa_ni}
is the truncation of the energy function at the second order.
Because the energy function $E(\{n_{p\sigma}\})$ has been known to be
very close to quadratic,
\cite{2005cococcioni035105,
2011zheng026403a,2015li053001a,
2017bajaj191101,
2018li215,
2018hait6288,
2020su1535} such a treatment is clean and reasonable,
which is a great advantage compared to the approximate correction
derived previously.
This new correction from Eq.~\ref{eq:exact_kappa_ni}
also surpasses the previous approximate ones for the case, in which
the energy function from a DFA is exactly quadratic.
Under this condition, the coefficient
$\frac{\partial^2 E(\{n_{m\tau}\})}{\partial n_{p\sigma}^2}$,
from the exact second-order correction
shows the correct behavior, that is
being a constant in $[N, N+1]$. However, the approximate one
varies when the electron number changes, because it is evaluated from
canonical orbitals that are from each $(N+n)$-electron systems.
In summary, the GSC with the exact second order correction shown
in Eq.~\ref{eq:exact_kappa_ni} is
capable of restoring the correct linear behavior exactly
when the DFA produces a quadratic error
in fractional charge, and approximately when the error
deviates from a quadratic behavior.

Note that the second-order derivatives of the energy with respect to
frontier orbital occupation numbers have already been well-established.
\cite{2012yang144110}
They are recognized as the chemical hardness generated from DFAs,
\cite{2012yang144110}
and an extension to fractional occupation numbers was also made.
\cite{2013peng184108}
By using the energy function as defined in Eq.~\ref{eqn:frac_energy},
with similar procedures through the Maxwell relationship and linear
response theory, we show that
the general second-order derivative of the energy with respect to any occupation number is
\begin{align}
    \frac{\partial^2 E(\{n_{m\tau}\})}{\partial n_{p\sigma}^2}
    \label{eq:gsc_eval_kxk}
    &=\langle \psi_{p\sigma} \psi_{p\sigma}^* | K^{\sigma\sigma} +
        \sum_{\tau\upsilon} K^{\sigma\tau} \chi^{\tau\upsilon} K^{\upsilon\sigma} |
        \psi_{p\sigma} \psi_{p\sigma}^*\rangle\\
    &=
    K_{pp\sigma, pp\sigma}
    - \sum_{ia\tau,jb\upsilon}
    \left( K_{pp\sigma, ia\tau} + K_{pp\sigma, ai\tau} \right)
    M^{-1}_{ia\tau,jb\upsilon}
    K_{jb\upsilon,pp\sigma},
    \label{eq:gsc_eval}
\end{align}
where $\chi^{\tau\upsilon}$ is the generalized linear response function,
\cite{2012yang144110,2013peng184108}
$K^{\sigma\tau}$ represents Hartree-exchange-correlation kernels,
defined as the functional derivative of \(H_{\text{s}}^{\sigma}\),
the (generalized) Kohn-Sham Hamiltonian for spin \(\sigma\) with
respect to \(\rho_{\text{s}}^{\tau}\), the Kohn-Sham density matrix for
spin \(\tau\):
\begin{align}
    K^{\sigma\tau} (\mathbf{r}_1, \mathbf{r}_2; \mathbf{r}_3, \mathbf{r}_4)
    &= \frac{\delta H_{\text{s}}^{\sigma}(\mathbf{r}_1, \mathbf{r}_2)}
            {\delta \rho_{\text{s}}^{\tau}(\mathbf{r}_3, \mathbf{r}_4)} \\
    &=
        \frac{\delta(\mathbf{r}_1, \mathbf{r}_2)
        \delta(\mathbf{r}_3, \mathbf{r}_4)}
        {|\mathbf{r}_1 - \mathbf{r}_3|}
        + \frac{\delta^2 E_{\rm{xc}}}
        {\delta \rho_{\text{s}}^{\sigma}(\mathbf{r}_2, \mathbf{r}_1)
         \delta \rho_{\text{s}}^{\tau}(\mathbf{r}_3, \mathbf{r}_4)},
\end{align}
in which $E_{\mathrm{xc}}$ is the exchange-correlation energy.
$K_{pp\sigma,qq\tau}$ represents the corresponding kernel matrix
\cite{1999hirata382}
\begin{align}
 K_{pq\sigma,mn\tau} &= \langle \psi_{p\sigma}\psi_{q\sigma}^* | K^{\sigma\tau} |
    \psi_{m\tau} \psi_{n\tau}^* \rangle \\
    &= \iiiint \dd \mathbf{r}_1 \dd \mathbf{r}_2
    \dd \mathbf{r}_3 \dd \mathbf{r}_4
    \psi_{p\sigma}^*(\mathbf{r}_1) \psi_{q\sigma}(\mathbf{r}_2)
    K^{\sigma\tau}(\mathbf{r}_1, \mathbf{r}_2; \mathbf{r}_3, \mathbf{r}_4)
    \psi_{m\tau}(\mathbf{r}_3) \psi_{n\tau}^*(\mathbf{r}_4),
\end{align}
and the matrix element $M_{ia\tau,jb\upsilon}$ is defined as
\begin{align}
    M_{ia\tau,jb\upsilon} = \delta_{\upsilon\tau}\delta_{ij}\delta_{ab}
    (\epsilon_{a\upsilon} - \epsilon_{i\upsilon}) + K_{ia\tau,jb\upsilon}
    + K_{ia\tau,bj\upsilon}.
\end{align}
The detailed derivation for the second-order derivative
of total energy with respect to occupation number can be found in the Supporting Information.
In connection to the chemical hardness, we here call
$\frac{\partial^2 E(\{n_{n_{m\tau}}\})}{\partial n_{p\sigma}^2}$
the \emph{orbital hardness}.
It is noteworthy that solving coupled perturbed Kohn-Sham equations can give
the same answer, which is also shown in the Supporting Information.

We now examine the physical meaning of the orbital hardness
(shown in the Supporting Information with details).
The associated 4-point generalized dielectric function \(\varepsilon^{\sigma\tau}\)
can be defined as
\begin{equation}
    \left(\varepsilon^{-1}\right)^{\sigma\tau}(\mathbf{r}_1, \mathbf{r}_2; \mathbf{r}_3, \mathbf{r}_4)
    =
    \frac{\delta H_{\text{s}}^{\sigma}(\mathbf{r}_1, \mathbf{r}_2)}
         {\delta v^{\tau}(\mathbf{r}_3, \mathbf{r}_4)},
\end{equation}
where \(v^{\tau}(\mathbf{r}_3, \mathbf{r}_4)\) is the generalized external potential,
which is nonlocal and spin-dependent, as an extension from the physical potential,
\(v(\mathbf{r})\), which is local and spin independent, first introduced in
Ref.~\cite{2013peng184108}. Then, the expression of Eq.~\ref{eq:gsc_eval_kxk}
for the orbital hardness leads to its interpretation as the interaction of orbitals
through the 4-point generalized screened interaction
\begin{equation}
    W^{\sigma\varsigma} = K^{\sigma\varsigma} + \sum_{\tau\upsilon} K^{\sigma\tau}
    \chi^{\tau\upsilon} K^{\upsilon\varsigma}
    = \sum_{\upsilon}\left(\varepsilon^{-1}\right)^{\sigma\upsilon} K^{\upsilon\varsigma}.
\end{equation}
The 4-point generalized functions, \(\chi^{\tau\upsilon}\), \(\left(\varepsilon^{-1}\right)^{\sigma\upsilon}\)
and \(W^{\sigma\varsigma}\) are the natural extensions of the corresponding spinless
two-point functions commonly used in many-body perturbation theory. \cite{martin2016interacting}

Note that the matrix $M_{ia\tau,jb\upsilon}$ in Eq.~\ref{eq:gsc_eval}
involves all the pairs of occupied and virtual orbitals, which gives the
dimension of
$\sum\limits_\tau N_{\rm{occ}}^\tau \times N_{\rm{vir}}^\tau$,
with $N_{\rm{occ}}^\tau$ being the number of occupied orbitals
and $N_{\rm{vir}}^{\tau}$ being the number of
virtual orbitals for $\tau$ spin. Therefore, directly
evaluating Eq.~\ref{eq:gsc_eval}
has the computational complexity of $\mathcal{O}\left((N_{\rm{occ}}N_{\rm{vir}})^3\right)$.
If only a few orbital energies, like the frontier orbitals, are of
interest in practice, one can evaluate the second order derivative
numerically to bypass the analytical expression, which is a shortcut to reduce
the computation to be several times of SCF calculations,
or one can use efficient iterative solution to linear equations
of the coupled perturbed Kohn-Sham approach, instead of matrix
inversion for the associated orbitals, which has lower computational
complexity of $\mathcal{O}\left((N_{\rm{occ}}N_{\rm{vir}})^2\right)$.

In the following, we present the results
to show the performance of the new expression for the GSC
with the analytical and exact second-order corrections (denoted as GSC2).
The original GSC
\cite{2011zheng026403a}
with the approximate coefficients $\kappa_{p\sigma}$
is denoted as GSC1.
Because the effective Hamiltonian from the GSC method for integer systems
is a projection operator consisting of canonical orbitals from the
associated DFA, the GSC method does not change the
canonical orbitals, the eigenstates for the associated DFA Hamiltonian,
and applying the GSC method with a post-SCF or SCF manner produces
identical results. Therefore, all the calculations of GSC1 and GSC2
are performed with the post-SCF calculation for the purpose of efficiency.
Details for the calculations, numerical
results and more clarification to the SCF calculations of the GSC method
are documented in the Supporting Information.

\begin{figure}[htbp]
    \centering
    \subfigure[$E(N)$]
    {\includegraphics[width=0.48\linewidth]{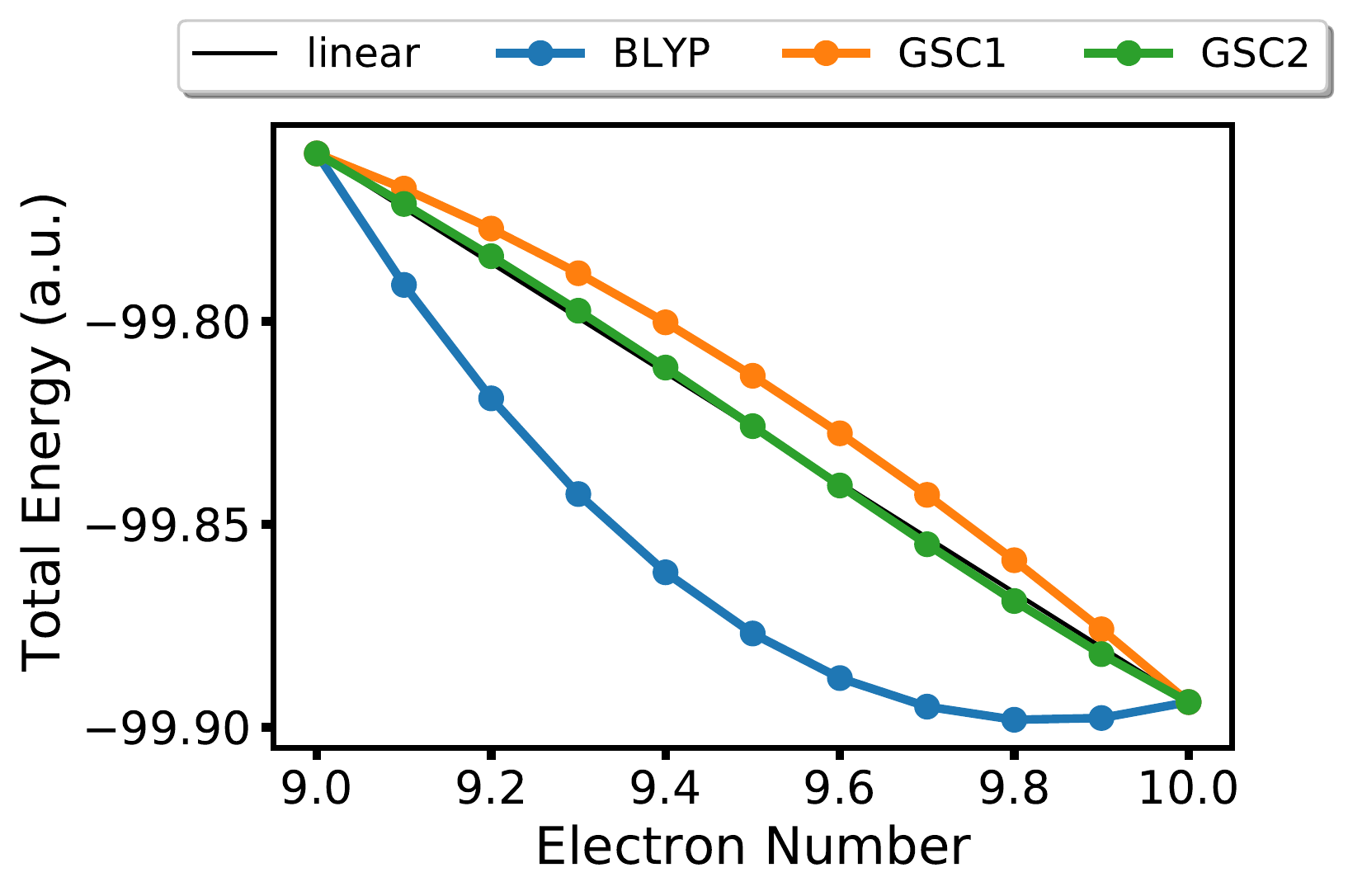}}
    \subfigure[$\Delta E(N)$]
    {\includegraphics[width=0.48\linewidth]{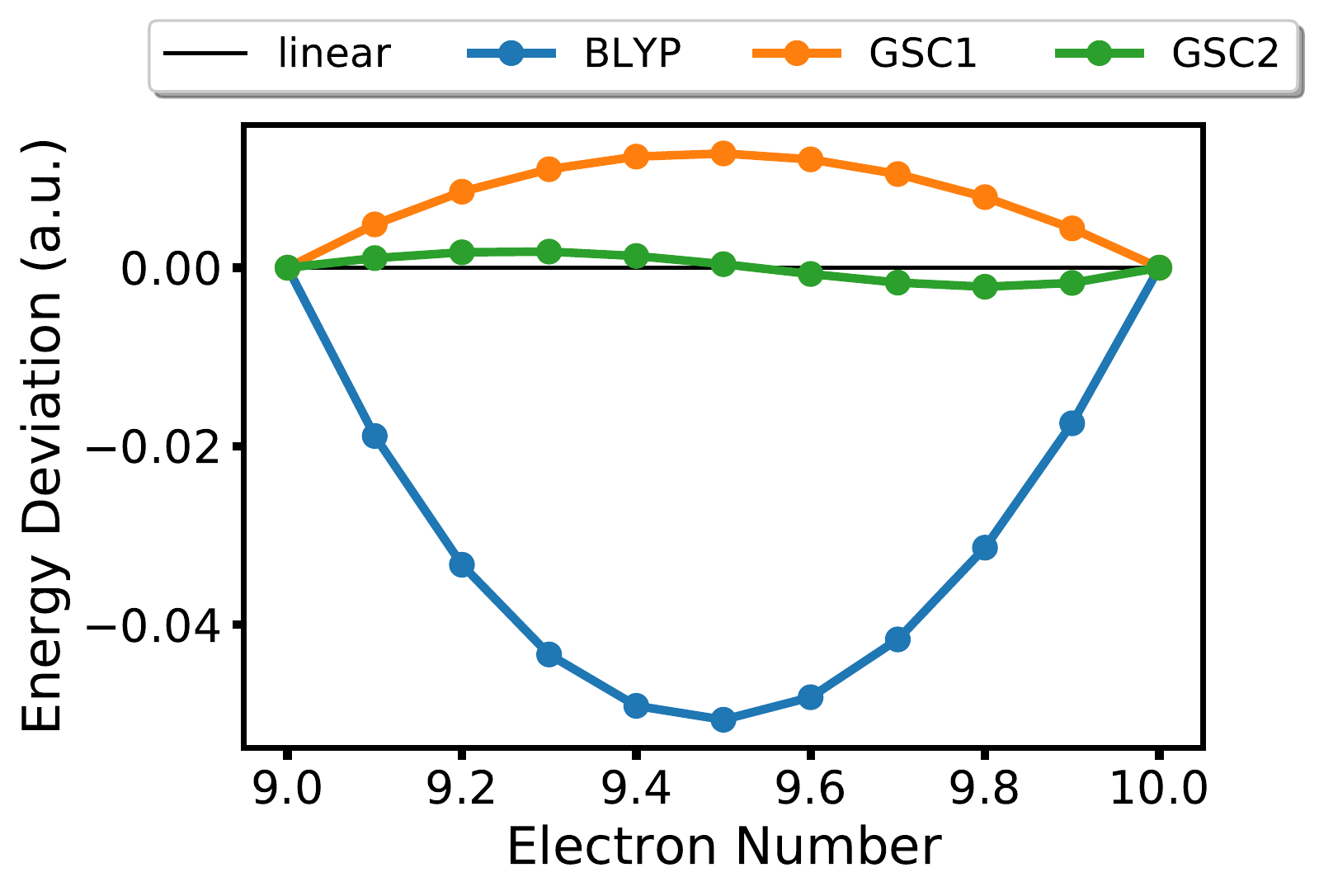}}
    \caption{Total energies of F atom from BLYP, GSC1-BLYP and GSC2-BLYP:
        (a) total energy versus electron number; (b) total energy
        difference versus electron number,
        $\Delta E(N) = E_{\rm{DFA}}(N) - E_{\rm{linear}}$.
        aug-cc-pVTZ is used as the basis set.
    }
    \label{fig:EN_curve}
\end{figure}

The $E(N)$ curve comes to our attention first as
the goal of the GSC method is to make it linear.
Figure~\ref{fig:EN_curve} shows the $E(N)$ curve
of the F atom with BLYP as the parent DFA for
the calculations of GSC1 and GSC2. The BLYP functional shows great delocalization
errors, while the GSC1-BLYP with the approximate correction
over-corrected and showed the localization errors.
With the exact second-order correction, GSC2-BLYP is capable of
producing mostly linear behavior for the $E(N)$ curve
with much smaller errors.

The first IPs and EAs are part of the main outcome from the GSC method,
as for integer systems only orbital energies get corrected.
The quality of IPs and EAs can reflect the performance of
the correction, and thus worthwhile to be examined.
The first IP and EA of an $N$-electron system can be evaluated with
the negative HOMO and LUMO energies of the $N$-electron system,
which are associated with one-electron removal and one-electron addition
processes respectively. \cite{2008cohen115123}
For the purpose of comparison, the $\Delta$-SCF method
is conducted as well to evaluate the first IPs and EAs, in which
the energy differences are calculated, i.e.,
$E_{\rm{IP}} = E(N-1) - E(N)$ for IPs and
$E_{\rm{EA}} = E(N) - E(N+1)$ for EAs.
The test sets for the first IPs and EAs are taken from Ref.~\citenum{2020mei10277}.
Experimental data are used as the reference to evaluate
the mean absolute error (MAE).

\begin{figure}[htbp]
    \centering
    \subfigure[IP]
    {\includegraphics[width=0.48\linewidth]{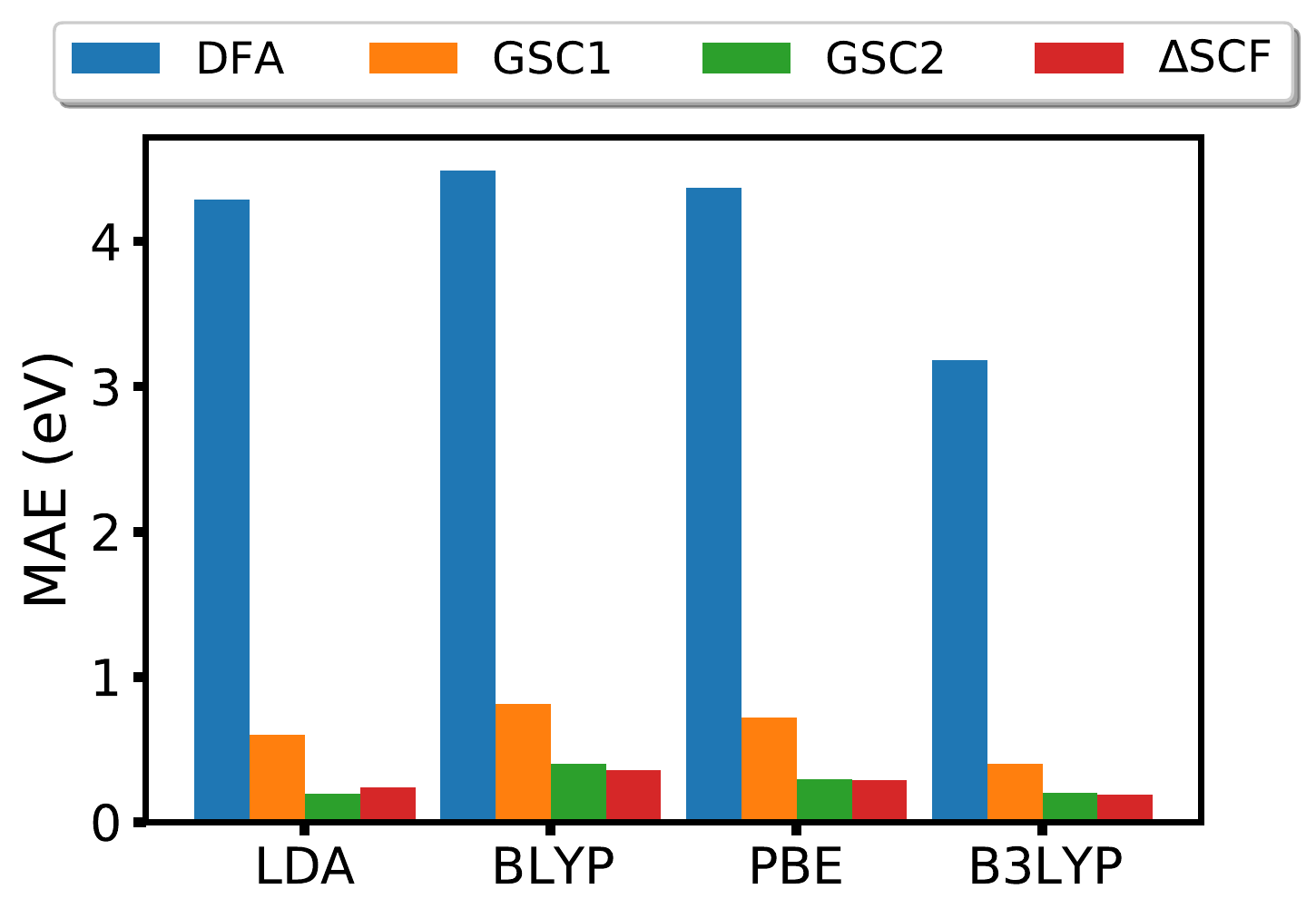}}
    \subfigure[EA]
    {\includegraphics[width=0.48\linewidth]{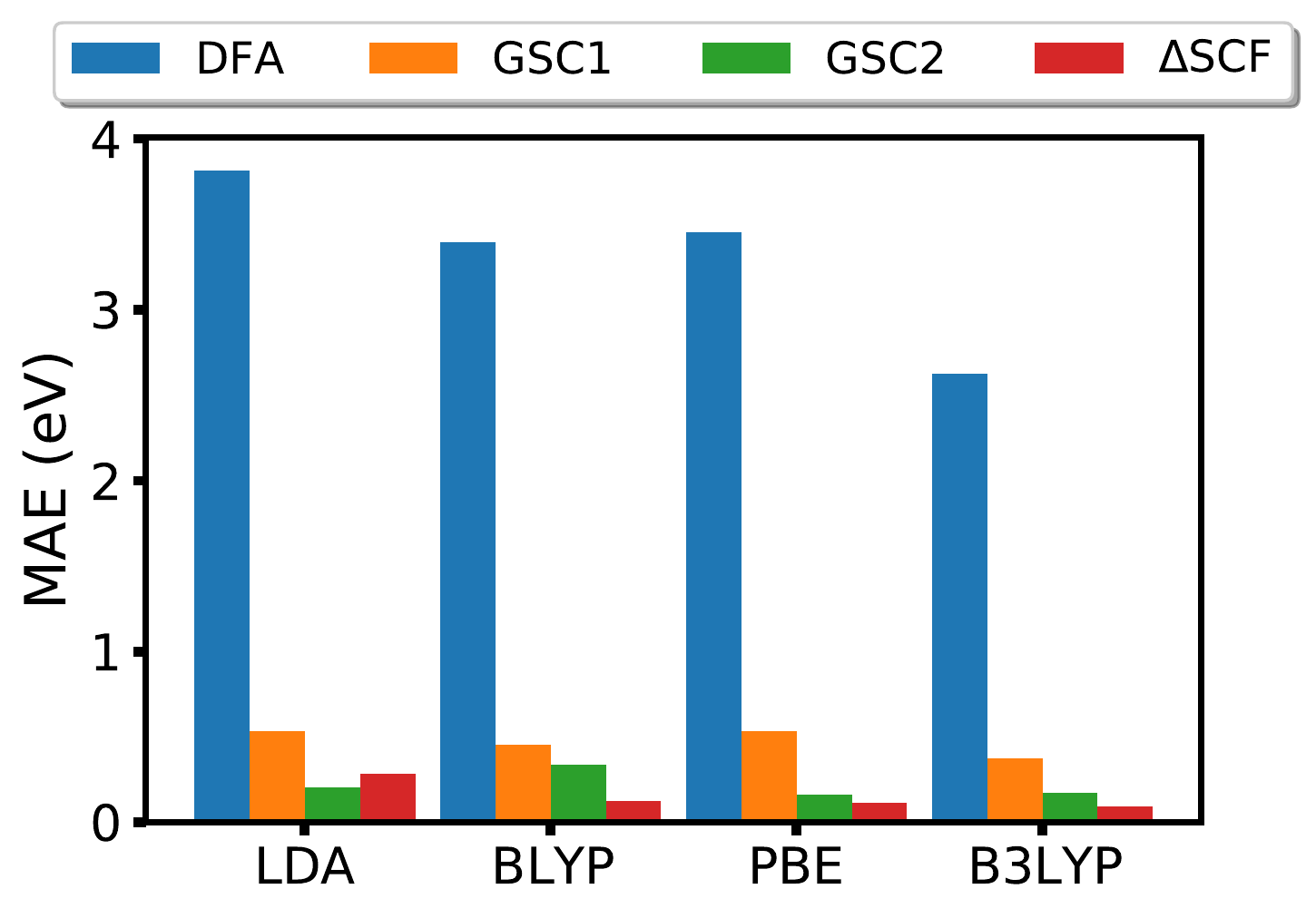}}
    \caption{The mean absolute error in eV for the first (a) IPs (64 cases)
        and (b) EAs (39 cases) obtained from different methods.
    }
    \label{fig:IP_EA}
\end{figure}

According to Figure~\ref{fig:IP_EA},
both GSC1 and GSC2
greatly improve the quality of the first IPs and EAs upon associated parent DFAs,
and the GSC2 apparently outperforms the approximate GSC1.
Taking B3LYP as an example,
the MAE of the first IPs/EAs predicted from the calculations
of $N$-electron systems
is 0.20/0.17 eV for GSC2 and 0.41/0.38 eV for GSC1,
where the errors get nearly halved in the new approach GSC2.
In addition, the GSC2 produces
results that are close to those from the $\Delta$-SCF method, while
it is not the case for the approximate GSC1 approach.
This indicates that the GSC method with exact second-order
derivatives has a more systematic and accurate correction
to the DFA \(E(N)\)
curve, producing mostly linear curves for all these systems
and thus good IPs/EAs.

Besides evaluating the first IP and EA of an $N$-electron system from
the negative HOMO and LUMO energy of an $N$-electron system, one can also approximate
the first IP with the negative LUMO energy of the $(N-1)$-electron system
(associated with an one-electron addition process),
and the first EA with the negative HOMO energy of the $(N+1)$-electron system
(associated with an one-electron removal process). Similar results are
observed from these two approaches and the detailed results are shown
in the Supporting Information.


\begin{figure}[htbp]
    \centering
    \subfigure[MAEs of different DFAs]
    {\includegraphics[width=0.48\linewidth]{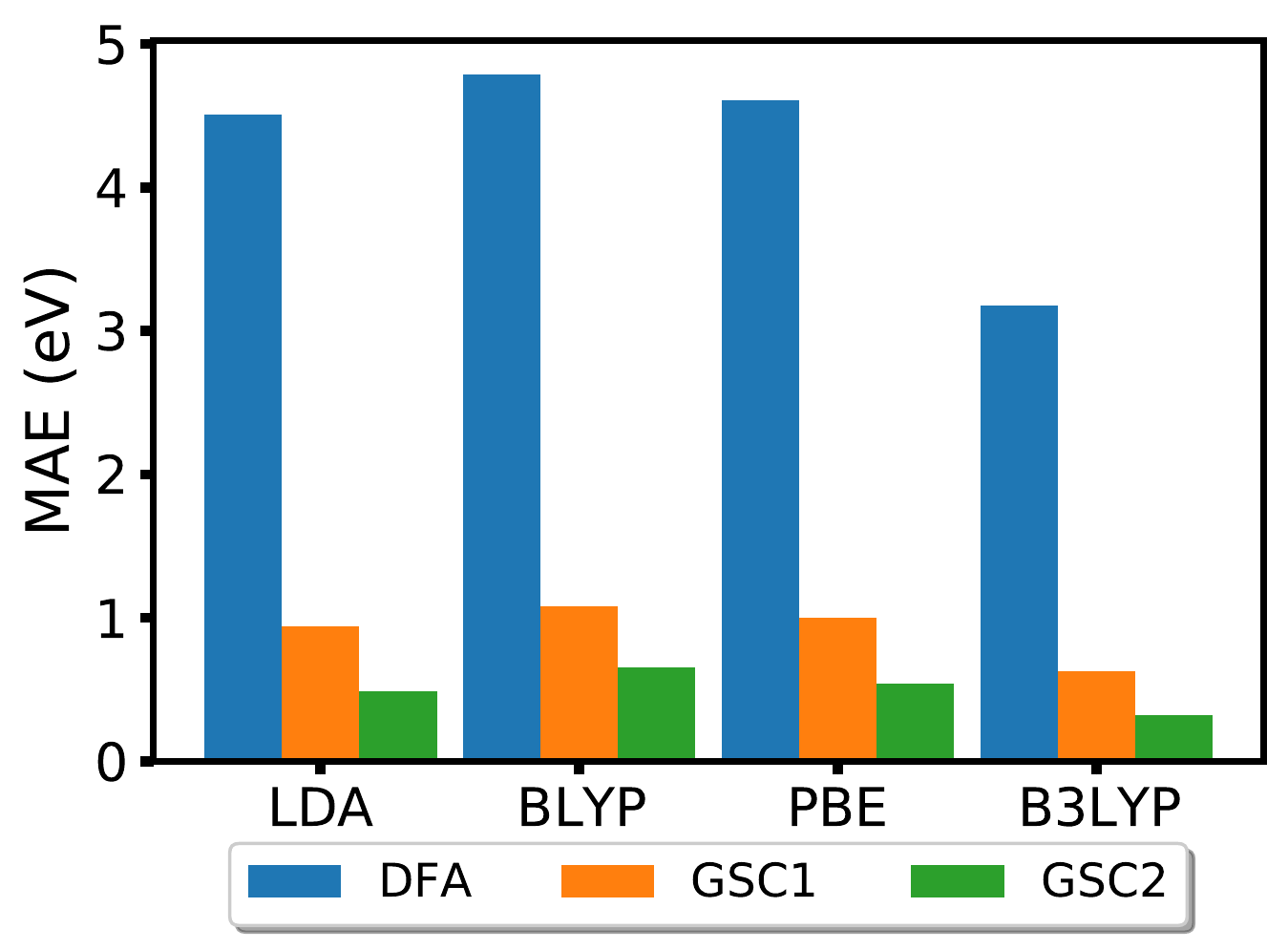}}
    \subfigure[B3LYP]
    {\includegraphics[width=0.50\linewidth]{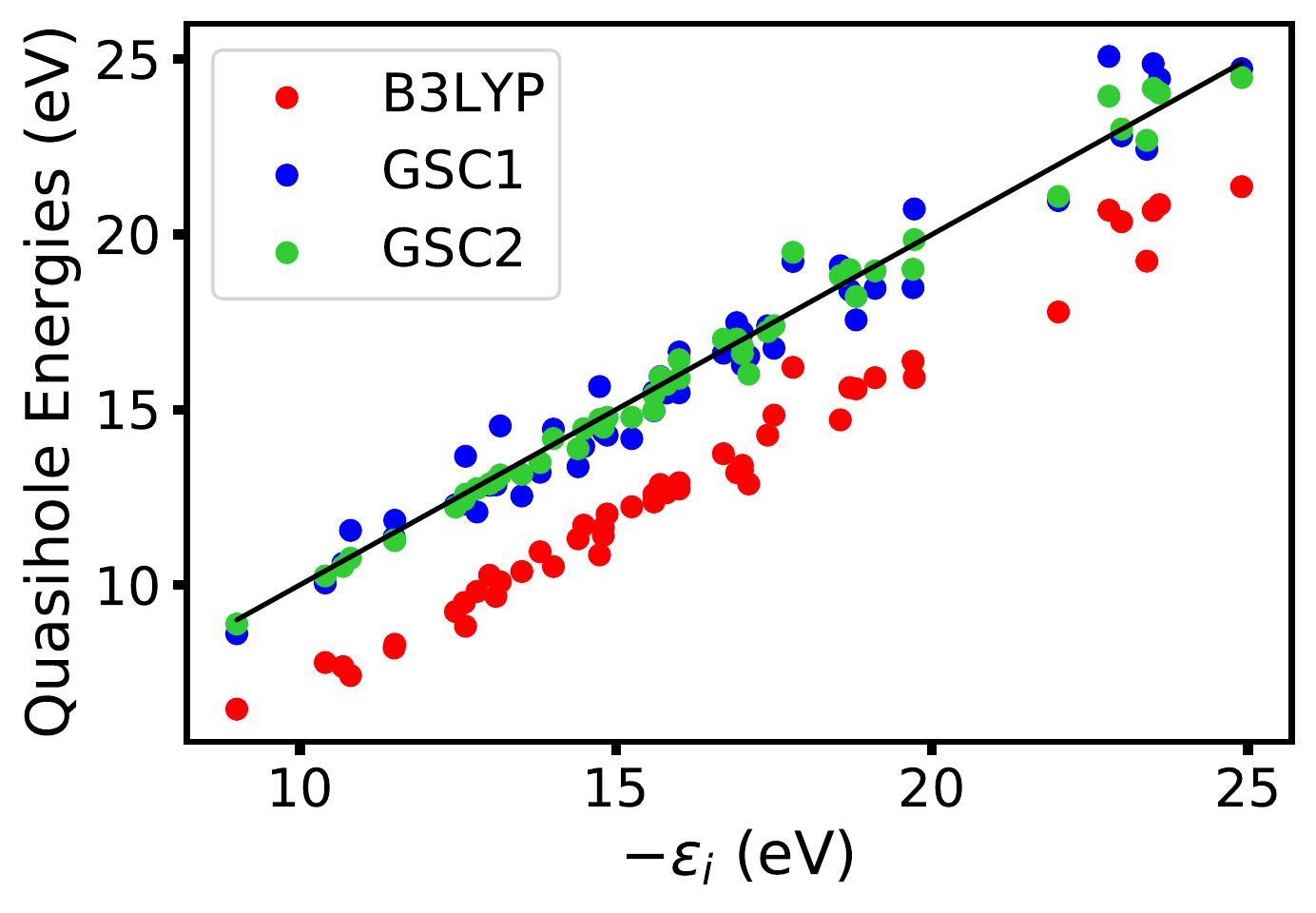}}
    \caption{
        Quasihole energies (52 cases from 11 small molecules)
        of the $N$-electron systems evaluated by the corresponding
        negative occupied orbital energies ($-\epsilon_i$) from different
        DFAs:
        (a) the MAEs for different DFAs, GSC1-DFAs and GSC2-DFAs.
        (b) Experimental quasihole energies versus the
        negative orbital energies from B3LYP, GSC1-B3LYP and
        GSC2-B3LYP.
    }
    \label{fig:quasihole}
\end{figure}

Next, we examine the prediction of other quasiparticle energies besides the
first IPs and EAs.
The PPLB linearity condition defines for the first IPs and EAs
only, it is still worthy to investigate quasihole
or quasiparticle energies other than the HOMO or LUMO.
A parallel extension to quasihole/quasiparticle
energies with orbital energies other than the HOMO or LUMO has been
assumed and applied with numerical success. \cite{2019mei673}
These quasihole energies are predicted
by the negative orbital energies of the corresponding
occupied orbitals of the $N$-electron systems from DFT calculations.
We use the same set of molecules as in Ref.~\citenum{2020yangc}
to test the performance of the GSC method.
Experimental data are used as references.
As shown in Figure~\ref{fig:quasihole},
conventional DFAs produce significant errors, while
the GSC method corrects the orbital energies to have an MAE that is below 1\,{eV}.
In addition, GSC2 outperforms GSC1 as expected.
For example, the MAE from GSC1 is 0.63\,{eV}, while
GSC2 gives 0.32\,{eV}.

\begin{figure}[htbp]
    \centering
    \subfigure[Maleic anhydride]
    {\includegraphics[width=0.48\linewidth]{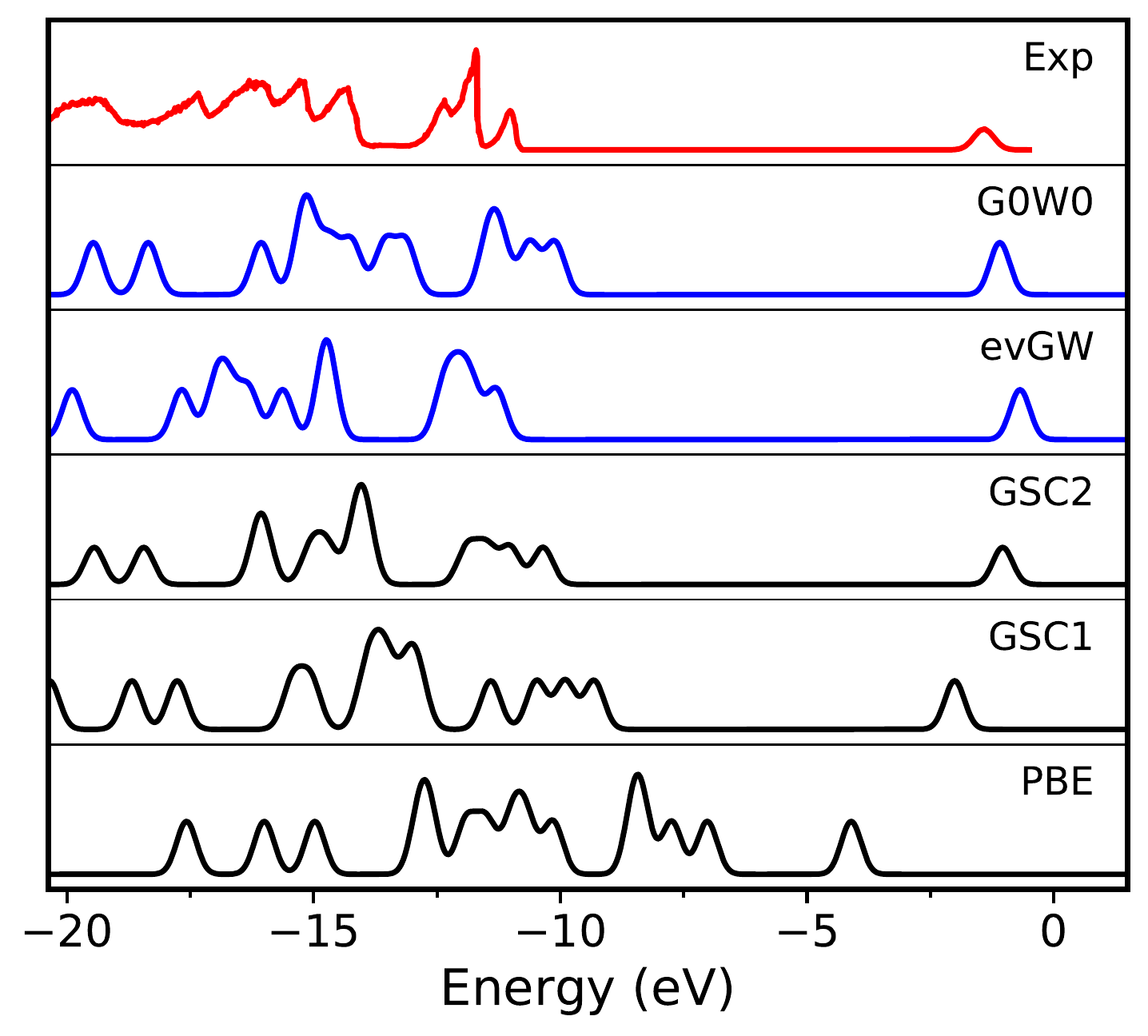}}
    \subfigure[Benzonitrile]
    {\includegraphics[width=0.48\linewidth]{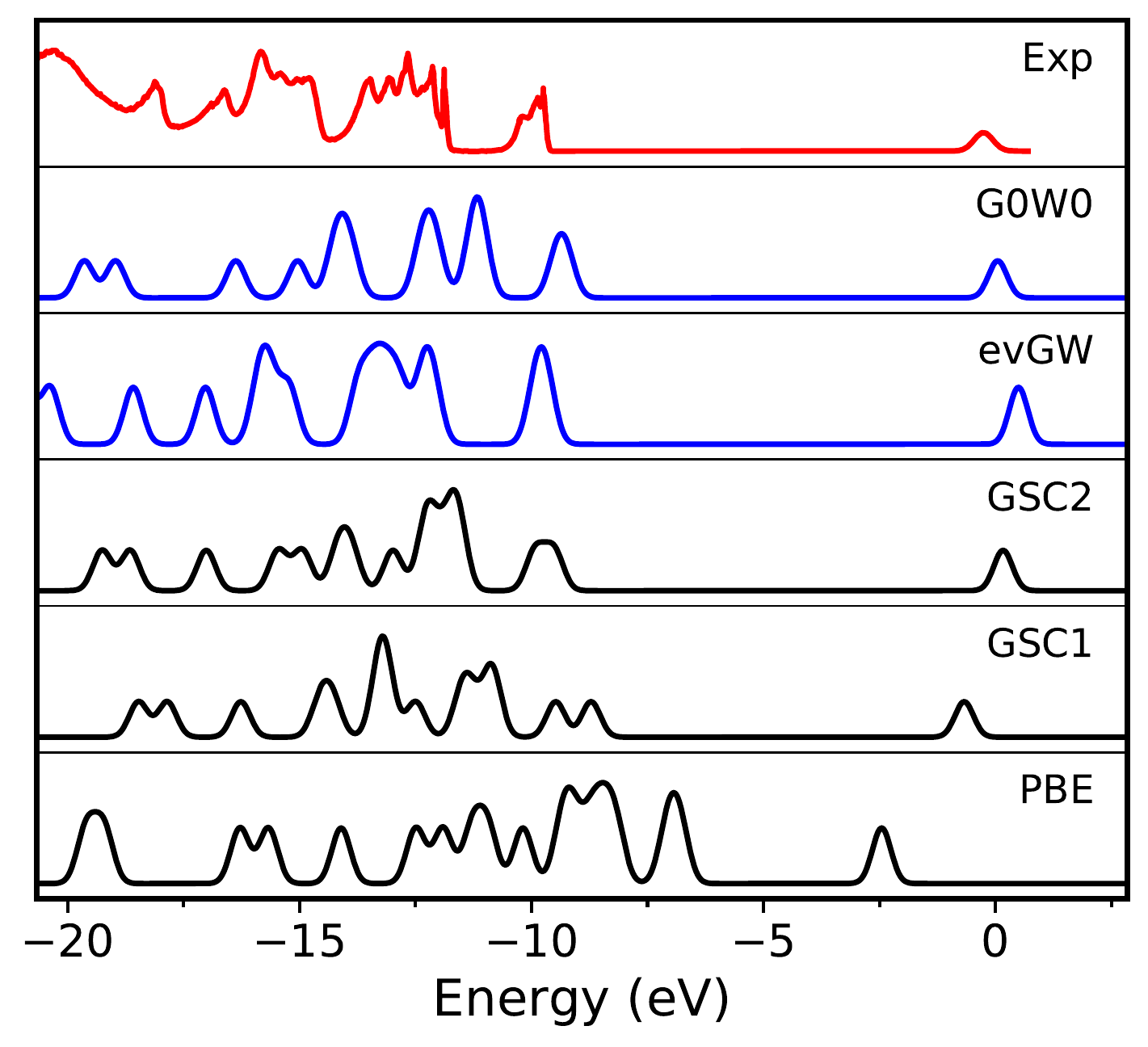}}
    \caption{Photoemission spectrum for (a) maleic anhydride and
    (b) benzonitrile.
    Experimental spectrum was obtained from Ref.~\citenum{2016knight626}
    for (a) and Ref.~\citenum{1981kimura} for (b).
    Calculations for GSC1, GSC2 and $GW$ were associated with PBE
    \cite{1996perdew3868} functional.
    }
    \label{fig:Spec}
\end{figure}

The photoemission spectra is another good source to
evaluate the quality of orbital energies.
We select 10 organic molecules with
small or moderate sizes from Ref.~\citenum{2019mei673}
to test. Experimental spectra are used as the reference.
Besides the DFT calculations,
we also conduct $GW$ ($G_0W_0$)
and eigenvalue self-consistent $GW$ (ev$GW$) calculations
to obtain the quasiparticle energies.
The calculated spectra is obtained from the Gaussian expansion
of calculated quasiparticle energies with a standard deviation of
0.2\,{eV}. Figure~\ref{fig:Spec} shows the results for
two representative molecules, maleic anhydride and benzonitrile.
It can be seen that GSC2 consistently outperforms the approximate GSC1,
as the peaks align better to
experimental references and have good agreement with those from
GW calculations.
The spectra of additional molecules tested can be found
in the Supporting Information.

\begin{figure}[htbp]
    \centering
    \includegraphics{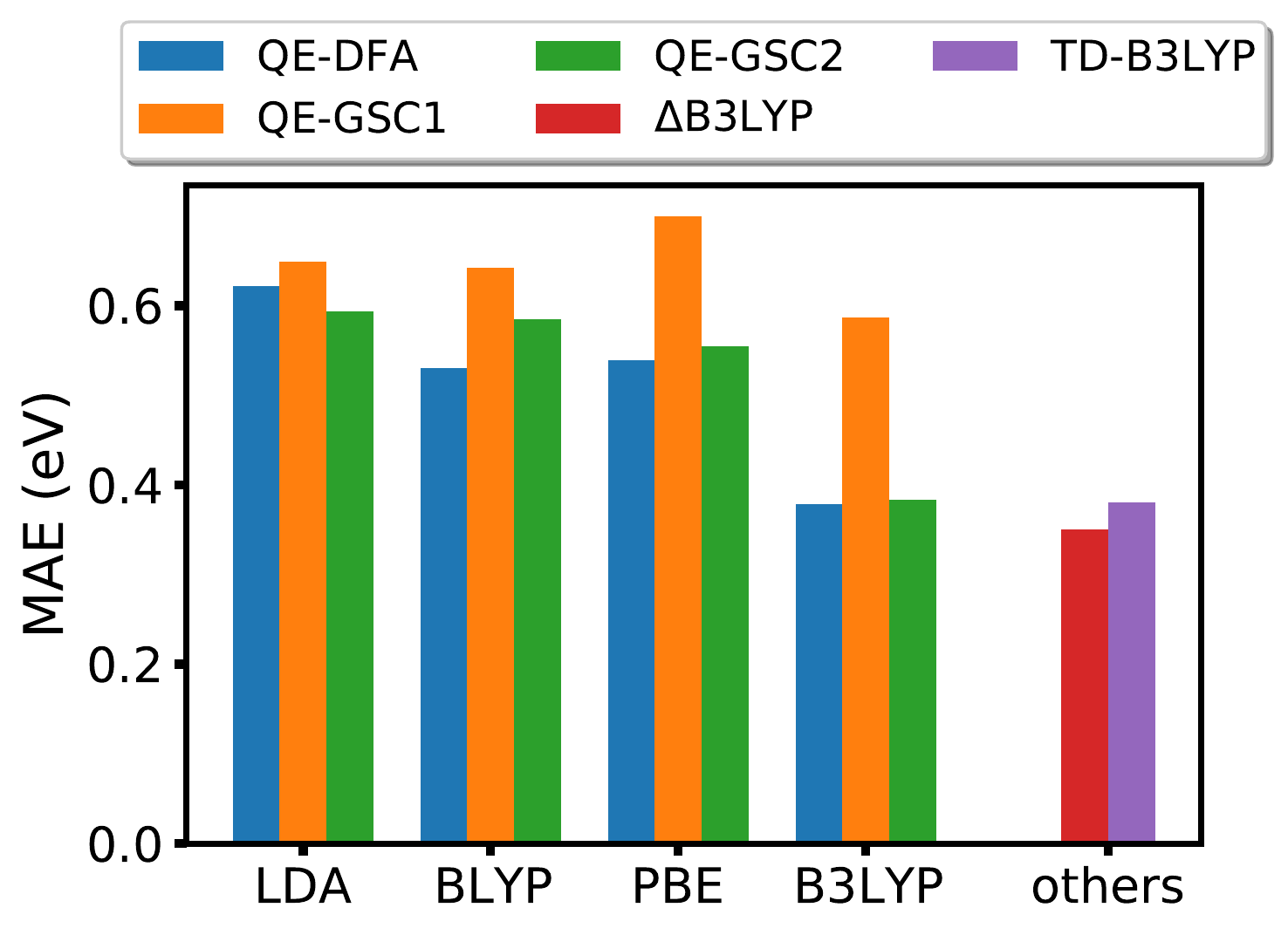}
    \caption{The mean absolute error (in eV) for the low-lying excitation
        energies from various QE-DFAs, $\Delta$-B3LYP and TD-B3LYP.
        The excitation includes 19 cases of triplet states and
        25 cases of singlet states.
        The reference data that were calculated from high-level
        wavefunction methods, and results for $\Delta$-B3LYP
        and TD-B3LYP were taken from Ref.~\citenum{2019mei673}.
    }
    \label{fig:exci_bar}
\end{figure}

The benchmark for low-lying excitation energies
from the QE-DFT method \cite{2019mei673,2019mei2545} is also carried out.
The excitation energy in QE-DFT is computed as the difference between two
corresponding (generalized) Kohn-Sham orbital energies.
We use the test set provided by Ref.~\citenum{2019mei673} to calculate
the first and second singlet and triplet excitation energies.
The results
from the $\Delta$-SCF method and time-dependent DFT
with B3LYP functional
are taken from Ref.~\citenum{2019mei673} for comparison.
As shown in Figure~\ref{fig:exci_bar},
the QE-GSC2 produces the improved excitation energies compared
to QE-GSC1. In particular, the results from QE-GSC2 are
comparable to those from $\Delta$-SCF-B3LYP.
In addition, GSC2-B3LYP shows comparable
performance to TD-B3LYP.
This shows a potential application for the original GSC
and the GSC2 form developed
in this work, as multiple excited states
are immediately accessible after only one calculation.

Although there are great improvements from the GSC method to the associated DFAs,
it should be kept in mind that the GSC method has its intrinsic limitations.
The delocalization error cannot be corrected effectively under
the framework of GSC method for systems with large size,
and there is no GSC correction at all for bulk systems,
which are clearly demonstrated with calculations for the
hydrogen chain and the helium cluster. \cite{2008mori-sanchez146401,2018li215}
For these scenarios, the LOSC method is designed by using the orbitalets
that can dynamically switch between the canonical orbitals and localized orbitals
to systematically reduce the delocalization error. \cite{2018li215,2020su1535}
Note that the formalism of LOSC is generalized based on the original GSC work
with the approximate corrections.
\cite{2018li215,2020su1535}
Therefore, if there is no localization
and the orbitalets are just the canonical orbitals, the
LOSC method would become the same as the original GSC approach. \cite{2018li215,2020su1535}

\begin{figure}[htbp]
    \centering
    \subfigure[GSC1 vs GSC2]
    {\includegraphics[width=0.48\linewidth]{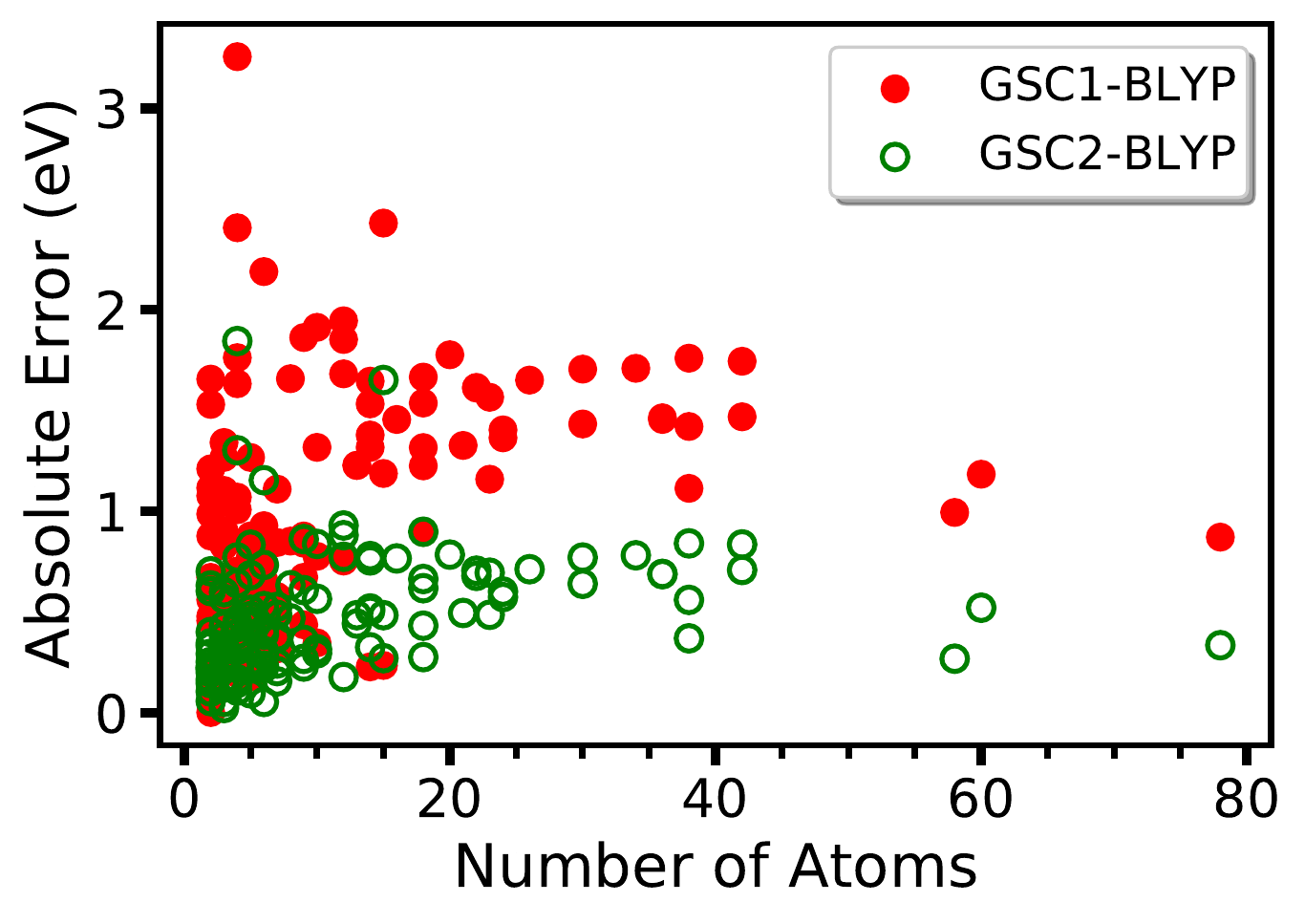}}
    \subfigure[GSC2 vs LOSC]
    {\includegraphics[width=0.50\linewidth]{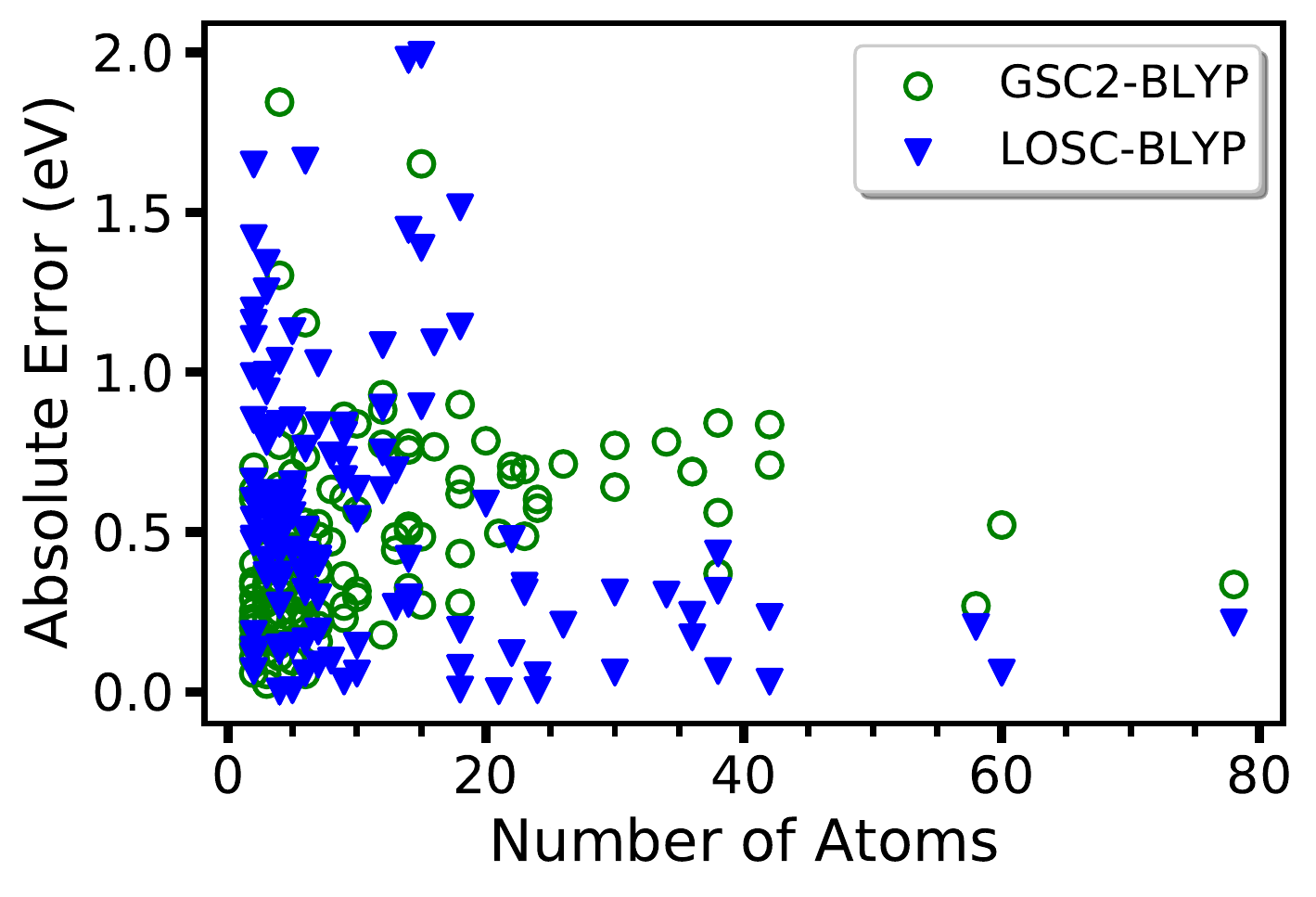}}
    \caption{The error distribution of first IPs obtained from GSC1, GSC2 and
    LOSC for molecules with different sizes.}
    \label{fig:size}
\end{figure}

To investigate the size dependence for the application of the GSC method,
we selected many real molecules with the equilibrium structures and
various sizes ranging from small systems, like water \ce{H2O}, to large systems,
like fullerene $\rm{C}_{60}$. Then we calculated the first IPs of these
molecules from GSC1, GSC2 and
the latest version of LOSC\cite{2020su1535}
for comparison.
The error distribution over the number of atoms is plotted in Figure~\ref{fig:size}.
According to Figure~\ref{fig:size}, we see GSC2 outperforms GSC1 regardless of
the molecular size, because of the exact second-order correction instead of the approximated one.
In addition, the comparison between GSC2 and LOSC clearly shows the size
dependence of the GSC method. For systems with less than about 20 atoms,
GSC2 shows better accuracy than LOSC. This is
because the orbitalets used in LOSC are more like canonical orbitals and
LOSC at these scenarios is close to the original GSC with approximate
second-order corrections. For systems larger than about 20 atoms, we observe the opposite
behavior that LOSC shows better accuracy.
This is because the LOSC at these scenarios characterizes the
delocalization error more effectively than the GSC method, in which
the orbitalets used in LOSC are more localized than
the canonical orbitals.
This observation also suggests that developing more accurate correction
in the scheme of LOSC, like the exact second-order correction in GSC2,
would further improve the performance of LOSC for large and complex systems.

In conclusion, we have developed further the global scaling correction
method with analytical and exact second-order energy corrections to better deal with
delocalization errors existing in conventional DFAs. With the application
of the exact second-order correction in the GSC method, we demonstrated
the excellent performance of the GSC2 approach to describe the first
IPs, EAs, other quasihole/quasiparticle energies and low-lying excitation energies,
which are all obtained from accurate (generalized) KS orbital energies
in ground state calculations.

\begin{acknowledgement}
Y. M. and Z.C. acknowledge the support from the National Institute
of General Medical Sciences of the National Institutes of Health under
award number R01-GM061870. W.Y. acknowledges the support from the
National Science Foundation (grant no. CHE-1900338). Y. M. was also
supported by the Shaffer-Hunnicutt Fellowship and Z.C. by
the Kathleen Zielik Fellowship from Duke University.
\end{acknowledgement}

\begin{suppinfo}
Supporting Information Available: mathematical derivations
for the coupled perturbed equations,
alternative derivations for energy second-order derivative,
comparison between using density matrix as the direct variable
and using occupation numbers as the direct variable,
computational details, and numerical results.
\end{suppinfo}

\bibliography{DFT,GSC,FL_LOSC,my,qm4d,SI,add}
\end{document}


\newpage
\tableofcontents

\section{More clarification on the GSC method}
\subsection{Derivation of the correction from GSC}
Following the notation in the main text, we derive the correction
$\Delta_{\rm{GSC}}([n_{p\sigma}])$ associated with the specific orbital
$\psi_{p\sigma}$. According to Eq.~10 in the main text, we evaluate the total energies
of integer systems up to the second order.
Consider the set of occupations $[n_{p\sigma}]$,
with one specific orbital, $\psi_{p\sigma}$, with
a fractional occupation at a time,
while the rest of orbitals are fully occupied with
$n_{i\sigma}=1, i\neq p$ or unoccupied with
$n_{a\sigma}=0, a\neq p$. When the orbital is
fully occupied, $n_{p\sigma}+\delta_{p\sigma}=1$,
using Eq.~10 of the main text, we have the energy of
the corresponding $(N+1)$-electron system as
\begin{align}
    \nonumber
    E([n_{p\sigma}+\delta_{p\sigma}=1])
        &= E([n_{p\sigma}]) +
        \frac{\partial E(\{n_{m\tau}\})}{\partial n_{p\sigma}} \delta_{p\sigma}
        + \frac{1}{2}
        \frac{\partial^2 E(\{n_{m\tau}\})}{\partial n_{p\sigma}^2} \delta_{p\sigma}^2
        + \mathcal{O}(\delta_{p\sigma}^3) \\
        &= E([n_{p\sigma}]) +
        \frac{\partial E(\{n_{m\tau}\})}{\partial n_{p\sigma}} (1 - n_{p\sigma})
        + \frac{1}{2}
        \frac{\partial^2 E(\{n_{m\tau}\})}{\partial n_{p\sigma}^2} (1-n_{p\sigma})^2
        + \mathcal{O}\left((1-n_{p\sigma})^3\right),
\end{align}
and when the orbital is fully unoccupied,
$n_{p\sigma} + \delta_{p\sigma} = 0$, we have
the energy of the corresponding $N$-electron system as
\begin{align}
    \nonumber
    E([n_{p\sigma}+\delta_{p\sigma}=0])
        &= E([n_{p\sigma}]) +
        \frac{\partial E(\{n_{m\tau}\})}{\partial n_{p\sigma}} \delta_{p\sigma}
        + \frac{1}{2}
        \frac{\partial^2 E(\{n_{m\tau}\})}{\partial n_{p\sigma}^2} \delta_{p\sigma}^2
        + \mathcal{O}(\delta_{p\sigma}^3) \\
        &= E([n_{p\sigma}]) -
        \frac{\partial E(\{n_{m\tau}\})}{\partial n_{p\sigma}} n_{p\sigma}
        + \frac{1}{2}
        \frac{\partial^2 E(\{n_{m\tau}\})}{\partial n_{p\sigma}^2} n_{p\sigma}^2
        + \mathcal{O}\left(n_{p\sigma}^3\right).
\end{align}
Substituting resulting energies of integer systems with
the truncation at the second order into
Eq.~2 in the main text,
we have the energy correction
from the GSC associated with the orbital \(\psi_{p\sigma}\) as
\begin{align}
    \nonumber
    \Delta_{\text{GSC}}([n_{p\sigma}])
        &=\Big\{
        (1-n_{p\sigma})E([n_{p\sigma}+\delta_{p\sigma}=0])
        + n_{p\sigma} E([n_{p\sigma} + \delta_{p\sigma}=1])
        \Big\} - E([n_{p\sigma}])
        \\ \nonumber
    &= \frac12
        \frac{\partial^2 E(\{n_{m\tau}\})}{\partial n_{p\sigma}^2}
    \left[\left(1-n_{p\sigma}\right) n_{p\sigma}^2 + n_{p\sigma}\left(1-n_{p\sigma}\right)^2 \right] \\
    &= \frac12
        \frac{\partial^2 E(\{n_{m\tau}\})}{\partial n_{p\sigma}^2}
    \left(n_{p\sigma} - n_{p\sigma}^2\right),
    \label{eqn:correction}
\end{align}
which is Eq.~11 of the main text. Note that all the derivatives with
respect to the occupation number are evaluated
at the fractional occupation $[n_{p\sigma}]$.

\subsection{Comments on the self-consistent calculations with GSC}
Based on the energy correction, the correction to the effective
Hamiltonian from GSC is
\begin{align}
    \Delta H_{\rm{GSC}}^{\sigma} (\mathbf{r}, \mathbf{r}') =
    \frac{\delta \Delta_{\rm{GSC}}}{\delta \rho_{\rm{s}}^{\sigma}(\mathbf{r}', \mathbf{r})}.
\end{align}
Based on
\(
    \hat{\rho_{\rm{s}}}^{\sigma} = \sum_{p\sigma}n_{p\sigma}
|\psi_{p\sigma}\rangle \langle \psi_{p\sigma} |
\)
and hence
$
n_{p\sigma} = \langle \psi_{p\sigma}| \rho_{\rm{s}}^{\sigma} | \psi_{p\sigma} \rangle
$,
using Eq.~\ref{eqn:correction}, we have
\begin{align}
    \label{eq:gsc_H}
    \Delta H_{\rm{GSC}}^{\sigma}(\mathbf{r}, \mathbf{r}') =
        \frac{1}{2}\sum_{p\sigma} \Bigg[
        \kappa_{p\sigma} (1 - 2 n_{p\sigma})
        |\psi_{p\sigma} \rangle \langle \psi_{p\sigma} |
     + \frac{\partial \kappa_{p\sigma}}{\partial \rho_{\rm{s}}(\mathbf{r}', \mathbf{r})}
     (n_{p\sigma} - n_{p\sigma}^2)
     \Bigg].
\end{align}
The first term in Eq.~\ref{eq:gsc_H} is the projection over the
canonical orbitals, which does not change the eigenstates from the
DFA Hamiltonian. The second term in Eq.~\ref{eq:gsc_H} involves the
response of the coefficient $\kappa_{p\sigma}$ to the density matrix.
For integer systems, the second term in Eq.~\ref{eq:gsc_H} does not
contribute because $(n_{p\sigma} - n_{p\sigma}^2) = 0$. Therefore,
for integer systems, applying GSC with a post-SCF or SCF approach
produces the identical canonical orbitals, electron density and
orbital energies. For fractional systems, post-SCF and SCF for GSC
are different and evaluating
$\frac{\partial \kappa_{p\sigma}}{\partial \rho_{\rm{s}}(\mathbf{r}, \mathbf{r}')}$
would involve higher order terms.

\section{Various derivations of the second-order energy derivatives with respect
to canonical occupation numbers}
In the following, we derive the analytical expression for
the exact second order derivatives of the total energy with respect to
canonical occupation number used in GSC.
We use indices $i, j, k$ for
occupied orbitals, indices $a, b$ for virtual orbitals,
indices $m, n, p, q, k, l$ for general orbitals,
$\sigma, \tau, \upsilon, \kappa$ for the spin of electrons,
$m_{p\sigma}, n_{p\sigma}$ for the $p$-th
canonical occupation numbers with spin $\sigma$,
and $\psi_{p\sigma}$ for the $p$-th canonical orbital (CO)
with spin $\sigma$.

\subsection{Derivation from the coupled perturbed equations}
The exact first derivative of the total energy with respect
to the canonical occupation number for an associated DFA is
\begin{align}
    \frac{\partial E} {\partial n_{p\sigma}}
    &=
        \sum_{\tau}\iint \dd \mathbf{r} \dd \mathbf{r}'
            \frac{\delta E}{\delta \rho_{\text{s}}^{\tau}(\mathbf{r}, \mathbf{r}')}
        \frac{\partial \rho_{\text{s}}^{\tau}(\mathbf{r}, \mathbf{r}')}
             {\partial n_{p\sigma}} \\
    &=
        \langle \psi_{p\sigma}  | H_{\text{s}}^{\sigma}  | \psi_{p\sigma} \rangle
        + \sum_{k\tau} n_{k\tau}
        \Big (
            \left \langle \frac{\partial \psi_{k\tau}}{\partial n_{p\sigma}}
            \middle | H_{\text{s}}^{\tau} \middle | \psi_{k\tau} \right \rangle
            + \left \langle \psi_{k\tau} \middle | H_{\text{s}}^{\tau} \middle |
            \frac{\partial \psi_{k\tau}}{\partial n_{p\sigma}} \right \rangle
        \Big) \\
    &=
        \langle \psi_{p\sigma}  | H_{\text{s}}^{\sigma}  | \psi_{p\sigma} \rangle
        + \sum_{q\tau} n_{q\tau} \epsilon_{q\tau} \frac{\partial}{\partial n_{p\sigma}}
        \langle \psi_{q\tau} | \psi_{q\tau} \rangle \\
    &=
        \langle \psi_{p\sigma}  | H_{\text{s}}^{\sigma}  | \psi_{p\sigma} \rangle,
        \label{eq:extended_Janak}
\end{align}
where \(H_{\text{s}}^{\sigma}\) is the Kohn-Sham (KS) or generalized KS (GKS) Hamiltonian and we used
\(\langle \psi_{q\tau} | \psi_{q\tau} \rangle = 1\). Eq.~\ref{eq:extended_Janak} is
a natural extension of the Janak theorem to the generalized KS case. \cite{1978janak7168}

The exact second order derivative of the total energy with respect to the
canonical occupation number for an associated DFA is
\begin{align}
    \nonumber
    \frac{\partial^2 E} {\partial n_{p\sigma}^2}
    &=  \frac{\partial}{\partial n_{p\sigma}}
        \sum_{\tau}\iint \dd \mathbf{r} \dd \mathbf{r}'
            \frac{\delta E}{\delta \rho_{\text{s}}^{\tau}(\mathbf{r}, \mathbf{r}')}
        \frac{\partial \rho_{\text{s}}^{\tau}(\mathbf{r}, \mathbf{r}')}
             {\partial n_{p\sigma}}
        \\
    \label{eq:d2E_d2n_3terms}
    &= \frac{\partial}{\partial n_{p\sigma}} \Bigg [
        \langle \psi_{p\sigma}  | H_{\text{s}}^{\sigma}  | \psi_{p\sigma} \rangle
        + \sum_{k\tau} n_{k\tau}
        \Big (
            \left \langle \frac{\partial \psi_{k\tau}}{\partial n_{p\sigma}}
            \middle | H_{\text{s}}^{\tau} \middle | \psi_{k\tau} \right \rangle
            + \left \langle \psi_{k\tau} \middle | H_{\text{s}}^{\tau} \middle |
            \frac{\partial \psi_{k\tau}}{\partial n_{p\sigma}} \right \rangle
        \Big)
    \Bigg]
\end{align}

We denote the last two terms in Eq.~\ref{eq:d2E_d2n_3terms} as
$T_2$, and we show that $T_2$ gives zero contribution in the end.
We use the property that the orbitals are eigenfunctions of the
(G)KS Hamiltonian. Thus, the last two terms in Eq.~\ref{eq:d2E_d2n_3terms}
turn into
\begin{align}
    \label{eq:d2E_d2n_T2_def}
    T_2 &= \frac{\partial}{\partial n_{p\sigma}} \Bigg [
        \sum_{k\tau} n_{k\tau}
        \Big (
            \left \langle \frac{\partial \psi_{k\tau}}{\partial n_{p\sigma}}
            \middle | H_{\text{s}}^{\tau} \middle | \psi_{k\tau} \right \rangle
            + \left \langle \psi_{k\tau} \middle | H_{\text{s}}^{\tau} \middle |
            \frac{\partial \psi_{k\tau}}{\partial n_{p\sigma}} \right \rangle
        \Big)
        \Bigg] \\
    &=  \frac{\partial}{\partial n_{p\sigma}}
        \left(
            \sum_{q\tau} n_{q\tau} \epsilon_{q\tau}
            \frac{\partial}{\partial n_{p\sigma}}
            \langle \psi_{q\tau} | \psi_{q\tau} \rangle
        \right) \\
    &= \epsilon_{p\sigma}
        \frac{\partial}{\partial n_{p\sigma}}
        \langle \psi_{p\sigma} | \psi_{p\sigma} \rangle
        + \sum_{q\tau} n_{q\tau} \frac{\partial \epsilon_{q\tau}}{\partial n_{p\sigma}}
        \frac{\partial}{\partial n_{p\sigma}}
        \langle \psi_{q\tau} | \psi_{q\tau} \rangle
        + \sum_{q\tau} n_{q\tau} \epsilon_{q\tau}
        \frac{\partial^2}{\partial n_{p\sigma}^2}
        \langle \psi_{q\tau} | \psi_{q\tau} \rangle \\
    & = 0.
    \label{eq:d2E_d2n_T2_zero}
\end{align}
Because the orbitals are always constrained to be orthonormal, namely
\(\langle \psi_{q\tau} | \psi_{q\tau} \rangle \equiv 1\),
the derivative of \(\langle \psi_{q\tau} | \psi_{q\tau} \rangle \)
to any order is always zeros. Therefore, we have $T_2$ equals to zero
at the end as shown in Eq.~\ref{eq:d2E_d2n_T2_zero}.

According to Eq.~\ref{eq:d2E_d2n_3terms} and \ref{eq:d2E_d2n_T2_zero},
the second order derivative of total energy with respect to canonical occupation number
is evaluated as
\begin{align}
    \label{eq:d2E_d2n_deriOrbE}
    \frac{\partial^2 E}
        {\partial n_{p\sigma}^2}
    &=
    \left \langle \psi_{p\sigma} \middle|
    \frac{\partial H_{\text{s}}^{\sigma}}{\partial n_{p\sigma}}
    \middle| \psi_{p\sigma} \right \rangle \\
    \label{eq:d2E_d2n_dHdrho_drhodn}
    &=
    \sum_{\tau} \iiiint \dd \mathbf{r}_1 \dd \mathbf{r}_2
    \dd \mathbf{r}_3 \dd \mathbf{r}_4
    \psi_{p\sigma}^*(\mathbf{r}_1)
    \psi_{p\sigma}(\mathbf{r}_2)
    \frac{\delta H_{\text{s}}^{\sigma}(\mathbf{r}_1, \mathbf{r}_2)}
         {\delta \rho_{\text{s}}^{\tau}(\mathbf{r}_3, \mathbf{r}_4)}
    \frac{\partial \rho_{\text{s}}^{\tau}(\mathbf{r}_3, \mathbf{r}_4)}
         {\partial n_{p\sigma}} \\
    &=
    \langle \psi_{p\sigma} \psi_{p\sigma}^* | K^{\sigma\sigma} |
    \psi_{p\sigma} \psi_{\sigma}^* \rangle
    +
    \sum_{k\tau} n_{k\tau} \Bigg(
    \left \langle
        \psi_{p\sigma} \psi_{p\sigma}^* \middle | K^{\sigma\tau}
        \middle | \frac{\partial \psi_{k\tau}}{\partial n_{p\sigma}} \psi_{k\tau}^*
    \right \rangle
    +
    \left \langle
        \psi_{p\sigma} \psi_{p\sigma}^* \middle | K^{\sigma\tau}
        \middle | \psi_{k\tau} \frac{\partial \psi_{k\tau}^*}{\partial n_{p\sigma}}
    \right \rangle
    \Bigg)
    \label{eq:d2E_d2n_Kernel_deriOrb}
\end{align}
in which the Hellmann-Feynman theorem is used in Eq.~\ref{eq:d2E_d2n_deriOrbE},
$K^{\sigma\tau}$ is the kernel operator, defined as the second-order functional
derivatives of the total energy functional with respect to \(\rho_{\text{s}}^{\tau}\)
\begin{align}
    K^{\sigma\tau} (\mathbf{r}_1, \mathbf{r}_2; \mathbf{r}_3, \mathbf{r}_4)
    &= \frac{\delta^2 E_v[\rho_{\text{s}}^{\sigma}]}
            {\delta \rho_{\text{s}}^{\tau}(\mathbf{r}_3, \mathbf{r}_4)
             \delta \rho_{\text{s}}^{\sigma}(\mathbf{r}_2, \mathbf{r}_1)}
        \label{eq:K_as2ndDeriv} \\
    &= \frac{\delta H_{\text{s}}^{\sigma}(\mathbf{r}_1, \mathbf{r}_2)}
            {\delta \rho_{\text{s}}^{\tau}(\mathbf{r}_3, \mathbf{r}_4)} \\
    &=
        \frac{\delta(\mathbf{r}_1, \mathbf{r}_2)
        \delta(\mathbf{r}_3, \mathbf{r}_4)}
        {|\mathbf{r}_1 - \mathbf{r}_3|}
        + \frac{\delta^2 E_{\text{xc}}}
        {\delta \rho_{\text{s}}^{\sigma}(\mathbf{r}_2, \mathbf{r}_1)
         \delta \rho_{\text{s}}^{\tau}(\mathbf{r}_3, \mathbf{r}_4)},
    \label{eq:kernel_op}
\end{align}
and
\begin{align}
    \langle \psi_{p\sigma}\psi_{q\sigma}^* | K^{\sigma\tau} |
    \psi_{m\tau} \psi_{n\tau}^* \rangle
    &= \iiiint \dd \mathbf{r}_1 \dd \mathbf{r}_2
    \dd \mathbf{r}_3 \dd \mathbf{r}_4
    \psi_{p\sigma}^*(\mathbf{r}_1) \psi_{q\sigma}(\mathbf{r}_2)
    K^{\sigma\tau}(\mathbf{r}_1, \mathbf{r}_2; \mathbf{r}_3, \mathbf{r}_4)
    \psi_{m\tau}(\mathbf{r}_3) \psi_{n\tau}^*(\mathbf{r}_4) \\
    &= K_{pq\sigma,mn\tau}. \label{eq:def_Kpqmn}
\end{align}

According to Eq.~\ref{eq:d2E_d2n_Kernel_deriOrb},
the evaluation of the second order derivative of total energy
involves the first order derivative of COs with respect to canonical occupation numbers, i.e.,
$\frac{\partial \psi_{k\tau}}{\partial n_{p\sigma}}$.
In the space of COs,
the relaxation of COs can be expanded under the basis of COs, i.e.,
\begin{align}
    \label{eq:deri_CO}
    \frac{\partial \psi_{k\tau}}{\partial n_{p\sigma}}
        &= \sum_{m} U_{km\tau}^{n_{p\sigma}} \psi_{m\tau},
\end{align}
with $U_{km\tau}^{n_{p\sigma}}$ being the expansion coefficients
associated with the occupation number $n_{p\sigma}$.
The CP equations are developed to solve the set of coefficients
$U_{km\tau}^{n_{p\sigma}}$ for all the orbitals.

The derivation of the CP equations with respect to the canonical occupation
numbers in this work is similar to the derivation of CP equations
for nuclear coordinates \cite{yamaguchi1994new,2019mei2545}.
It is based on two stationary conditions
for the (G)KS calculation at the self-consistent
(SCF) solution: (1) the COs are orthonormal to each other;
(2) the (G)KS Hamiltonian is diagonal under the
representation of COs; that is,
\begin{align}
    \label{eq:cond_CO}
    \langle\psi_{m\tau} | \psi_{n\tau}\rangle &= \delta_{mn}, \\
    \label{eq:cond_H}
    \langle\psi_{m\tau} | H_{\text{s}}^{\tau} | \psi_{n\tau}\rangle &=
        \delta_{mn} \epsilon_{m\tau},
\end{align}
with $\delta_{mn}$ being the Kronecker delta function.
Taking derivatives of Eq. \ref{eq:cond_CO} and \ref{eq:cond_H} with respect to
canonical occupation numbers gives a set of constraints for
the relaxation of COs with respect to canonical occupation numbers,
which are,
\begin{align}
    \label{eq:constrain_orthonormal}
    \frac{\partial \langle\psi_{m\tau} | \psi_{n\tau}\rangle}
        {\partial n_{p\sigma}} &= 0, \quad m\ge n, \\
    \label{eq:constrain_diag}
    \frac{\partial \langle\psi_{m\tau} | H_{\text{s}}^{\tau} |
        \psi_{n\tau}\rangle}{\partial n_{p\sigma}} &= 0, \quad m > n.
\end{align}

\subsubsection{Interger systems}
First, we focus on the derivation for integer systems,
in which the canonical occupation numbers
are either 1 for occupied orbitals or 0 for virtual
orbitals.

From Eq. \ref{eq:constrain_orthonormal}, we have
\begin{align}
    \nonumber
    0 &=
        \frac{\partial \langle\psi_{m\tau} | \psi_{n\tau}\rangle}
        {\partial n_{p\sigma}}\\ \nonumber
      & = \sum_{k} U_{mk\tau}^{n_{p\sigma}}
        \langle \psi_{k\tau} | \psi_{n\tau} \rangle
        + \sum_{l} U_{nl\tau}^{n_{p\sigma}}
        \langle \psi_{m\tau} | \psi_{l\tau} \rangle \\
        \label{eq:U_relation}
      &= U_{mn\tau}^{n_{p\sigma}} + U_{nm\tau}^{n_{p\sigma}} = 0.
\end{align}

From Eq. \ref{eq:constrain_diag}, we have
\begin{align}
    \nonumber
    0 &=
        \frac{\partial \langle\psi_{m\tau} | H_{\text{s}}^{\tau} |
        \psi_{n\tau}\rangle}{\partial n_{p\sigma}} \\
     &= \left \langle
        \psi_{m\tau} \middle| \frac{\partial H_{\text{s}}^{\tau}}{\partial n_{p\sigma}}
        \middle| \psi_{n\tau}
        \right \rangle
     + \left(
        \left \langle
        \frac{\partial \psi_{m\tau}}{\partial n_{p\sigma}}
        \middle|
        H_{\text{s}}^{\tau}
        \middle| \psi_{n\tau}
        \right \rangle
        +
        \left \langle
        \psi_{m\tau}
        \middle|
        H_{\text{s}}^{\tau}
        \middle|
        \frac{\partial \psi_{n\tau}}{\partial n_{p\sigma}}
        \right \rangle
       \right)
      \label{eq:dH_off_diag_3terms}
\end{align}
Following the similar steps shown in Eqs.~\ref{eq:d2E_d2n_deriOrbE}
-~\ref{eq:d2E_d2n_Kernel_deriOrb} to derive the derivative of (G)KS
Hamiltonian, and using relations shown in Eq.~\ref{eq:deri_CO} and Eq.~\ref{eq:U_relation},
we can express Eq.~\ref{eq:dH_off_diag_3terms} as
\begin{align}
        \nonumber
    0 &=\Big(
        K_{mn\tau, pp\sigma} + \sum_{iq\upsilon}
        n_{i\upsilon} U_{iq\upsilon}^{n_{p\sigma}}
        \left(
        K_{mn\tau, qi\upsilon} +  K_{mn\tau, iq\upsilon}
        \right)
        \Big)
        + \sum_{r} U_{mr\tau}^{n_{p\sigma}}
        \langle \psi_{r\tau} | H_{\text{s}}^{\tau} | \psi_{n\tau} \rangle
        + \sum_{s} U_{ns\tau}^{n_{p\sigma}}
        \langle \psi_{m\tau} | H_{\text{s}}^{\tau} | \psi_{s\tau} \rangle \\
        \nonumber
      &=
        K_{mn\tau, pp\sigma} + \sum_{iq\upsilon}
        n_{i\upsilon} U_{iq\upsilon}^{n_{p\sigma}}
        \left(
        K_{mn\tau, qi\upsilon} +  K_{mn\tau, iq\upsilon}
        \right)
        + U_{mn\tau}^{n_{p\sigma}} \epsilon_{n\tau}
        + U_{nm\tau}^{n_{p\sigma}} \epsilon_{m\tau}\\
        &=
        K_{mn\tau, pp\sigma} + \sum_{iq\upsilon}
        n_{i\upsilon} U_{iq\upsilon}^{n_{p\sigma}}
        \left(
        K_{mn\tau, qi\upsilon} +  K_{mn\tau, iq\upsilon}
        \right)
        + U_{mn\tau}^{n_{p\sigma}}
        \left( \epsilon_{n\tau} - \epsilon_{m\tau} \right),
    \label{eq:dH_off_diag_sum_iq}
\end{align}
where $K_{mn\tau, pq\sigma}$ is the matrix representation of the kernel operator
as defined in Eq.~\ref{eq:kernel_op}.

The summation over index $q$ for general COs
in Eq. \ref{eq:dH_off_diag_sum_iq} can be separated into the occupied and
virtual part for integer systems, i.e.,
\begin{align}
    \label{eq:summation_sep}
    \sum\limits_{q} = \sum\limits_{j} + \sum\limits_{a}.
\end{align}
Based on the separation shown in Eq. \ref{eq:summation_sep}
and the relation of $\{U_{mn\tau}^{n_{p\sigma}}\}$ shown in Eq.~\ref{eq:U_relation},
the summation term in Eq.~\ref{eq:dH_off_diag_sum_iq} can be
simplified as
\begin{align}
    \nonumber
    \sum_{iq\upsilon}
        n_{i\upsilon} U_{iq\upsilon}^{n_{p\sigma}}
        \left(
        K_{mn\tau, qi\upsilon} +  K_{mn\tau, iq\upsilon}
        \right)
    & =
        \sum_{ij\upsilon} n_{i\upsilon} U_{ij\upsilon}^{n_{p\sigma}}
        \left( K_{mn\tau, ij\upsilon} + K_{mn\tau, ji\upsilon} \right) \\
        &\quad +
         \sum_{ia\upsilon} n_{i\upsilon} U_{ia\upsilon}^{n_{p\sigma}}
        \left( K_{mn\tau, ia\upsilon} + K_{mn\tau, ai\upsilon} \right)
        \\
        \nonumber
        & =
        \frac12 \sum_{ij\upsilon} n_{i\upsilon}
        \left( U_{ij\upsilon}^{n_{p\sigma}} + U_{ji\upsilon}^{n_{p\sigma}} \right)
        \left( K_{mn\tau, ij\upsilon} + K_{mn\tau, ji\upsilon} \right) \\
        &\quad + \sum_{ia\upsilon} n_{i\upsilon}
        U_{ia\upsilon}^{n_{p\sigma}}
        \left( K_{mn\tau, ia\upsilon} + K_{mn\tau, ai\upsilon} \right)
        \\
        & =
        \sum_{\upsilon ia} n_{i\upsilon}
        U_{ia\upsilon}^{n_{p\sigma}}
        \left( K_{mn\tau, ia\upsilon} + K_{mn\tau, ai\upsilon} \right).
        \label{eq:dH_off_diag_sum_iq_final}
\end{align}
Note that the contribution from the occupied-occupied
block of $U_{ij\upsilon}^{n_{p\sigma}}$ vanishes and only the occupied-virtual
block of $U_{ia\upsilon}^{n_{p\sigma}}$ remains.

Combining Eq.~\ref{eq:dH_off_diag_sum_iq} and
Eq.~\ref{eq:dH_off_diag_sum_iq_final}, we have
a set of linear equations,
\begin{align}
    \label{eq:cp_eq_0}
        &U_{mn\tau}^{n_{p\sigma}}
        \left( \epsilon_{n\tau} -\epsilon_{m\tau} \right) +
        \sum_{ia\upsilon} n_{i\upsilon} U_{ia\upsilon}^{n_{p\sigma}}
        \left( K_{mn\tau, ia\upsilon} + K_{mn\tau, ai\upsilon} \right)
        = -K_{mn\tau, pp\sigma}, \quad m > n.
\end{align}
The summation in Eq. \ref{eq:cp_eq_0} is over only pairs of
occupied and virtual orbitals (occupied-virtual block for
$\{U_{ia\upsilon}^{n_{p\sigma}}\}$), which reduces the dimension
of the set of equations in matrix form to
$\sum\limits_\sigma N_{\text{occ}}^\sigma \times N_{\text{vir}}^\sigma$,
with $N_{\rm{occ}}^\sigma$ being the number of occupied orbitals
and $N_{\rm{vir}}^{\sigma}$ being the number of
virtual orbitals for $\sigma$ spin.
Expressing Eq. \ref{eq:cp_eq_0} in matrix form, we have
the set of linear matrix equations with respect to all the canonical occupied
numbers $n_{p\sigma}$,
\begin{align}
    \label{eq:cp_eq}
    \mathbf{M U}^{n_{p\sigma}} &= \mathbf{b}^{n_{p\sigma}}, \\
    \label{eq:cp_A_matrix}
    {M}_{ia\tau, jb\upsilon} &=
        \left( \epsilon_{a\tau} - \epsilon_{i\tau} \right)
        \delta_{ij} \delta_{ab} \delta_{\tau\upsilon} +
        n_{j\upsilon} \left( K_{ia\tau, jb\upsilon} + K_{ia\tau, bj\upsilon} \right), \\
    {b}_{ia\tau}^{n_{p\sigma}} &= -K_{ia\tau, pp\sigma}.
    \label{eq:cp_eq_end}
\end{align}
Eqs. \ref{eq:cp_eq} - \ref{eq:cp_eq_end} are the CP equations
with respect to canonical occupied numbers, which give solutions to
the expansion coefficients for the occupied-virtual block
of $\{U_{ia\tau}^{n_{p\sigma}}\}$.

With solutions of $\{U_{ia\tau}^{n_{p\sigma}}\}$
from the CP equations, Eq. \ref{eq:cp_eq},
we can evaluate the analytical second order derivative of
total energy with respect to canonical occupation numbers for integer
systems as
\begin{align}
    \frac{\partial^2 E}{\partial n_{p\sigma}^2}
    &= K_{pp\sigma, pp\sigma}
    + \sum_{ia\tau} U_{ia\tau}^{n_{p\sigma}}
    \left( K_{pp\sigma, ia\tau} + K_{pp\sigma, ai\tau} \right)\\
    &= K_{pp\sigma, pp\sigma}
    - \sum_{ia\tau,jb\upsilon}
    \left( K_{pp\sigma, ia\tau} + K_{pp\sigma, ai\tau} \right)
    M_{ia\tau,jb\upsilon}^{-1} K_{jb\upsilon,pp\sigma}.
    \label{eq:2nd_deri_final}
\end{align}
This equation is in a convenient form for applications
when only a few orbitals are needed for
the GSC correction. The corresponding terms
$\{U_{ia\tau}^{p\sigma}\}$ can be directly obtained from
numerical iterative  solution to the CP equations.

\subsubsection{Fractional systems}
For fractional systems with fractional occupation numbers,
the CP equations become a little more complicated. For fractional
systems, we clarify orbitals to be fully occupied with $n_i = 1$,
fractionally occupied with $0 < n_f < 1$, and virtual with
$n_a = 0$. Therefore, the summation over all the orbitals is
divided into three blocks,
i.e.,
\begin{align}
    \label{eq:summation_sep_frac}
    \sum\limits_{q} = \sum\limits_{i} + \sum\limits_{f} + \sum\limits_{a},
\end{align}
with index $f$ for fractionally occupied orbitals.
Using the similar treatment to the derivation of the integer systems,
we have the following set of linear equations
for the fractional systems as,
\begin{align}
    \label{eq:cp_eq_frac_0}
    K_{mn\tau, pp\sigma} &=
        U_{mn\tau}^{n_{p\sigma}}
        \left( \epsilon_{m\tau} -\epsilon_{n\tau} \right)
        - \sum_{\upsilon}
        \left( \sum_{m=i} + \sum_{m=f} \right)
        \left( \sum_{n=i} + \sum_{n=f} + \sum_{n=a} \right)
         n_{m\upsilon} U_{mn\upsilon}^{n_{p\sigma}}
        (K_{pp\tau, mn\upsilon} + K_{pp\tau, nm\upsilon})
        \\
        \nonumber
        &= U_{mn\tau}^{n_{p\sigma}}
        \left( \epsilon_{m\tau} -\epsilon_{n\tau} \right)
        \\
        & \quad - \sum_{\upsilon}
        \left(
            \sum_{m=i, n=f}
            + \sum_{m=i, n=a}
            + \sum_{m=f, n=i}
            + \sum_{m=f, n=f, m > n}
            + \sum_{m=f, n=a}
        \right)
            n_{m\upsilon} U_{mn\upsilon}^{n_{p\sigma}}
            \left( K_{pp\tau, mn\upsilon} + K_{pp\tau, nm\upsilon} \right)
    \label{eq:cp_eq_frac_1}
\end{align}
The dimension of the CP equations (Eq. \ref{eq:cp_eq_frac_1}) is
$\sum\limits_\sigma N_{\text{occ}}^\sigma \times
(N_{\text{frac}}^\sigma + N_{\text{vir}}^\sigma)
+ N_{\text{frac}}^\sigma \times (N_{\text{occ}}^\sigma
+ N_{\text{vir}}^\sigma) + N_{\text{frac}}^\sigma
\times (N_{\text{frac}}^\sigma - 1) / 2$, with
$N_{\text{frac}}^\sigma$ being the number of fractionally
occupied orbitals with spin $\sigma$.

\subsubsection{The cross terms in the second derivatives}
In GSC2, we only need the second
order derivatives of total energy with respect to the same occupation number.
The cross terms with respect to two different occupation numbers are not needed,
because we only consider one specific orbital with the fractional
occupation. Here, we present the analytical expression
of the cross terms for future purpose. Take the integer system
as the example. The expression of cross terms can be generalized
easily based on Eq.~\ref{eq:d2E_d2n_Kernel_deriOrb} and
Eq.~\ref{eq:2nd_deri_final}, which is shown as
\begin{align}
    \nonumber
    \frac{\partial^2 E}{\partial n_{p\sigma} \partial n_{q\tau}}
    &= K_{pp\sigma, qq\tau}
    - \sum_{ia\upsilon}
    U_{ia\upsilon}^{n_{q\tau}}
    \left( K_{pp\sigma, ia\upsilon} + K_{pp\sigma, ai\upsilon} \right)\\
    &= K_{pp\sigma, qq\tau}
    - \sum_{ia\upsilon,jb\kappa}
    \left( K_{pp\sigma, ia\upsilon} + K_{pp\sigma, ai\upsilon} \right)
    M_{ia\upsilon,jb\kappa}^{-1}K_{jb\kappa,qq\tau}.
    \label{eq:2nd_deri_gen}
\end{align}

\subsection{Derivation from the Maxwell relation and linear response}
It is possible to derive the second derivatives from
the Maxwell relation following the previous work
\cite{2012yang144110,2013peng184108}.
Consider the generalized KS calculations, in which
the set of canonical occupation numbers is given as $\{n_{p\sigma}\}$
and the GKS spin density matrix is
$\rho_{\text{s}}^{\sigma}=\sum_{p\sigma} n_{p\sigma} |\psi_{p\sigma}\rangle \langle \psi_{p\sigma}|$.
The total energy functional is defined as
\begin{equation}
    E_v[\rho_{\text{s}}^{\sigma}] = \text{Tr}(\hat{T}\rho_{\text{s}}^{\sigma})
    + E_{\text{H}}[\rho_{\text{s}}^{\sigma}] + E_{\text{xc}}[\rho_{\text{s}}^{\sigma}]
    + \sum_{\tau} \iint \dd \mathbf{r} \dd \mathbf{r}'
    \rho_{\text{s}}^{\tau}(\mathbf{r}', \mathbf{r}) v^{\tau}(\mathbf{r}, \mathbf{r}'),
\end{equation}
where \(v^{\tau}(\mathbf{r}, \mathbf{r}')\) is the generalized external potential,
which is nonlocal and spin-dependent, as an extension from the physical potential,
\(v(\mathbf{r})\), which is local and spin independent. Thus, \(v(\mathbf{r})\)
is a special case of \(v^{\tau}(\mathbf{r}, \mathbf{r}')\).
In the energy expression, note that \(v^{\tau}(\mathbf{r}, \mathbf{r}')\) is dual
to the generalized KS density matrix \(\rho_{\text{s}}^{\tau}(\mathbf{r}', \mathbf{r})\),
not the physical one-particle density matrix.
Such a generalized external potential was first introduced in Ref.~\citenum{2013peng184108}.

For a given set of \(\{n_{p\sigma}\}\) and \(v^{\sigma}(\mathbf{r}, \mathbf{r}')\),
the energy is defined as the following minimum over the space of orthonormal orbitals
\(\{\psi_{m\sigma}\}\):
\begin{align}
    E(v^{\sigma}, \{n_{p\sigma}\})
        = \min_{\{\psi_{m\sigma}\}} E_{v} [\rho_{\rm{s}}^{\sigma}].
\end{align}
Under the perturbation \(\delta v^{\tau}(\mathbf{r}, \mathbf{r}')\)
and \(\{\dd n_{p\sigma}\}\), the first order change in the energy is
\begin{align}
  \dd E &= \sum_{\tau} \iint \dd \mathbf{r} \dd \mathbf{r}'
           \frac{\delta E}{\delta v^{\tau}(\mathbf{r}, \mathbf{r}')}
           \delta v^{\tau}(\mathbf{r}, \mathbf{r}')
           + \sum_{p\sigma} \frac{\partial E}{\partial n_{p\sigma}}
           \dd n_{p\sigma} \\
        &= \sum_{\tau} \iint \dd \mathbf{r} \dd \mathbf{r}'
           \rho_{\text{s}}^{\tau}(\mathbf{r}', \mathbf{r})
           \delta v^{\tau}(\mathbf{r}, \mathbf{r}')
           + \sum_{p\sigma} \epsilon_{p\sigma} \dd n_{p\sigma},
           \label{eq:1st_order_dE}
\end{align}
where we have used the equality
\begin{equation}
    \frac{\delta E(v^{\sigma}, \{n_{p\sigma}\})}{\delta v^{\tau}(\mathbf{r}, \mathbf{r}')}
    = \rho_{\text{s}}^{\tau}(\mathbf{r}', \mathbf{r}),
    \label{eq:Hellman-Feynman}
\end{equation}
which is the extension of the Hellman-Feynman theorem for electron density.
We also used the extended Janak's theorem (Eq.~\ref{eq:extended_Janak}),
\(\frac{\partial E}{\partial n_{p\sigma}}=\epsilon_{p\sigma}\),
where \(\epsilon_{p\sigma}\) is the corresponding eigenvalue of generalized KS Hamiltonian
\(H_{\text{s}}^{\sigma}\).
The Maxwell relationship from Eq.~\ref{eq:1st_order_dE} reads
\begin{equation}
    \frac{\partial \rho_{\text{s}}^{\tau}(\mathbf{r}', \mathbf{r})}
         {\partial n_{p\sigma}}
    =
    \frac{\delta \epsilon_{p\sigma}}{\delta v^{\tau}(\mathbf{r}, \mathbf{r}')}.
\end{equation}
Following Eq.~\ref{eq:d2E_d2n_dHdrho_drhodn} and using the Maxwell relations, we
have the second-order derivative of total energy with respect to the canonical
occupation number as
\begin{align}
    \frac{\partial^2 E}{\partial n_{p\sigma} \partial n_{q\kappa}}
    &=
    \left \langle \psi_{q\kappa} \middle|
        \frac{\partial H_{\text{s}}^{\kappa}}{\partial n_{p\sigma}}
    \middle| \psi_{q\kappa} \right \rangle \\
    &=
    \sum_{\tau} \iiiint \dd \mathbf{r}_1 \dd \mathbf{r}_2
    \dd \mathbf{r}_3 \dd \mathbf{r}_4
    \psi_{q\kappa}^*(\mathbf{r}_1)
    \frac{\delta H_{\text{s}}^{\kappa}(\mathbf{r}_1, \mathbf{r}_2)}
         {\delta \rho_{\text{s}}^{\tau}(\mathbf{r}_3, \mathbf{r}_4)}
    \frac{\partial \rho_{\text{s}}^{\tau}(\mathbf{r}_3, \mathbf{r}_4)}
         {\partial n_{p\sigma}}
    \psi_{q\kappa}(\mathbf{r}_2) \\
    &=
    \sum_{\tau} \iiiint \dd \mathbf{r}_1 \dd \mathbf{r}_2
    \dd \mathbf{r}_3 \dd \mathbf{r}_4
    \psi_{q\kappa}^*(\mathbf{r}_1)
    \psi_{q\kappa}(\mathbf{r}_2)
    K^{\kappa\tau}(\mathbf{r}_1, \mathbf{r}_2; \mathbf{r}_3, \mathbf{r}_4)
    \frac{\delta \epsilon_{p\sigma}}{\delta v^{\tau}(\mathbf{r}_4, \mathbf{r}_3)},
    \label{eq:maxwell_d2E_d2n}
\end{align}
where $K^{\sigma\tau}$ is the kernel operator as defined in Eq.~\ref{eq:kernel_op}.

Now we evaluate the response of the orbital energy with respect to the external potential
\begin{align}
\frac{\delta \epsilon_{p\sigma}}{\delta v^{\tau}(\mathbf{r}_3, \mathbf{r}_4)}
    &= \iint \dd\mathbf{r}_1 \dd \mathbf{r}_2
        \psi_{p\sigma}^*(\mathbf{r}_1) \frac{\delta H_{\rm{s}}^{\sigma}(\mathbf{r}_1, \mathbf{r}_2)}
        {\delta v^{\tau}(\mathbf{r}_3, \mathbf{r}_4)}
        \psi_{p\sigma}(\mathbf{r}_2).
\end{align}
Based on
\begin{equation}
  H_{\text{s}}^{\sigma}(\mathbf{r}, \mathbf{r}')
  = -\frac{1}{2}\nabla^2
  + v^{\sigma}(\mathbf{r}, \mathbf{r}')
  + v_{J}^{\sigma}(\mathbf{r})\delta(\mathbf{r}-\mathbf{r}')
  + v_{\text{xc}}^{\sigma}(\mathbf{r}, \mathbf{r}'),
\end{equation}
thus, we have
\begin{align}
    \frac{\delta H_{\text{s}}^{\sigma}(\mathbf{r}_1, \mathbf{r}_2)}
       {\delta v^{\tau}(\mathbf{r}_3, \mathbf{r}_4)}
    &=
    \delta(\mathbf{r}_1, \mathbf{r}_3) \delta(\mathbf{r}_2, \mathbf{r}_4)
    \delta_{\sigma\tau}
    + \sum_{\upsilon}
    \iint \dd \mathbf{r}_5 \dd \mathbf{r}_6
    \frac{\delta v_{\text{Hxc}}^{\sigma}(\mathbf{r}_1, \mathbf{r}_2)}
         {\delta \rho_{\text{s}}^{\upsilon}(\mathbf{r}_5, \mathbf{r}_6)}
    \frac{\delta \rho_{\text{s}}^{\upsilon}(\mathbf{r}_5, \mathbf{r}_6)}
         {\delta v^{\tau}(\mathbf{r}_3, \mathbf{r}_4)} \\
    &=
    \delta(\mathbf{r}_1, \mathbf{r}_3) \delta(\mathbf{r}_2, \mathbf{r}_4)
    \delta_{\sigma\tau} \nonumber \\
    &\quad+ \sum_{\upsilon}
    \iint \dd \mathbf{r}_5 \dd \mathbf{r}_6
    \left[
        \delta(\mathbf{r}_1, \mathbf{r}_2)
        \delta(\mathbf{r}_5, \mathbf{r}_6)
        \frac{1}{|\mathbf{r}_1 - \mathbf{r}_5|}
        +
    \frac{\delta v_{\text{xc}}^{\sigma}(\mathbf{r}_1, \mathbf{r}_2)}
         {\delta \rho_{\text{s}}^{\upsilon}(\mathbf{r}_5, \mathbf{r}_6)}
    \right]
    \chi^{\upsilon\tau}(\mathbf{r}_5, \mathbf{r}_6; \mathbf{r}_3, \mathbf{r}_4) \\
    \label{eq:dH_dv}
    &=
    \delta(\mathbf{r}_1, \mathbf{r}_3) \delta(\mathbf{r}_2, \mathbf{r}_4)
    \delta_{\sigma\tau}
    + \sum_{\upsilon}
    \iint \dd \mathbf{r}_5 \dd \mathbf{r}_6
    K^{\sigma\upsilon}(\mathbf{r}_1, \mathbf{r}_2; \mathbf{r}_5, \mathbf{r}_6)
    \chi^{\upsilon\tau}(\mathbf{r}_5, \mathbf{r}_6; \mathbf{r}_3, \mathbf{r}_4),
\end{align}
where
$v_{\text{Hxc}}^{\sigma}(\mathbf{r}_1, \mathbf{r}_2)
=v_{J}^{\sigma} (\mathbf{r}_1, \mathbf{r}_2)
+ v_{\text{xc}}^{\sigma}(\mathbf{r}_1, \mathbf{r}_2)$.

The generalized linear response function \(\chi^{\upsilon\tau}\) in Eq.~\ref{eq:dH_dv}
has been defined as the linear response of the Kohn-Sham density matrix
\(\rho_{\text{s}}^{\sigma}(\mathbf{r}_1, \mathbf{r}_2)\)
with respect to the change of \(v^{\tau}(\mathbf{r}_3, \mathbf{r}_4)\),
the generalized nonlocal external potential as in Ref.~\citenum{2013peng184108}:
\begin{align}
    \chi^{\sigma\tau}(\mathbf{r}_1, \mathbf{r}_2; \mathbf{r}_3, \mathbf{r}_4)
    &= \frac{\delta \rho_{\text{s}}(\mathbf{r}_1, \mathbf{r}_2)}
            {\delta v^{\tau}(\mathbf{r}_3, \mathbf{r}_4)} \label{eq:X_def} \\
    &= \frac{\delta^2 E[v^\sigma, \{n_{p\sigma}\}]}
            {\delta v^{\tau}(\mathbf{r}_3, \mathbf{r}_4)
             \delta v^{\sigma}(\mathbf{r}_2, \mathbf{r}_1)}, \label{eq:X_as2ndDeriv}
\end{align}
where we used Eq.~\ref{eq:Hellman-Feynman}. The analytical expression for \(\chi^{\sigma\tau}\)
has been derived in Ref.~\citenum{2013peng184108}. It is related to the inverse of the matrix
\(\mathbf{M}\) as follows\cite{2012yang144110,2013peng184108}
\begin{align}
    \label{eq:lr_anal_expression}
    \chi^{\upsilon\tau}(\mathbf{r}_5, \mathbf{r}_6; \mathbf{r}_3, \mathbf{r}_4)
    &=
    -\sum_{ia,jb} M_{ia\upsilon,jb\tau}^{-1}
    \left[
        \psi_{i\upsilon}(\mathbf{r}_5) \psi_{a\upsilon}^*(\mathbf{r}_6)
        \psi_{j\tau}^*(\mathbf{r}_4) \psi_{b\tau}(\mathbf{r}_3)
        +
        \psi_{i\upsilon}^*(\mathbf{r}_6) \psi_{a\upsilon}(\mathbf{r}_5)
        \psi_{j\tau}(\mathbf{r}_3) \psi_{b\tau}^*(\mathbf{r}_4)
    \right],
\end{align}
in which \(\mathbf{M}\) is given in Eq.~\ref{eq:cp_A_matrix} as
\begin{align}
    \label{eq:lr_M_matrix}
    M_{ia\upsilon,jb\tau}
    &= \delta_{\upsilon\tau}\delta_{ij}\delta_{ab}(\epsilon_{a\upsilon} - \epsilon_{i\upsilon})
       + K_{ia\upsilon,jb\tau} + K_{ia\upsilon,bj\tau}.
\end{align}
Combining Eq.~\ref{eq:dH_dv} and Eq.~\ref{eq:lr_anal_expression},
we have the response of the orbital energy with respect to
the generalized external potential as
\begin{align}
    \frac{\delta \epsilon_{p\sigma}}{\delta v^{\tau}(\mathbf{r}_3, \mathbf{r}_4)}
    \nonumber
    &=
    \delta_{\sigma\tau} \psi_{p\sigma}^*(\mathbf{r}_3) \psi_{p\sigma}(\mathbf{r}_4)
    \\
    & \quad + \sum_{\upsilon}
    \iiiint \dd \mathbf{r}_1 \dd \mathbf{r}_2 \dd \mathbf{r}_5 \dd \mathbf{r}_6
    \psi_{p\sigma}^*(\mathbf{r}_1)
    \psi_{p\sigma}(\mathbf{r}_2)
    K^{\sigma\upsilon}(\mathbf{r}_1, \mathbf{r}_2; \mathbf{r}_5, \mathbf{r}_6)
    \chi^{\upsilon\tau}(\mathbf{r}_5, \mathbf{r}_6; \mathbf{r}_3, \mathbf{r}_4)
    \label{eq:dorbE_dv_KX}
    \\
    \nonumber
    &=
    \delta_{\sigma\tau} \psi_{p\sigma}^*(\mathbf{r}_3) \psi_{p\sigma}(\mathbf{r}_4)
    \\
    &\quad - \sum_{\upsilon} \sum_{ia,jb} M^{-1}_{ia\upsilon,jb\tau}
    \left(
        K_{pp\sigma,ia\upsilon} \psi_{b\tau}(\mathbf{r}_3) \psi_{j\tau}^*(\mathbf{r}_4)
        +
        K_{pp\sigma,ai\upsilon} \psi_{j\tau}(\mathbf{r}_3) \psi_{b\tau}^*(\mathbf{r}_4)
    \right).
    \label{eq:dorbE_dv}
\end{align}
Combining Eqs.~\ref{eq:maxwell_d2E_d2n}, \ref{eq:dorbE_dv_KX} and
\ref{eq:dorbE_dv},
we have the second order derivative of the total energy with respect to
the canonical occupation numbers as
\begin{align}
    \frac{\partial^2 E}{\partial n_{p\tau}\partial n_{q\varsigma}}
    &= \int \dd 1 \dd 2 \dd 3 \dd 4
    \psi_{q\varsigma}^*(1)\psi_{q\varsigma}(2)
    \sum_{\sigma}
        \frac{\delta H_{\text{s}}^{\varsigma}(1,2)}
             {\delta\rho_{\text{s}}^{\sigma}(3,4)}
        \frac{\delta\varepsilon_{p\tau}}{\delta v^{\sigma}(4,3)} \\
    &= \int \dd 1 \dd 2 \dd 3 \dd 4
    \psi_{q\varsigma}^*(1)\psi_{q\varsigma}(2)
    \sum_{\sigma}
        \frac{\delta H_{\text{Hxc}}^{\varsigma}(1,2)}
             {\delta\rho_{s}^{\sigma}(3,4)} \times \nonumber \\
    &\quad \left[\psi_{p\tau}^*(4) \psi_{p\tau}(3) \delta_{\tau\sigma} +
    \sum_{\upsilon}
    \int \dd 5 \dd 6 \dd 7 \dd 8
    \psi_{p\tau}^*(5)\psi_{p\tau}(6)
        \frac{\partial H_{\text{Hxc}}^{\tau}(5,6)}
             {\delta\rho_{s}^{\upsilon}(7,8)}
    \chi^{\upsilon\sigma}(7,8;4,3)\right] \\
    &= \int \dd 1 \dd 2 \dd 3 \dd 4
    \psi_{q\varsigma}^*(1)\psi_{q\varsigma}(2)
        \frac{\delta H_{\text{Hxc}}^{\varsigma}(1,2)}
             {\delta\rho_{s}^{\tau}(3,4)}
    \psi_{p\tau}^*(4)\psi_{p\tau}(3) \nonumber \\
    &\quad+
    \int \dd 1 \dd 2 \dd 3 \dd 4
    \psi_{q\varsigma}^*(1)\psi_{q\varsigma}(2)
    \sum_{\sigma}\sum_{\upsilon}
    \int \dd 5 \dd 6 \dd 7 \dd 8
        \frac{\delta H_{\text{Hxc}}^{\varsigma}(1,2)}
             {\delta\rho_{s}^{\sigma}(3,4)}
    \chi^{\upsilon\sigma}(7,8;4,3) \times \nonumber \\ &\quad
        \frac{\partial H_{\text{Hxc}}^{\tau}(5,6)}
             {\delta\rho_{s}^{\upsilon}(7,8)}
    \psi_{p\tau}^*(5)\psi_{p\tau}(6) \\
    &= \int \dd 1 \dd 2 \dd 3 \dd 4
    \psi_{q\varsigma}^*(1)\psi_{q\varsigma}(2)K^{\varsigma\tau}(1,2;3,4)
    \psi_{p\tau}^*(4)\psi_{p\tau}(3) \nonumber \\
    &\quad+\sum_{\sigma}\sum_{\upsilon}
    \int \dd 1 \dd 2 \dd 3 \dd 4 \dd 7 \dd 8 \dd 5 \dd 6
    \psi_{q\varsigma}^*(1)\psi_{q\varsigma}(2)
    K^{\varsigma\sigma}(1,2;3,4)\chi^{\upsilon\sigma}(7,8;4,3) \times \nonumber \\ &\quad
    K^{\tau\upsilon}(5,6;7,8)\psi_{p\tau}^*(5)\psi_{p\tau}(6) \\
    &= \int \dd 1 \dd 2 \dd 3 \dd 4
    \psi_{q\varsigma}^*(1)\psi_{q\varsigma}(2)K^{\varsigma\tau}(1,2;3,4)
    \psi_{p\tau}^*(4)\psi_{p\tau}(3) \nonumber \\
    &\quad+\sum_{\sigma}\sum_{\upsilon}
    \int \dd 1 \dd 2 \dd 3 \dd 4 \dd 7 \dd 8 \dd 5 \dd 6
    \psi_{q\varsigma}^*(1)\psi_{q\varsigma}(2)K^{\varsigma\sigma}(1,2;3,4)
    \chi^{\sigma\upsilon}(3,4;7,8) \times \nonumber \\ &\quad K^{\upsilon\tau}(7,8;6,5)
    \psi_{p\tau}^*(5)\psi_{p\tau}(6) \\
    &=\left\langle \psi_{q\varsigma}\psi_{q\varsigma}^* \middle|
    K^{\varsigma\tau}+\sum_{\sigma\upsilon}K^{\varsigma\sigma}
    \chi^{\sigma\upsilon}K^{\upsilon\tau} \middle| \psi_{p\tau}\psi_{p\tau}^*\right\rangle,
    \label{eq:KKXK}
\end{align}
where we used \(\chi^{\upsilon\sigma}(1,2;3,4)=\chi^{\sigma\upsilon}(4,3;2,1)\) based on
Eq.~\ref{eq:X_as2ndDeriv} and \(K^{\tau\upsilon}(1,2;3,4)=K^{\upsilon\tau}(4,3;2,1)\)
based on Eq.~\ref{eq:K_as2ndDeriv}. In addition,
we also implied spacial integration for any product of two 4-point quantities
and used a short-hand notation: \(1\) for \(\mathbf{r}_1\).
We defined the matrix of the kernel \(K\) as
\begin{align}
    K_{uv\tau,st\varsigma}
    &= \int \dd 1 \dd 2 \dd 3 \dd 4
    \psi_{u\tau}^*(2)\psi_{v\tau}(1)
    \frac{\delta^2 E_{\text{Hxc}}}{\delta\rho_{\text{s}}^{\tau}(1,2)\delta\rho_{\text{s}}^{\varsigma}(3,4)}
    \psi_{s\varsigma}(3)\psi_{t\varsigma}^*(4) \\
    &= \int \dd 1 \dd 2 \dd 3 \dd 4
    \psi_{u\tau}^*(2)\psi_{v\tau}(1) K^{\tau\varsigma}(2,1;3,4)
    \psi_{s\varsigma}(3)\psi_{t\varsigma}^*(4) \\
    &= \int \dd 1 \dd 2 \dd 3 \dd 4
    \psi_{u\tau}^*(2)\psi_{v\tau}(1)
    \frac{\partial H_{\text{Hxc}}^{\tau}(2,1)}{\delta\rho_{\text{s}}^{\varsigma}(3,4)}
    \psi_{s\varsigma}(3)\psi_{t\varsigma}^*(4) \\
    &=\left\langle \psi_{u\tau}\psi_{v\tau}^*\middle|K^{\tau\varsigma}\middle|\psi_{s\varsigma}\psi_{t\varsigma}^*\right\rangle.
\end{align}
Thus, for a GKS system,
\begin{align}
    \label{eq:GKSKernel}
    K_{uv\tau,st\varsigma}
    &= \int \dd\mathbf{r}_{1} \dd\mathbf{r}'_{1} \dd\mathbf{r}_{2} \dd\mathbf{r}_{2}'
    \psi_{u\tau}^*(\mathbf{r}_{1}')\psi_{v\tau}(\mathbf{r}_{1})
    \left[\frac{\delta(\mathbf{r}_{1}-\mathbf{r}_{1}')\delta(\mathbf{r}_{2}-\mathbf{r}_{2}')}
               {|\mathbf{r}_{1}-\mathbf{r}'_{2}|}
         +\frac{\delta^2 E_{\text{xc}}}
               {\delta\rho_{\text{s}}^{\tau}(\mathbf{r}_{1},\mathbf{r}_{1}')
                \delta\rho_{\text{s}}^{\varsigma}(\mathbf{r}_{2},\mathbf{r}_{2}')}
    \right] \times \nonumber \\ &\quad
    \psi_{t\varsigma}^*(\mathbf{r}_{2})\psi_{s\varsigma}(\mathbf{r}_{2}') \\
    &= \int \dd 1 \dd 2 \dd 3 \dd 4
    \psi_{u\tau}^*(2)\psi_{v\tau}(1)
    \left[\frac{\delta(1,2)\delta(3,4)}{|\mathbf{r}_{1}-\mathbf{r}_{3}|}
    +\frac{\delta^{2}E_{\text{xc}}}{\delta\rho_{\text{s}}^{\tau}(1,2)\delta\rho_{\text{s}}^{\varsigma}(3,4)}\right]
    \psi_{t\varsigma}^*(4)\psi_{s\varsigma}(3) \\
    &= \int \dd 1 \dd 2 \dd 3 \dd 4
    \psi_{u\tau}^*(2)\psi_{v\tau}(1)K^{\tau\varsigma}(1,2;3,4)
    \psi_{s\varsigma}(3)\psi_{t\varsigma}^*(4) \\
    &=\left\langle \psi_{u\tau}\psi_{v\tau}^*\middle|K^{\varsigma\tau}\middle|\psi_{s\varsigma}\psi_{t\varsigma}^*\right\rangle,
\end{align}
and as a special case for a KS system,
\begin{align}
    \label{eq:KSKernel}
    K_{uv\tau,st\varsigma}
    &= \int \dd 1 \dd 2
    \psi_{u\tau}^*(1)\psi_{v\tau}(1)
    \left[\frac{1}{|\mathbf{r}_{1}-\mathbf{r}_{2}|}
    +\frac{\delta^2 E_{\text{xc}}}{\delta\rho_{\text{s}}^{\tau}(1)\delta\rho_{\text{s}}^{\varsigma}(2)}\right]
    \psi_{t\varsigma}^*(2)\psi_{s\varsigma}(2) \\
    &= \left\langle \psi_{u\tau}\psi_{v\tau}^*\middle|K^{\varsigma\tau}\middle|\psi_{s\varsigma}\psi_{t\varsigma}^*\right\rangle.
\end{align}
Using Eq.~\ref{eq:lr_anal_expression}, we obtain
\begin{equation}
    \frac{\partial^2 E}{\partial n_{p\tau}\partial n_{q\varsigma}}
    = K_{qq\varsigma,pp\tau}
    -\sum_{jb\sigma,kc\upsilon}
    \left(K_{qq\varsigma,jb\sigma}+K_{qq\varsigma,bj\sigma}\right)
    M_{jb\sigma,kc\upsilon}^{-1} K_{kc\upsilon,pp\tau}.
    \label{eq:maxwell_d2E_d2n_final}
\end{equation}
In the end, the expression (Eq.~\ref{eq:maxwell_d2E_d2n_final})
derived from using the Maxwell relation and linear response theory
agrees with the expression (Eq.~\ref{eq:2nd_deri_final})
from the CP equations.
The final result, Eq.~\ref{eq:maxwell_d2E_d2n_final},
is applicable to systems with integer number of
electrons \cite{2012yang144110} and also
to systems with fractional number of electrons
with the use of occupation-scaled orbitals and the
extended definition of occupied and virtual space
\cite{2013yang104114,2013peng184108}.
\clearpage

\subsection{Derivation using the density matrix as the direct variable}
Here we will show that, using the density matrix as the direct
variable will produce the same result as using occupation number
as the direct variable, with a few extra terms that are
essentially zero. In addition, we will show that second order
derivative of orbitals with respect to the occupation number is not required, even
the density relaxation is indeed truncated at the second order.

For simplicity, we ignore the notation for the spacial coordinates and
the explicit integral. Following the main text,
we consider only changing the occupation number of one specific orbital
$\psi_{n_{p\sigma}}$, and denote the set of occupation number as $[n_{p\sigma}]$.

The relaxation of the density matrix expanding on the fractional system is
\begin{equation}
    \rho_{\text{s}}^{\tau}([n_{p\sigma} + \delta]) =
    \rho_{\text{s}}^{\tau}([n_{p\sigma}]) + f^{\tau}_{p\sigma} \delta
    + \gamma^{\tau}_{p\sigma} \delta^2
    + \xi^{\tau}_{p\sigma} \delta^3 + \cdots,
\end{equation}
where
\begin{equation}
    f^{\tau}_{p\sigma}
    = \frac{\partial \rho_{\text{s}}^{\tau}}{\partial n_{p\sigma}},
\end{equation}
\begin{equation}
    \gamma^{\tau}_{p\sigma}
    = \frac12 \frac{\partial^2 \rho_{\text{s}}^{\tau}}{\partial n_{p\sigma}^2},
\end{equation}
\begin{equation}
    \xi^{\tau}_{p\sigma}
    = \frac16 \frac{\partial^3 \rho_{\text{s}}^{\tau}}{\partial n_{p\sigma}^3}.
\end{equation}

Expanding the energy of integer systems around the fractional system and
keeping all the terms up the second order, we have
\begin{align}
    \label{eq:E0_expand}
    E(0) &\approx
        E([n_{p\sigma}]) + \sum_{\tau} \frac{\delta E}{\delta \rho_{\rm{s}}^{\tau}}
        \left( -n_{p\sigma} f^{\tau}_{p\sigma} + n_{p\sigma}^2 \gamma^{\tau}_{p\sigma} \right)
        + \frac{1}{2} \sum_{\tau\upsilon} \frac{\delta^2 E}{\partial \rho_{\rm{s}}^{\tau} \partial \rho_{\rm{s}}^{\upsilon}}
        n_{p\sigma}^2 f^{\tau}_{p\sigma} f^{\upsilon}_{p\sigma},\\
    \label{eq:E1_expand}
    E(1) &\approx
        E([n_{p\sigma}]) + \sum_{\tau} \frac{\delta E}{\delta \rho_{\rm{s}}^{\tau}}
        \left[ (1 - n_{p\sigma}) f^{\tau}_{p\sigma} + (1 - n_{p\sigma})^2 \gamma^{\tau}_{p\sigma} \right]
        + \frac{1}{2} \sum_{\tau\upsilon} \frac{\delta^2 E}{\partial \rho_{\rm{s}}^{\tau} \partial \rho_{\rm{s}}^{\upsilon}}
        (1 - n_{p\sigma} )^2 f^{\tau}_{p\sigma} f^{\upsilon}_{p\sigma},
\end{align}
where $E(0)$ and $E(1)$ are the total energies of the corresponding $N$-electron
and $(N+1)$-electron system respectively.
In the above equations, to keep the notation simple, we have implied
the integration with respect to the spatial coordinates.

Based on Eq.~\ref{eq:E0_expand} and Eq.~\ref{eq:E1_expand},
the energy correction from GSC reads
\begin{align}
    \Delta_{\rm{GSC}}([n_{p\sigma}]) &= (1 - n_{p\sigma}) E(0) + n_{p\sigma} E(1) - E([n_{p\sigma}]) \\
    \nonumber
        &\approx \sum_{\tau}
        \frac{\delta E}{\delta \rho_{\rm{s}}^{\tau}}
            \Big[
                \Big(-(1-n_{p\sigma})n_{p\sigma} + n_{p\sigma}(1 - n_{p\sigma})\Big) f^{\tau}_{p\sigma}
                +
                \Big((1-n_{p\sigma})n_{p\sigma}^2 + n_{p\sigma}(1-n_{p\sigma})^2\Big) \gamma^{\tau}_{p\sigma}
            \Big] \\
        &\quad + \frac{1}{2}\sum_{\tau\upsilon}
        \frac{\delta^2 E}{\delta \rho_{\rm{s}}^{\tau} \delta \rho_{\rm{s}}^{\upsilon}}
        \Big(
            (1-n_{p\sigma})n_{p\sigma}^2 + n_{p\sigma}(1-n_{p\sigma})^2
        \Big) f^{\tau}_{p\sigma}f^{\upsilon}_{p\sigma} \\
        &= \sum_{\tau} (n_{p\sigma} - n_{p\sigma}^2) \frac{\delta E}{\delta \rho_{\rm{s}}^{\tau}} \gamma^{\tau}_{p\sigma}
         + \frac{1}{2} \sum_{\tau\upsilon} (n_{p\sigma} - n_{p\sigma}^2) \frac{\delta^2 E}{\delta \rho_{\rm{s}}^{\tau}
         \rho_{\rm{s}}^{\upsilon}}
         f^{\tau}_{p\sigma} f^{\upsilon}_{p\sigma} \\
        \label{eq:gsc_corr_in_density_relaxation}
        &= \frac{1}{2}(n_{p\sigma} - n_{p\sigma}^2)
        \Bigg(
        \sum_{\tau}
            \frac{\delta E}{\delta \rho_{\rm{s}}^{\tau}} \frac{\partial^2 \rho_{\rm{s}}^{\tau}}{\partial n_{p\sigma}^2}
         +
        \sum_{\tau\upsilon}
         \frac{\delta^2 E}{\delta \rho_{\rm{s}}^{\tau} \rho_{\rm{s}}^{\upsilon}}
         \frac{\partial \rho_{\rm{s}}^{\tau}}{\partial n_{p\sigma}}
         \frac{\partial \rho_{\rm{s}}^{\upsilon}}{\partial n_{p\sigma}}
         \Bigg)\\
        &= \frac{1}{2}(n_{p\sigma} - n_{p\sigma}^2) (\kappa^1_{p\sigma} + \kappa^2_{p\sigma}),
\end{align}
where
\begin{align}
    \label{eq:gsc_corr_K1}
    \kappa^1_{p\sigma} &= \sum_{\tau}
            \frac{\delta E}{\delta \rho_{\rm{s}}^{\tau}} \frac{\partial^2 \rho_{\rm{s}}^{\tau}}{\partial n_{p\sigma}^2},
\end{align}
and
\begin{align}
    \label{eq:gsc_corr_K2}
    \kappa^2_{p\sigma} &= \sum_{\tau\upsilon}
         \frac{\delta^2 E}{\delta \rho_{\rm{s}}^{\tau} \rho_{\rm{s}}^{\upsilon}}
         \frac{\partial \rho_{\rm{s}}^{\tau}}{\partial n_{p\sigma}}
         \frac{\partial \rho_{\rm{s}}^{\upsilon}}{\partial n_{p\sigma}}.
\end{align}
According to Eq.~\ref{eq:gsc_corr_in_density_relaxation}, the correction from
GSC involves the relaxation of density matrix up to the second order, which
involves the derivatives of orbitals with respect to occupation number up to the second
order.

Next, we are going to derive the analytical expression for
the coefficients $\kappa^1_{p\sigma}$ and $\kappa^2_{p\sigma}$,
and show Eq.~\ref{eq:gsc_corr_in_density_relaxation}
gives the identical results to Eq.~\ref{eq:d2E_d2n_Kernel_deriOrb},
which is derived from using occupation number as the direct variables.
The first order relaxation of the density matrix with respect to the occupation
number $n_{p\sigma}$ is
\begin{align}
    \frac{\partial \rho_{\text{s}}^{\tau}}{\partial n_{p\sigma}}
    = \delta_{\sigma\tau} \psi_{p\sigma}\psi_{p\sigma}^*
    + \sum_k n_{k\tau}
    \left[\psi_{k\tau} \frac{\partial \psi_{k\tau}^*}{\partial n_{p\sigma}}+
    \frac{\partial \psi_{k\tau}}{\partial n_{p\sigma}}
    \psi_{k\tau}^*\right].
    \label{eq:deri1_drho_dn}
\end{align}
The second order relaxation of the density matrix with respect to the occupation
number $n_{p\sigma}$ is
\begin{align}
    \frac{\partial^2 \rho_{\text{s}}^{\tau}}{\partial n_{p\sigma}^2}
    &=
    \frac{\partial}{\partial n_{p\sigma}}
    \left\{\delta_{\sigma\tau} \psi_{p\sigma} \psi_{p\sigma}
    + \sum_k n_{k\tau}
    \left[\psi_{k\tau} \frac{\partial \psi_{k\tau}^*}{\partial n_{p\sigma}}+
    \frac{\partial \psi_{k\tau}}{\partial n_{p\sigma}}
    \psi_{k\tau}^*\right]
    \right\} \\
    &=
    2\delta_{\sigma\tau}
    \left[
    \psi_{p\sigma}
    \frac{\partial \psi_{p\sigma}^*}{\partial n_{p\sigma}}
    +
    \frac{\partial \psi_{p\sigma}}{\partial n_{p\sigma}}
    \psi_{p\sigma}^*
    \right] + \sum_k n_{k\tau}
    \left[
    \psi_{k\tau}
    \frac{\partial^2 \psi_{k\tau}^*}{\partial n_{p\sigma}^2}
    +
    2\frac{\partial \psi_{k\tau}}{\partial n_{p\sigma}}
    \frac{\partial \psi_{k\tau}^*}{\partial n_{p\sigma}}
    +
    \frac{\partial^2 \psi_{k\tau}}{\partial n_{p\sigma}^2}
    \psi_{k\tau}^*
    \right].
    \label{eq:deri2_drho_dn}
\end{align}
Substituting Eq.~\ref{eq:deri1_drho_dn} into Eq.~\ref{eq:gsc_corr_K1},
we have $\kappa^1_{p\sigma}$ expressed as
\begin{align}
    \kappa^1_{p\sigma}
    &=
    2 \left(
        \left \langle
        \psi_{p\sigma}
        \middle|
        H_{\text{s}}^{\sigma}
        \middle|
        \frac{\partial \psi_{p\sigma}}{\partial n_{p\sigma}}
        \right \rangle
        +
        \left \langle
        \frac{\partial \psi_{p\sigma}}{\partial n_{p\sigma}}
        \middle|
        H_{\text{s}}^{\sigma}
        \middle|
        \psi_{p\sigma}
        \right \rangle
    \right) \nonumber \\
    &\quad + \sum_{k\tau} n_{k\tau}
    \left(
        \left \langle
        \psi_{k\tau}
        \middle|
        H_{\text{s}}^{\tau}
        \middle|
        \frac{\partial^2 \psi_{k\tau}}{\partial n_{p\sigma}^2}
        \right \rangle
        + 2
        \left \langle
        \frac{\partial \psi_{k\tau}}{\partial n_{p\sigma}}
        \middle|
        H_{\text{s}}^{\tau}
        \middle|
        \frac{\partial \psi_{k\tau}}{\partial n_{p\sigma}}
        \right \rangle
        +
        \left \langle
        \frac{\partial^2 \psi_{k\tau}}{\partial n_{p\sigma}^2}
        \middle|
        H_{\text{s}}^{\tau}
        \middle|
        \psi_{k\tau}
        \right \rangle
    \right) \\
    &=\sum_{k\tau} n_{k\tau}
    \left(
        \left \langle
        \psi_{k\tau}
        \middle|
        H_{\text{s}}^{\tau}
        \middle|
        \frac{\partial^2 \psi_{k\tau}}{\partial n_{p\sigma}^2}
        \right \rangle
        + 2
        \left \langle
        \frac{\partial \psi_{k\tau}}{\partial n_{p\sigma}}
        \middle|
        H_{\text{s}}^{\tau}
        \middle|
        \frac{\partial \psi_{k\tau}}{\partial n_{p\sigma}}
        \right \rangle
        +
        \left \langle
        \frac{\partial^2 \psi_{k\tau}}{\partial n_{p\sigma}^2}
        \middle|
        H_{\text{s}}^{\tau}
        \middle|
        \psi_{k\tau}
        \right \rangle
    \right),
    \label{eq:K1_final}
\end{align}
in which we use the property that the orbital is the eigenfunction of the
(generalized) KS Hamiltonian and
$\frac{\partial}{\partial n_{p\sigma}}
\left \langle \psi_{p\sigma} | \psi_{p\sigma} \right \rangle\equiv 0
$ because of the constraint of orthonormality.
Substituting Eq.~\ref{eq:deri2_drho_dn} into Eq.~\ref{eq:gsc_corr_K2},
we have $\kappa^2_{p\sigma}$ expressed as
\begin{align}
    \nonumber
    \kappa^2_{p\sigma} &=
        \langle \psi_{p\sigma}\psi_{p\sigma} | K^{\sigma\sigma}|
        \psi_{p\sigma}\psi_{p\sigma} \rangle
        + 2 \sum_{k\tau} n_{k\tau}
        \Bigg[
        \left \langle \psi_{p\sigma} \psi_{p\sigma}
        \middle | K^{\sigma\tau} \middle | \frac{\partial \psi_{k\tau}}{\partial n_{p\sigma}} \psi_{k\tau}
        \right \rangle
        +
        \left \langle \psi_{p\sigma} \psi_{p\sigma}
        \middle | K^{\sigma\tau} \middle | \psi_{k\tau} \frac{\partial \psi_{k\tau}}{\partial n_{p\sigma}}
        \right \rangle
        \Bigg] \\
        \nonumber
        &\quad
        + \sum_{k\tau,l\upsilon} n_{k\tau} n_{l\upsilon}
        \Bigg[
        \left \langle
        \frac{\partial \psi_{k\tau}}{\partial n_{p\sigma}} \psi_{k\tau}
        \middle | K^{\sigma\tau} \middle |
        \frac{\partial \psi_{l\upsilon}}{\partial n_{p\sigma}} \psi_{l\upsilon}
        \right \rangle
        +
        \left \langle
        \frac{\partial \psi_{k\tau}}{\partial n_{p\sigma}} \psi_{k\tau}
        \middle | K^{\sigma\tau} \middle |
        \psi_{l\upsilon} \frac{\partial \psi_{l\upsilon}}{\partial n_{p\sigma}}
        \right \rangle \\
        &\quad +
        \left \langle
        \psi_{k\tau} \frac{\partial \psi_{k\tau}}{\partial n_{p\sigma}}
        \middle | K^{\sigma\tau} \middle |
        \psi_{l\upsilon} \frac{\partial \psi_{l\upsilon}}{\partial n_{p\sigma}}
        \right \rangle
        +
        \left \langle
        \psi_{k\tau} \frac{\partial \psi_{k\tau}}{\partial n_{p\sigma}}
        \middle | K^{\sigma\tau} \middle |
        \frac{\partial \psi_{l\upsilon}}{\partial n_{p\sigma}} \psi_{l\upsilon}
        \right \rangle
        \Bigg]
        \label{eq:K2_final}
\end{align}
Combining Eq.~\ref{eq:K1_final} and Eq.~\ref{eq:K2_final},
we get the complete coefficients $\kappa^1_{p\sigma} + \kappa^2_{p\sigma}$
that are derived from density relaxation.
We use \(\kappa(\rho)\) to represent the one we obtain here, which
uses the density matrix as the direct variable, and
\(\kappa(n)\) to represent the one we derived in the main text, which uses
the occupation number as the direct variable.
Comparing $\kappa(n)$ (Eq.~\ref{eq:d2E_d2n_Kernel_deriOrb})
and $\kappa(\rho)$ (Eq.~\ref{eq:K1_final} and Eq.~\ref{eq:K2_final}),
we see there are additional terms that are expressed as
\begin{align}
    \nonumber
    \kappa(\rho) - \kappa(n) &=
        \sum_{k\tau} n_{k\tau}
    \Bigg[
        \left \langle
        \psi_{k\tau}
        \middle|
        H_{\text{s}}^{\tau}
        \middle|
        \frac{\partial^2 \psi_{k\tau}}{\partial n_{p\sigma}^2}
        \right \rangle
        + 2
        \left \langle
        \frac{\partial \psi_{k\tau}}{\partial n_{p\sigma}}
        \middle|
        H_{\text{s}}^{\tau}
        \middle|
        \frac{\partial \psi_{k\tau}}{\partial n_{p\sigma}}
        \right \rangle
        +
        \left \langle
        \frac{\partial^2 \psi_{k\tau}}{\partial n_{p\sigma}^2}
        \middle|
        H_{\text{s}}^{\tau}
        \middle|
        \psi_{k\tau}
        \right \rangle
    \Bigg] \\
    \nonumber
    &\quad + \sum_{k\tau} n_{k\tau}
        \Bigg[
        \left \langle \psi_{p\sigma} \psi_{p\sigma}
        \middle | K^{\sigma\tau} \middle | \frac{\partial \psi_{k\tau}}{\partial n_{p\sigma}} \psi_{k\tau}
        \right \rangle
        +
        \left \langle \psi_{p\sigma} \psi_{p\sigma}
        \middle | K^{\sigma\tau} \middle | \psi_{k\tau} \frac{\partial \psi_{k\tau}}{\partial n_{p\sigma}}
        \right \rangle
        \Bigg] \\
        \nonumber
        &\quad + \sum_{k\tau,l\upsilon} n_{k\tau} n_{l\upsilon}
        \Bigg[
        \left \langle
        \frac{\partial \psi_{k\tau}}{\partial n_{p\sigma}} \psi_{k\tau}
        \middle | K^{\sigma\tau} \middle |
        \frac{\partial \psi_{l\upsilon}}{\partial n_{p\sigma}} \psi_{l\upsilon}
        \right \rangle
        +
        \left \langle
        \frac{\partial \psi_{k\tau}}{\partial n_{p\sigma}} \psi_{k\tau}
        \middle | K^{\sigma\tau} \middle |
        \psi_{l\upsilon} \frac{\partial \psi_{l\upsilon}}{\partial n_{p\sigma}}
        \right \rangle \\
        &\quad +
        \left \langle
        \psi_{k\tau} \frac{\partial \psi_{k\tau}}{\partial n_{p\sigma}}
        \middle | K^{\sigma\tau} \middle |
        \psi_{l\upsilon} \frac{\partial \psi_{l\upsilon}}{\partial n_{p\sigma}}
        \right \rangle
        +
        \left \langle
        \psi_{k\tau} \frac{\partial \psi_{k\tau}}{\partial n_{p\sigma}}
        \middle | K^{\sigma\tau} \middle |
        \frac{\partial \psi_{l\upsilon}}{\partial n_{p\sigma}} \psi_{l\upsilon}
        \right \rangle
        \Bigg].
\end{align}
However, these additional terms are zero.
Recall the $T_2$ term defined in Eq.~\ref{eq:d2E_d2n_T2_def}, which has been
shown to be zero.
Following the similar steps shown in Eqs.~\ref{eq:d2E_d2n_deriOrbE}
-~\ref{eq:d2E_d2n_Kernel_deriOrb} to derive the derivative of (G)KS
Hamiltonian, one can evaluate the derivatives term by term involved in
Eq.~\ref{eq:d2E_d2n_T2_def}. Then it is
easy to find that $\kappa(\rho) - \kappa(n)$ agrees to the $T_2$ term,
namely,
\begin{align}
    \kappa(\rho) - \kappa(n) = T_2 = 0.
\end{align}
Therefore, we prove that the coefficients in GSC2
derived from using density matrix or using occupation number as the
variables are identical. In addition, we show that the second order derivative
of orbitals with respect to occupation is not needed, even though
the density relaxation is truncated at the second order.

\clearpage

\section{Physical interpretation of the orbital hardness matrix $\frac{\partial^2 E}{\partial n_{p\tau}\partial n_{q\varsigma}}$}

We now examine the physical meaning of the \emph{orbital hardness matrix},
$\frac{\partial^{2}E}{\partial n_{p\tau}\partial n_{q\varsigma}}$,
starting from the analytical expression Eq.~\ref{eq:maxwell_d2E_d2n_final}.
First, we define the generalized 4-point dielectric function $\varepsilon^{\sigma\tau}$
through its inverse, $\left(\varepsilon^{-1}\right)^{\sigma\tau}$,
as
\begin{align}
    \left(\varepsilon^{-1}\right)^{\sigma\tau}
    (\mathbf{r}_{1},\mathbf{r}_{2};\mathbf{r}_{3},\mathbf{r}_{4})
    &=\frac{\delta H_{\text{s}}^{\sigma}(\mathbf{r}_{1},\mathbf{r}_{2})}
           {\delta v^{\tau}(\mathbf{r}_{3},\mathbf{r}_{4})} \\
	&=\frac{\delta v_{\text{s}}^{\sigma}(\mathbf{r}_{1},\mathbf{r}_{2})}
           {\delta v^{\tau}(\mathbf{r}_{3},\mathbf{r}_{4})}.
    \label{eq:4PInverseEpsilon}
\end{align}
This 4-point dielectric function extends the 2-point dielectric function
commonly used many-body perturbation theory \cite{martin2016interacting}
in three ways: (1) The nonlocal potential for the generalized Kohn-Sham
noninteracting system is used instead of the local KS potential;
(2) The entire one-body potential $v_{\text{s}}^{\sigma}(\mathbf{r}_{1},\mathbf{r}_{2})$,
with classical Coulomb and exchange-correlation contributions,
is used instead of only the classical Coulomb potential only;
(3) $v^{\sigma}(\mathbf{r},\mathbf{r}')$, the generalized external potential,
which is nonlocal and spin-dependent, is used as an extension from the physical potential,
$v(\mathbf{r})$, which is local and spin independent. \cite{2013peng184108}
Since
\begin{equation}
    v_{\text{s}}^{\sigma}(\mathbf{r}_{1},\mathbf{r}_{2})
    =v^{\sigma}(\mathbf{r}_{1},\mathbf{r}_{2})+v_{\text{Hxc}}^{\sigma}(\mathbf{r}_{1},\mathbf{r}_{2}),
\end{equation}
we have
\begin{align}
    \left(\varepsilon^{-1}\right)^{\sigma\tau}(\mathbf{r}_{1},\mathbf{r}_{2};\mathbf{r}_{3},\mathbf{r}_{4})
    &=
    \delta(\mathbf{r}_{1},\mathbf{r}_{3})\delta(\mathbf{r}_{2},\mathbf{r}_{4})\delta_{\sigma\tau}
        +\frac{\delta v_{\text{Hxc}}^{\sigma}(\mathbf{r}_{1},\mathbf{r}_{2})}
              {\delta v^{\tau}(\mathbf{r}_{3},\mathbf{r}_{4})} \\
	&=
    \delta(\mathbf{r}_{1},\mathbf{r}_{3})\delta(\mathbf{r}_{2},\mathbf{r}_{4})\delta_{\sigma\tau}
    + \sum_{\upsilon}\frac{\delta v_{\text{Hxc}}^{\sigma}(\mathbf{r}_{1},\mathbf{r}_{2})}
           {\delta\rho_{\text{s}}^{\upsilon}(\mathbf{r}_{5},\mathbf{r}_{6})}
      \frac{\delta\rho_{\text{s}}^{\upsilon}(\mathbf{r}_{5},\mathbf{r}_{6})}
           {\delta v^{\tau}(\mathbf{r}_{3},\mathbf{r}_{4})} \\
	&=
    \delta(\mathbf{r}_{1},\mathbf{r}_{3})\delta(\mathbf{r}_{2},\mathbf{r}_{4})\delta_{\sigma\tau}
    + \sum_{\upsilon}K^{\sigma\upsilon}(\mathbf{r}_{1},\mathbf{r}_{2};\mathbf{r}_{5},\mathbf{r}_{6})
    \chi^{\upsilon\tau}(\mathbf{r}_{5},\mathbf{r}_{6};\mathbf{r}_{3},\mathbf{r}_{4}),
\end{align}
where spacial integration is implied. Then, from Eq.~\ref{eq:KKXK},
\begin{align}
    \frac{\partial^{2}E}{\partial n_{p\tau}\partial n_{q\varsigma}}
    &=\left\langle \psi_{q\varsigma}\psi_{q\varsigma}^{*}\right|K^{\varsigma\tau}
    +\sum_{\sigma\upsilon}K^{\varsigma\sigma}\chi^{\sigma\upsilon}K^{\upsilon\tau}
    \left|\psi_{p\tau}\psi_{p\tau}^{*}\right\rangle \\
	&=\left\langle \psi_{q\varsigma}\psi_{q\varsigma}^{*}\right|
    \sum_{\upsilon}\left(\varepsilon^{-1}\right)^{\varsigma\upsilon}K^{\upsilon\tau}
    \left|\psi_{p\tau}\psi_{p\tau}^{*}\right\rangle \\
	&=\left\langle \psi_{q\varsigma}\psi_{q\varsigma}^{*}\right|
    W^{\varsigma\tau}
    \left|\psi_{p\tau}\psi_{p\tau}^{*}\right\rangle,
    \label{eq:HardnessAsScreenedInteraction}
\end{align}
where we defined the 4-point generalized screened interaction as
\begin{equation}
    W^{\varsigma\tau}=\sum_{\upsilon}\left(\varepsilon^{-1}\right)^{\varsigma\upsilon}K^{\upsilon\tau}.
    \label{eq:W_def}
\end{equation}
If we set the nonlocal exchange-correlation contributions in
$v_{\text{Hxc}}^{\sigma}(\mathbf{r}_{1},\mathbf{r}_{2})$ to zero, the 4-point kernel
$K^{\upsilon\tau}$ would be reduced to the 2-point spinless classical Coulomb potential
and the 4-point generalized screened interaction $W^{\varsigma\tau}$ would be reduced the
commonly used 2-point spinless screened interaction.
Therefore, the 4-point generalized functions, $\chi^{\upsilon\tau}$,
$\left(\varepsilon^{-1}\right)^{\sigma\tau}$ and $W^{\sigma\varsigma}$
are the natural extensions of the corresponding spinless two-point functions
commonly used in many-body perturbation theory. \cite{martin2016interacting}

We can now interpret the orbital hardness matrix element
$\frac{\partial^{2}E}{\partial n_{p\tau}\partial n_{q\varsigma}}$
as the generalized screened interaction between orbital $q$ and orbital $p$.
This should have important consequence for systems where screening is critical,
such as in 2-dimension and 3-dimension extended systems.

\clearpage

\section{Computational details and results}
\begin{figure}[htbp]
    \centering
    \includegraphics[width=0.6\linewidth]{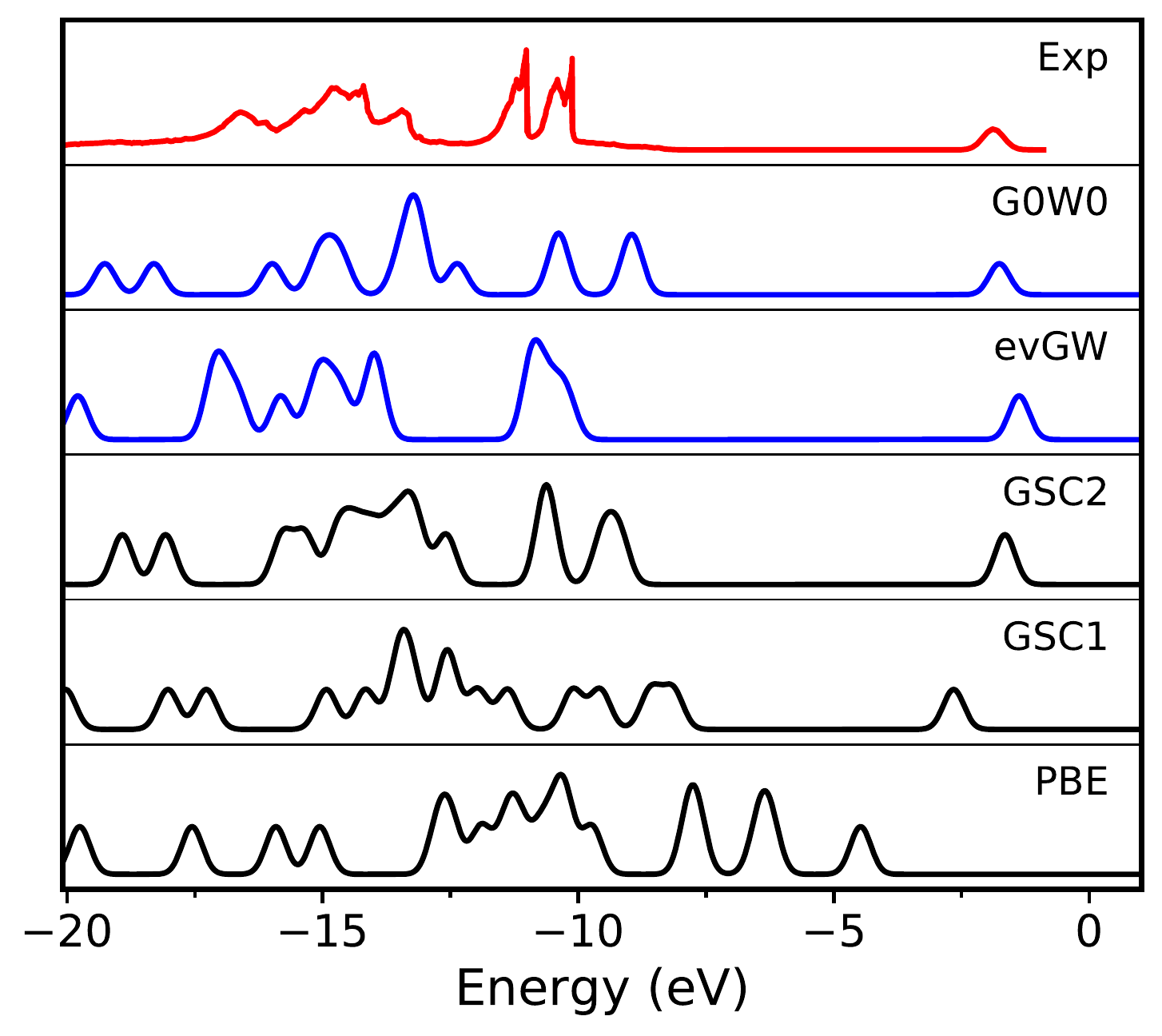}
    \caption{Photoemission spectrum of benzoquinone. The experimental
    spectrum was obtained from Ref. \citenum{1972brundle1475}}
    \label{fig:spec_2}
\end{figure}

\begin{figure}[htbp]
    \centering
    \includegraphics[width=0.6\linewidth]{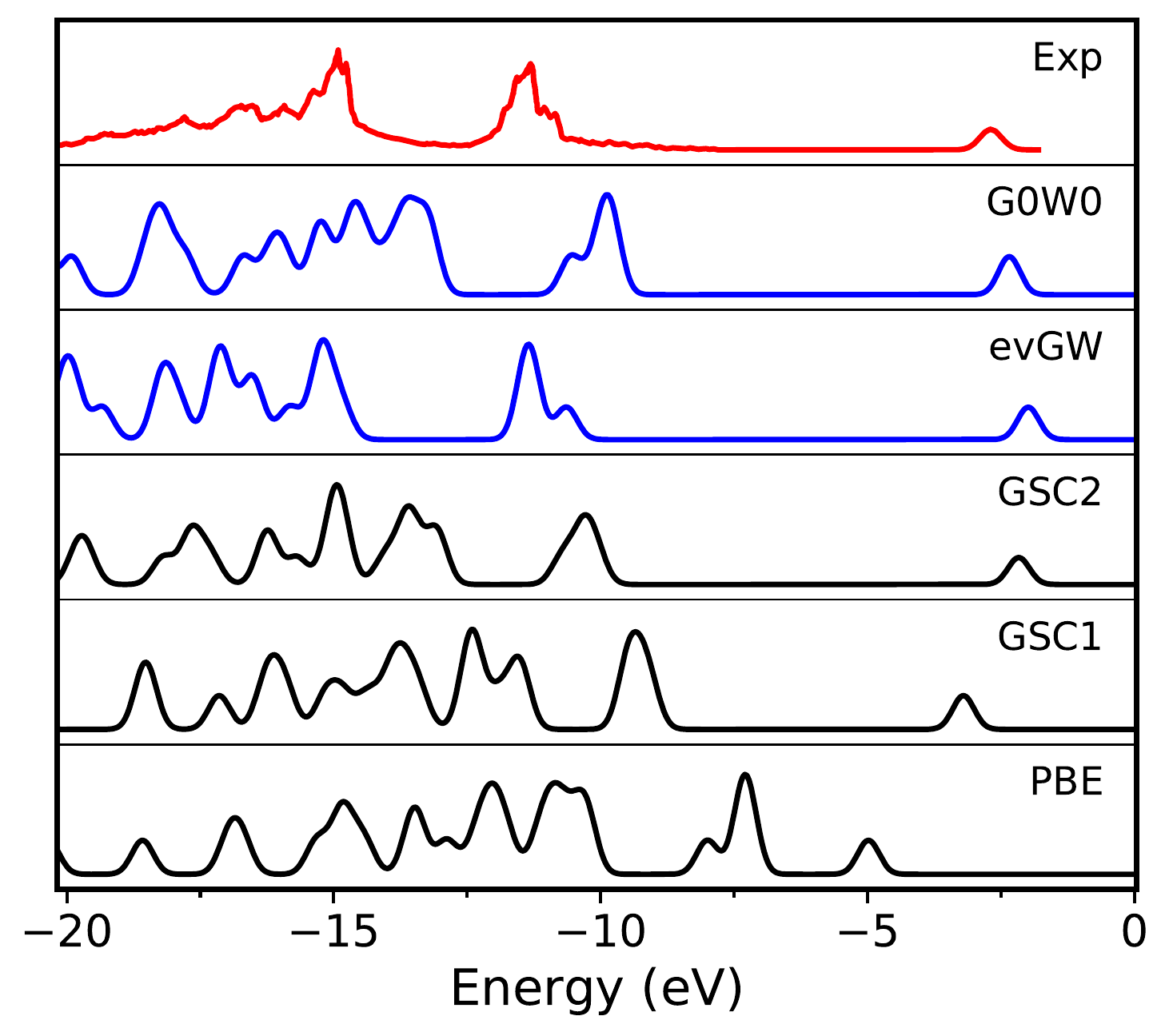}
    \caption{Photoemission spectrum of tetrafluorobenzoquinone.
    The experimental spectrum was obtained from Ref.
    \citenum{1972brundle1475}}
    \label{fig:spec_4}
\end{figure}

\begin{figure}[htbp]
    \centering
    \includegraphics[width=0.6\linewidth]{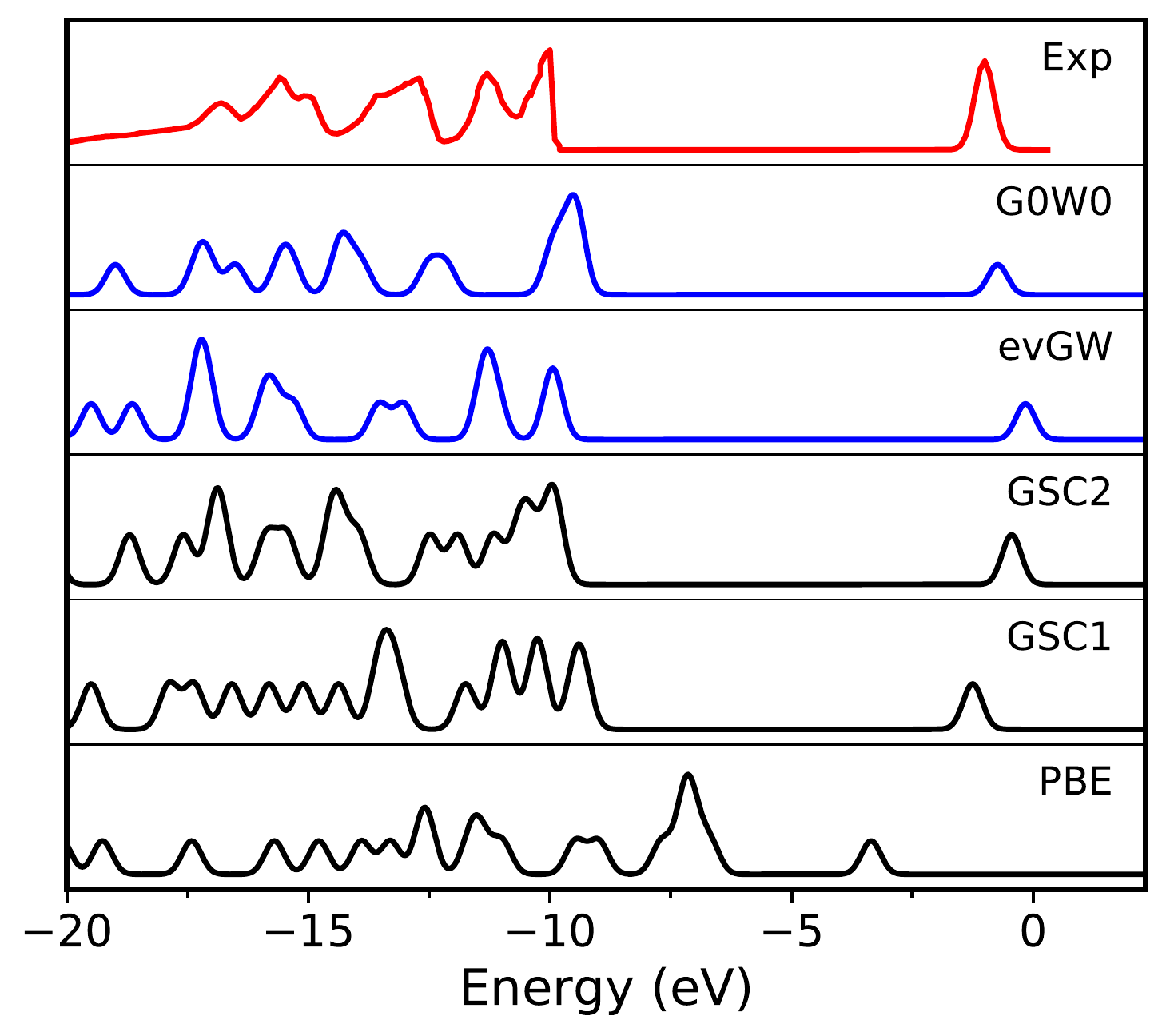}
    \caption{Photoemission spectrum of nitrobenzene.
    The experimental spectrum was obtained from Ref.
    \citenum{2003rabalais960}}
    \label{fig:spec_5}
\end{figure}

\begin{figure}[htbp]
    \centering
    \includegraphics[width=0.6\linewidth]{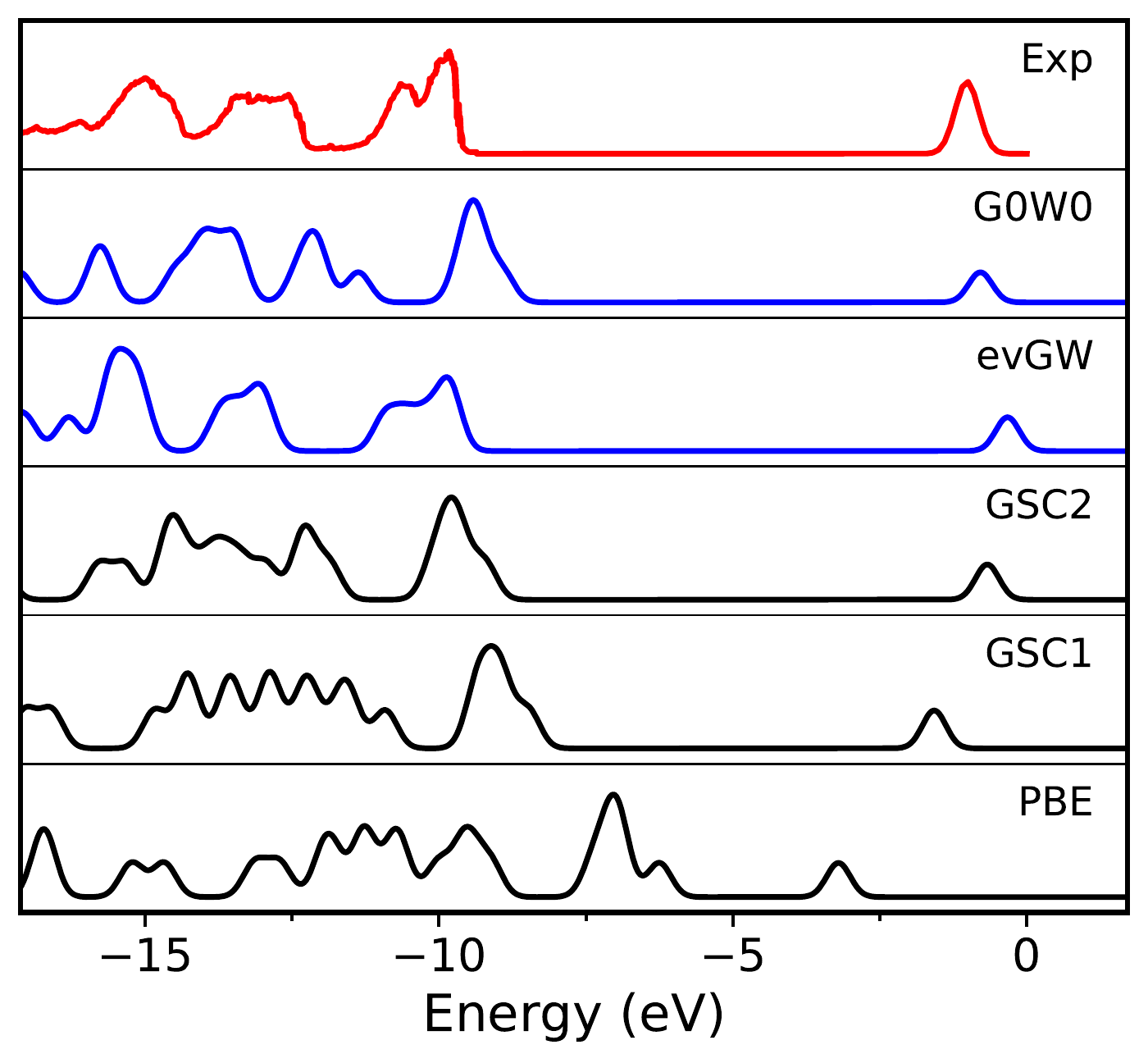}
    \caption{Photoemission spectrum of Phthalimide.
    The experimental spectrum was obtained from Ref.
    \citenum{1977galasso237}}
    \label{fig:spec_6}
\end{figure}

\begin{figure}[htbp]
    \centering
    \includegraphics[width=0.6\linewidth]{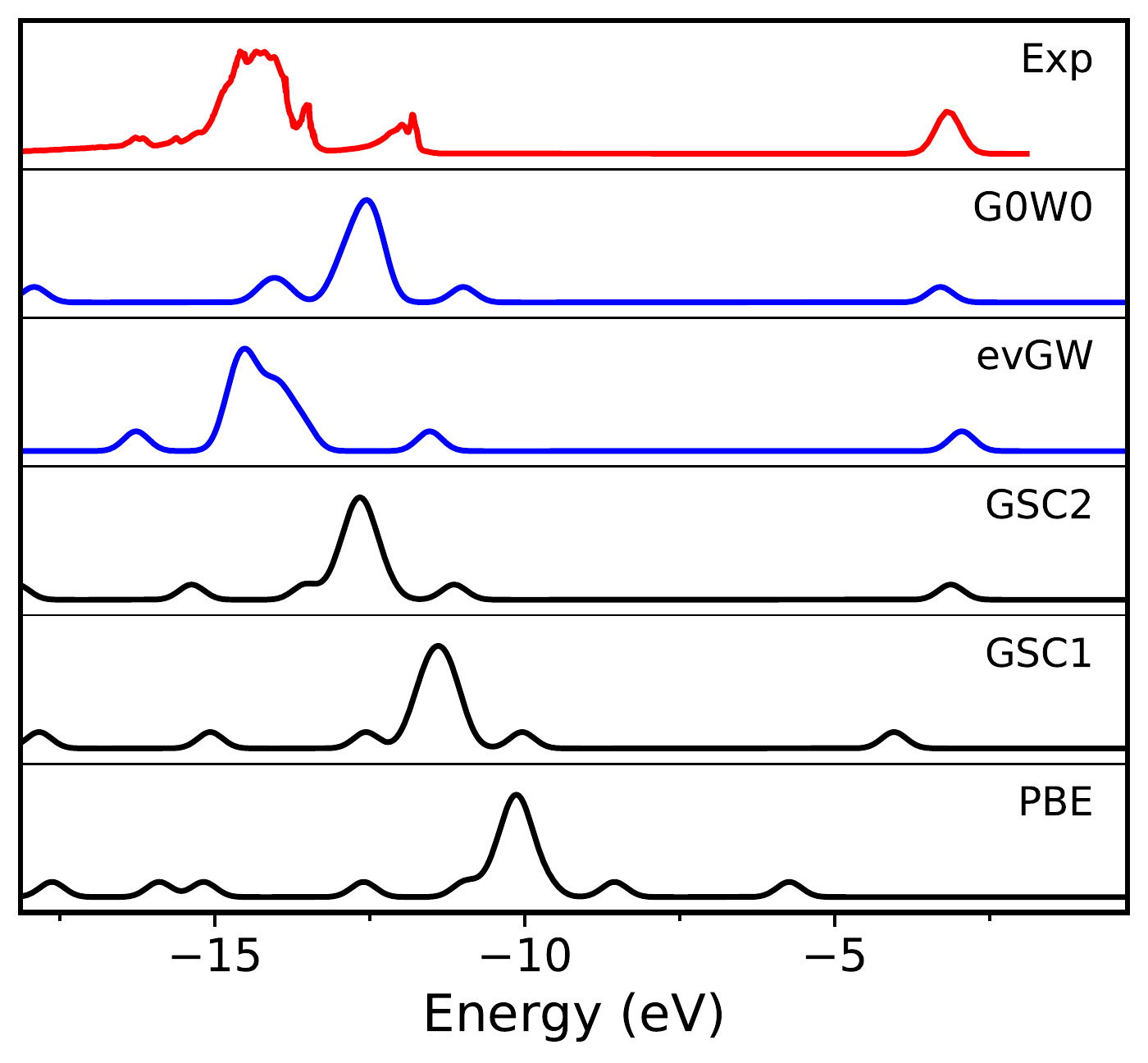}
    \caption{Photoemission spectrum of TCNE.
    The experimental spectrum was obtained from Ref.
    \citenum{1974ikemoto790}}
    \label{fig:spec_7}
\end{figure}

\begin{figure}[htbp]
    \centering
    \includegraphics[width=0.6\linewidth]{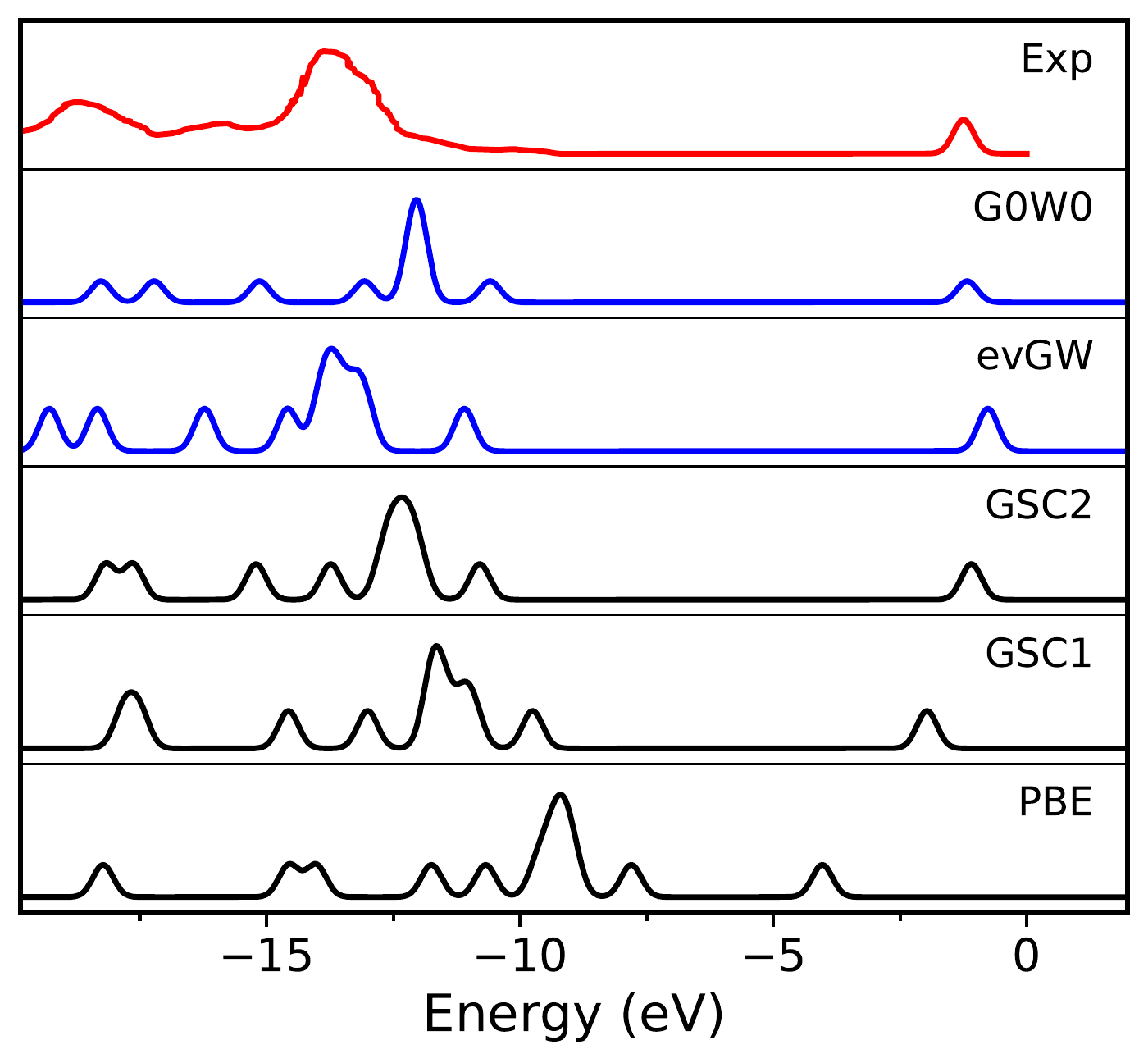}
    \caption{Photoemission spectrum of fumaronitrile.
    The experimental spectrum was obtained from Ref.
    \citenum{1976fujikawa1492}}
    \label{fig:spec_10}
\end{figure}

\begin{figure}[htbp]
    \centering
    \includegraphics[width=0.6\linewidth]{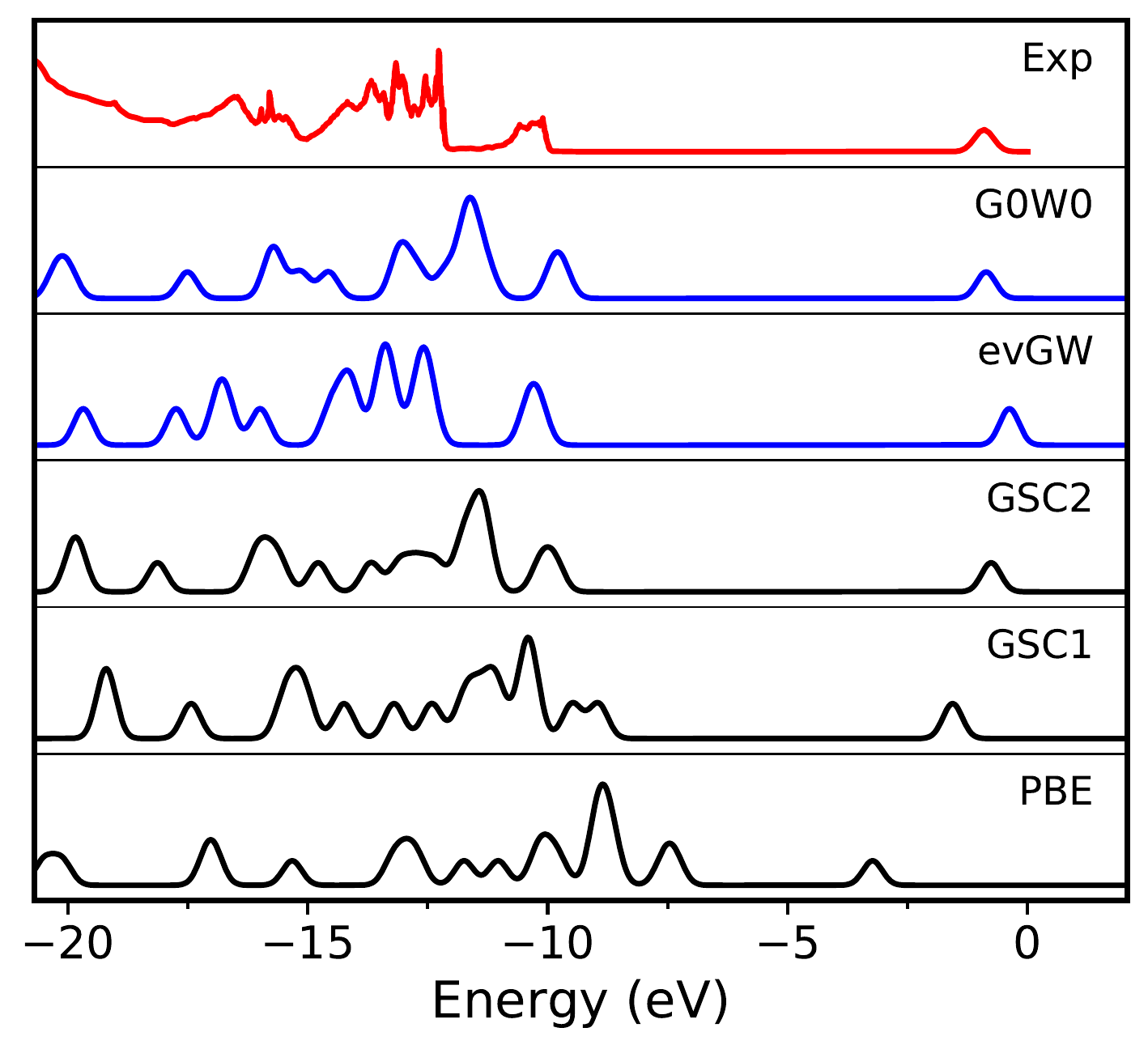}
    \caption{Photoemission spectrum of mDCNB.
    The experimental spectrum was obtained from Ref.
    \citenum{1978neijzen199}}
    \label{fig:spec_11}
\end{figure}

\begin{figure}[htbp]
    \centering
    \includegraphics[width=0.6\linewidth]{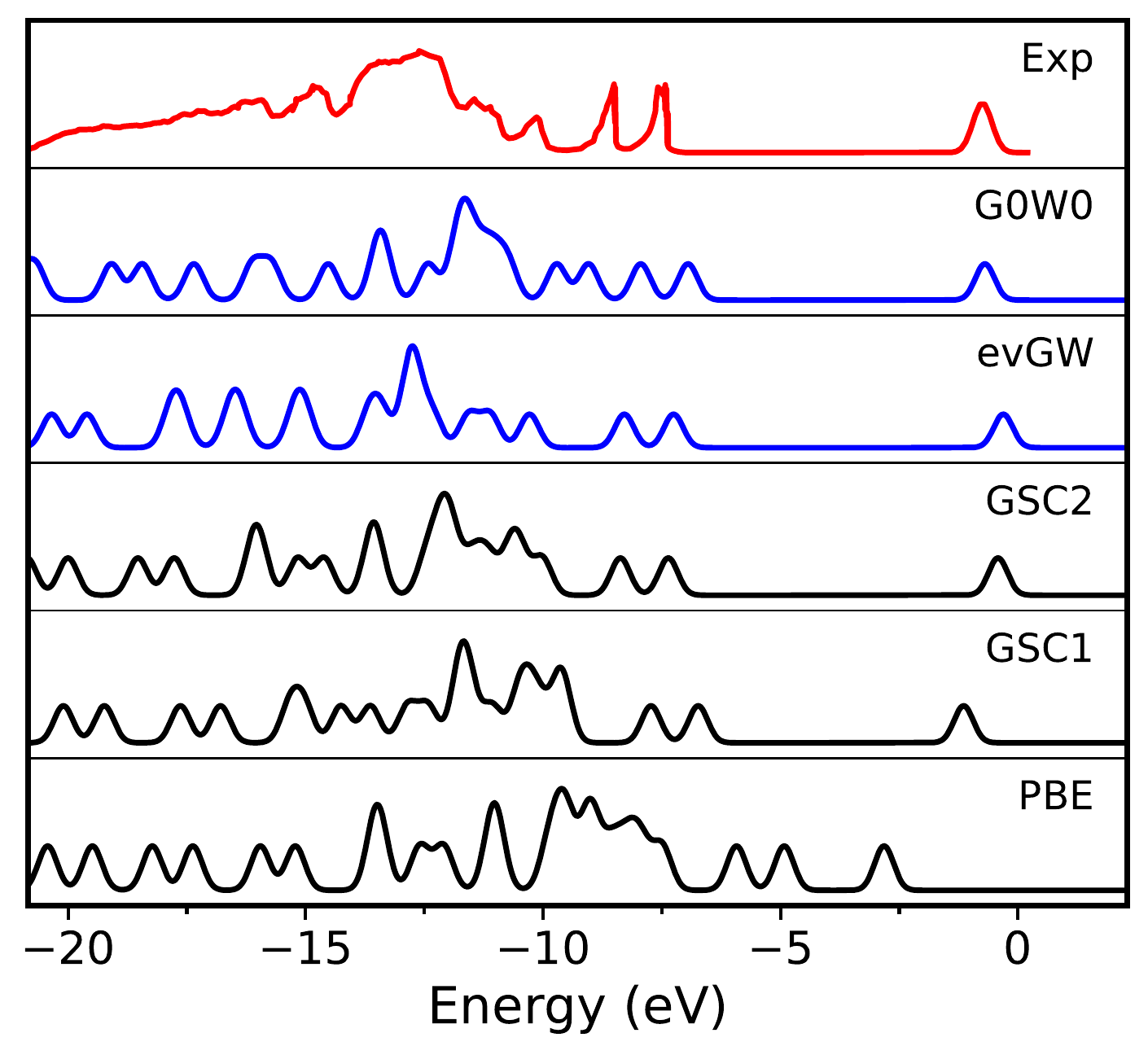}
    \caption{Photoemission spectrum of azulene.
    The experimental spectrum was obtained from Ref.
    \citenum{1980dougherty33}}
    \label{fig:spec_14}
\end{figure}

Four conventional DFAs (LDA,
\cite{1980vosko1211,1992perdew13249}
PBE,
\cite{1996perdew3868}
BLYP
\cite{1988becke3100,1988lee789}
and B3LYP
\cite{1994stephens11627,1994kim10094,1996perdew3868}) were tested as the
parent DFAs for GSC method. In the following test,
all the DFT and $\Delta$-SCF calculations were performed from using the QM4D package
\cite{qm4d}, and the GW calculations were performed from using
FHI-aims package.
\cite{blumInitioMolecularSimulations2009,renResolutionofidentityApproachHartree2012}
The mean absolute error (MAE) and mean signed error (MSE)
were calculated to evaluate the performances of tested methods.
Table \ref{tab:F_EN_curve} shows the results for the
$E(N)$ curve of F atom.
Table \ref{tab:IP_N} - \ref{tab:IP_delta_scf} show the results
for the first ionization potentials (IPs).
Table \ref{tab:EA_N} - \ref{tab:EA_delta_scf} show the results
for the first electron affinities (EAs).
Table \ref{tab:quasihole} shows the results
for the quasihole energies.
Table \ref{tab:exci} shows the results for the
low-lying excitation energies.
Notation NA in all the tables indicates that
the results are not available because of the failures in the
self-consistent calculations.
Figure \ref{fig:spec_2} - \ref{fig:spec_14} show the
photoemission spectra for the tested molecules.

\subsection{$E(N)$ curve of F atom}
\begin{table}[htbp]
    \caption{The total energies for F atom with number of electrons being
        in $[9, 10]$ from BLYP, GSC1-BLYP, and GSC2-BLYP.
        Linear reference is interpolated linearly between
        the two integer points.}
    \label{tab:F_EN_curve}
\begin{tabular}{@{}rrrrr@{}}
\toprule
N    & BLYP           & GSC1-BLYP       & GSC2-BLYP      & Linear Reference \\ \midrule
9.0  & $-$99.7586697141 & $-$99.7586697141 & $-$99.7586697141 & $-$99.7586697141   \\
9.1  & $-$99.7910387558 & $-$99.7673278223 & $-$99.7711047567 & $-$99.7721799060   \\
9.2  & $-$99.8189805356 & $-$99.7771717987 & $-$99.7839621508 & $-$99.7856900980   \\
9.3  & $-$99.8425470780 & $-$99.7881629614 & $-$99.7973966677 & $-$99.7992002899   \\
9.4  & $-$99.8618153550 & $-$99.8002586975 & $-$99.8113965817 & $-$99.8127104818   \\
9.5  & $-$99.8768836804 & $-$99.8134172286 & $-$99.8258227644 & $-$99.8262206738   \\
9.6  & $-$99.8878684040 & $-$99.8275911985 & $-$99.8404339680 & $-$99.8397308657   \\
9.7  & $-$99.8949014513 & $-$99.8427447254 & $-$99.8549088429 & $-$99.8532410576   \\
9.8  & $-$99.8981267240 & $-$99.8588411122 & $-$99.8688855091 & $-$99.8667512495   \\
9.9  & $-$99.8976971851 & $-$99.8758533756 & $-$99.8819674247 & $-$99.8802614415   \\
10.0 & $-$99.8937716334 & $-$99.8937716334 & $-$99.8937716334 & $-$99.8937716334   \\ \bottomrule
\end{tabular}
\end{table}

\clearpage
\newgeometry{left=0.5in, right=0.5in, top=1in, bottom=1in}
\begin{landscape}
\subsection{The ionization potentials (IPs)}
The same test set in Ref. \citenum{2020mei10277} was used in this work
to test the IPs. See Ref. \citenum{2020mei10277} for details
about the molecular geometries and the experimental energies.
6-311++G(3df, 3pd) is used as the basis set for all the calculations.

\renewcommand\arraystretch{0.58}
\small{
\begin{longtable}{lr|rrrr|rrrr|rrrr}
    \caption{The experimental first ionization potentials of $N$-electron systems
            versus the negative HOMO energies of $N$-electron systems from different DFAs.
            Units are in eV.}
    \label{tab:IP_N} \\
    \toprule
             &       & \multicolumn{4}{c|}{DFAs}          & \multicolumn{4}{c|}{GSC1-DFAs}                & \multicolumn{4}{c}{GSC2-DFAs}                \\ \midrule
Mol          & Exp.  & LDA    & PBE    & BLYP   & B3LYP  & LDA      & PBE      & BLYP      & B3LYP      & LDA      & PBE      & BLYP      & B3LYP      \\ \midrule
    \endfirsthead
    \multicolumn{14}{l}{Table \ref{tab:IP_N} continued: IPs from $N$-systems.} \\ \toprule
             &       & \multicolumn{4}{c|}{DFAs}          & \multicolumn{4}{c|}{GSC1-DFAs}                & \multicolumn{4}{c}{GSC2-DFAs}                \\ \midrule
Mol          & Exp.  & LDA    & PBE    & BLYP   & B3LYP  & LDA      & PBE      & BLYP      & B3LYP      & LDA      & PBE      & BLYP      & B3LYP      \\ \midrule
    \endhead
    \midrule
    \endfoot
    \endlastfoot
          H4C & 14.40 &  9.45 &  9.43 &  9.37 & 10.75 & 13.77 & 13.72 & 13.71 & 14.30 & 14.11 & 14.03 & 13.98 & 14.23 \\
          H3N & 11.03 &  6.33 &  6.24 &  6.17 &  7.56 & 11.49 & 11.42 & 11.34 & 11.72 & 11.08 & 10.89 & 10.81 & 10.93 \\
           HO & 13.07 &  7.40 &  7.36 &  7.35 &  8.99 & 13.54 & 13.54 & 13.53 & 13.95 & 13.16 & 13.03 & 13.01 & 13.11 \\
          H2O & 12.74 &  7.37 &  7.22 &  7.18 &  8.81 & 13.38 & 13.25 & 13.21 & 13.66 & 12.84 & 12.57 & 12.51 & 12.60 \\
           HF & 16.20 &  9.78 &  9.60 &  9.59 & 11.51 & 16.98 & 16.81 & 16.80 & 17.30 & 16.35 & 16.02 & 15.98 & 16.05 \\
         H4Si & 12.84 &   NA &  8.51 &  8.43 &  9.65 &   NA & 11.57 & 11.96 & 12.48 &   NA & 12.22 & 12.33 & 12.63 \\
           HP & 10.18 &   NA &  6.18 &  5.89 &  7.02 &   NA &  9.60 &  9.30 &  9.76 &   NA & 10.14 &  9.85 & 10.10 \\
          H2P &  9.82 &  5.98 &  6.04 &  5.78 &   NA &  9.35 &  9.43 &  9.16 &   NA &  9.86 &  9.90 &  9.65 &   NA \\
          H3P & 10.61 &  6.77 &  6.72 &  6.59 &  7.67 & 10.19 & 10.15 & 10.02 & 10.43 & 10.57 & 10.47 & 10.36 & 10.55 \\
           HS & 10.41 &   NA &  6.21 &  6.09 &  7.25 &   NA &  9.99 &  9.85 & 10.26 &   NA & 10.29 & 10.18 & 10.37 \\
     H2S(2B1) & 10.48 &  6.39 &  6.29 &  6.15 &  7.31 & 10.11 & 10.03 &  9.87 & 10.30 & 10.46 & 10.30 & 10.16 & 10.35 \\
          HCl & 12.82 &   NA &   NA &   NA &  9.22 &   NA &   NA &   NA & 12.64 &   NA &   NA &   NA & 12.68 \\
         H2C2 & 11.51 &  7.37 &  7.19 &  7.04 &  8.19 & 11.31 & 11.14 & 10.97 & 11.34 & 11.60 & 11.32 & 11.16 & 11.27 \\
         H4C2 & 10.74 &  6.96 &  6.77 &  6.61 &  7.66 & 10.64 & 10.46 & 10.28 & 10.61 & 10.90 & 10.61 & 10.45 & 10.52 \\
           CO & 14.08 &  9.15 &  9.06 &  9.03 & 10.56 & 13.97 & 13.86 & 13.85 & 14.45 & 14.04 & 13.90 & 13.86 & 14.20 \\
      N2(2Sg) & 15.61 & 10.42 & 10.27 & 10.26 & 11.97 & 14.65 & 14.51 & 14.49 & 15.37 & 15.44 & 15.24 & 15.21 & 15.72 \\
           O2 & 12.49 &   NA &  6.89 &  6.88 &  8.80 &   NA & 11.51 & 11.50 & 12.52 &   NA & 12.41 & 12.38 & 12.97 \\
           P2 & 10.82 &  7.23 &  7.11 &  6.88 &  7.82 & 10.10 &  9.99 &  9.74 & 10.12 & 10.62 & 10.44 & 10.22 & 10.36 \\
           S2 &  9.56 &   NA &  5.91 &  5.72 &  6.91 &   NA &  8.55 &  8.35 &  9.03 &   NA &  9.47 &  9.27 &  9.66 \\
          Cl2 & 11.77 &   NA &  7.43 &  7.29 &  8.59 &   NA & 10.27 & 10.11 & 10.86 &   NA & 11.22 & 11.07 & 11.49 \\
          FCl & 12.95 &   NA &  7.96 &  7.86 &  9.34 &   NA & 11.51 & 11.42 & 12.26 &   NA & 12.42 & 12.32 & 12.75 \\
           CS & 11.51 &  7.49 &  7.43 &  7.34 &  8.71 & 11.66 & 11.58 & 11.51 & 12.04 & 11.48 & 11.37 & 11.26 & 11.54 \\
          BF3 & 16.18 & 10.33 & 10.08 & 10.03 & 11.97 & 13.22 & 12.97 & 12.92 & 14.28 & 14.72 & 14.40 & 14.34 & 15.19 \\
         BCl3 & 11.91 &  7.87 &  7.74 &  7.57 &  8.88 &  9.80 &  9.68 &  9.50 & 10.43 & 10.94 & 10.79 & 10.61 & 11.21 \\
          CO2 & 13.90 &  9.33 &  9.08 &  8.99 & 10.46 & 12.90 & 12.66 & 12.56 & 13.32 & 13.92 & 13.58 & 13.47 & 13.76 \\
          CF2 & 12.40 &  7.48 &  7.37 &  7.35 &  8.82 & 11.67 & 11.52 & 11.52 & 12.21 & 12.04 & 11.86 & 11.83 & 12.24 \\
          COS & 11.36 &   NA &  7.48 &  7.32 &  8.45 &   NA & 10.43 & 10.26 & 10.83 &   NA & 11.19 & 11.02 & 11.26 \\
          CS2 & 10.18 &   NA &  6.82 &  6.62 &  7.62 &   NA &  9.11 &  8.91 &  9.46 &   NA & 10.00 &  9.80 & 10.03 \\
          H2C & 10.40 &  5.76 &  5.73 &  5.49 &  6.81 & 10.24 & 10.24 &  9.99 & 10.43 & 10.43 & 10.37 & 10.12 & 10.37 \\
          H3C &  9.78 &  5.40 &  5.42 &  5.21 &  6.47 &  9.74 &  9.80 &  9.58 &  9.99 &  9.87 &  9.88 &  9.67 &  9.88 \\
         H5C2 &  8.60 &  4.55 &  4.61 &  4.43 &  5.66 &  8.31 &  8.40 &  8.26 &  8.72 &  8.47 &  8.52 &  8.35 &  8.62 \\
           CN & 14.22 &  9.54 &  9.38 &  9.25 & 10.53 & 14.03 & 13.88 & 13.74 & 14.13 & 14.42 & 14.15 & 14.02 & 14.07 \\
          HCO &  9.37 &  4.96 &  5.00 &  4.94 &  6.32 &  8.50 &  8.52 &  8.47 &  9.17 &  9.35 &  9.36 &  9.32 &  9.74 \\
        CH2OH &  8.18 &  3.97 &  3.97 &  3.85 &  5.18 &  7.63 &  7.64 &  7.54 &  8.17 &  8.09 &  8.06 &  7.95 &  8.32 \\
         CH3O & 10.94 &  6.04 &  6.06 &  6.03 &  7.59 & 10.72 & 10.77 & 10.76 & 11.25 & 10.57 & 10.53 & 10.49 & 10.78 \\
         H4CO & 11.17 &  6.41 &  6.30 &  6.26 &  7.76 & 10.83 & 10.81 & 10.80 & 11.28 & 10.87 & 10.70 & 10.64 & 10.86 \\
         H3CF & 13.47 &   NA &  8.07 &  8.04 &  9.66 &   NA & 12.20 & 12.20 & 12.77 &   NA & 12.68 & 12.64 & 12.98 \\
         H2CS &  9.47 &  5.62 &  5.54 &  5.40 &  6.54 &  8.95 &  8.89 &  8.77 &  9.24 &  9.36 &  9.21 &  9.08 &  9.28 \\
        CH2SH &  7.79 &  4.19 &  4.20 &  4.03 &  5.20 &  7.18 &  7.17 &  7.01 &  7.55 &  7.72 &  7.70 &  7.53 &  7.87 \\
        H3CCl & 11.49 &   NA &  7.10 &  6.97 &  8.26 &   NA & 10.80 & 10.69 & 11.23 &   NA & 11.14 & 11.01 & 11.26 \\
        H6C2O & 10.89 &  6.32 &  6.22 &  6.17 &  7.65 & 10.41 & 10.44 & 10.45 & 10.90 & 10.46 & 10.33 & 10.28 & 10.52 \\
        H4C2O & 10.38 &  6.09 &  5.97 &  5.92 &  7.35 & 10.17 & 10.09 & 10.06 & 10.59 & 10.16 &  9.97 &  9.89 & 10.13 \\
        H3COF & 11.68 &  6.76 &  6.63 &  6.62 &  8.32 & 10.84 & 10.74 & 10.75 & 11.57 & 11.16 & 10.98 & 10.95 & 11.40 \\
        H4C2S &  9.15 &  5.47 &  5.37 &  5.23 &  6.37 &  8.88 &  8.80 &  8.67 &  9.13 &  9.09 &  8.92 &  8.77 &  8.98 \\
         C2N2 & 13.59 &  9.55 &  9.31 &  9.15 & 10.34 & 12.24 & 11.99 & 11.83 & 12.48 & 13.31 & 12.99 & 12.82 & 13.10 \\
         H4B2 & 10.17 &  6.47 &  6.36 &  6.22 &  7.29 & 10.06 &  9.96 &  9.82 & 10.16 & 10.23 & 10.06 &  9.94 & 10.08 \\
           HN & 13.48 &  7.97 &  7.91 &  7.73 &   NA & 13.58 & 13.53 & 13.36 &   NA & 13.64 & 13.52 & 13.33 &   NA \\
          H2N & 12.12 &  7.26 &  7.27 &  7.22 &  8.61 & 12.49 & 12.54 & 12.48 & 12.83 & 12.19 & 12.14 & 12.10 & 12.19 \\
         H2N2 & 10.28 &  5.71 &  5.66 &  5.64 &  7.03 &  9.29 &  9.23 &  9.21 &  9.90 & 10.00 &  9.89 &  9.85 & 10.22 \\
         H3N2 &  8.34 &  4.16 &  4.15 &  4.07 &  5.51 &  7.94 &  7.91 &  7.85 &  8.53 &  8.38 &  8.33 &  8.24 &  8.63 \\
          HOF & 13.03 &  7.48 &  7.31 &  7.31 &  9.14 & 12.11 & 11.94 & 11.94 & 12.89 & 12.71 & 12.45 & 12.43 & 12.91 \\
         H2Si &  9.55 &  5.90 &  5.86 &  5.76 &  6.73 &  8.87 &  8.84 &  8.72 &  9.13 &  9.41 &  9.33 &  9.24 &  9.43 \\
         H3Si &  8.86 &  5.30 &  5.36 &  5.12 &  6.14 &  8.40 &  8.49 &  8.26 &  8.66 &  8.85 &  8.92 &  8.71 &  8.95 \\
        H2Si2 &  8.22 &  5.21 &  5.11 &  4.92 &  5.80 &  7.50 &  7.41 &  7.21 &  7.64 &  8.10 &  7.96 &  7.77 &  7.99 \\
        H4Si2 &  8.36 &  5.53 &  5.42 &  5.24 &  6.01 &  7.92 &  7.83 &  7.63 &  7.94 &  8.37 &  8.21 &  8.05 &  8.16 \\
        H5Si2 &  8.37 &  5.21 &  5.26 &  5.01 &  5.99 &  8.01 &  8.09 &  7.85 &  8.27 &  8.33 &  8.39 &  8.16 &  8.41 \\
        H6Si2 & 10.73 &  7.36 &  7.30 &  7.16 &  8.21 & 10.04 & 10.00 &  9.88 & 10.41 & 10.47 & 10.38 & 10.26 & 10.54 \\
         B2F4 & 13.30 &  8.63 &  8.44 &  8.39 &  9.93 & 11.37 & 11.14 & 11.11 & 12.20 & 12.44 & 12.19 & 12.15 & 12.82 \\
  H4C3(cyclo) & 10.04 &  6.23 &  6.11 &  5.98 &  7.04 &  9.14 &  9.04 &  8.93 &  9.43 &  9.82 &  9.65 &  9.51 &  9.71 \\
 H4C3(allene) & 10.31 &  6.72 &  6.57 &  6.42 &  7.51 &  9.99 &  9.87 &  9.73 & 10.17 & 10.31 & 10.08 &  9.94 & 10.09 \\
         H7C3 &  7.80 &  4.02 &  4.10 &  3.94 &  5.14 &  7.46 &  7.57 &  7.45 &  7.94 &  7.58 &  7.65 &  7.50 &  7.78 \\
         H4CS &  9.55 &  5.70 &  5.61 &  5.48 &  6.63 &  9.11 &  9.05 &  8.92 &  9.39 &  9.44 &  9.29 &  9.16 &  9.38 \\
        H4C4O &  9.09 &  5.90 &  5.68 &  5.53 &  6.51 &  8.58 &  8.37 &  8.21 &  8.65 &  9.17 &  8.88 &  8.73 &  8.88 \\
        H5C4N &  8.42 &  5.37 &  5.17 &  5.01 &  5.96 &  8.01 &  7.81 &  7.64 &  8.07 &  8.54 &  8.26 &  8.10 &  8.25 \\ \midrule
          MAE &       &  4.28 &  4.37 &  4.49 &  3.18 &  0.60 &  0.72 &  0.81 &  0.41 &  0.20 &  0.30 &  0.41 &  0.20 \\
          MSE &       & $-$4.28 & $-$4.37 & $-$4.49 & $-$3.18 & $-$0.49 & $-$0.64 & $-$0.74 & $-$0.20 & $-$0.14 & $-$0.29 & $-$0.41 & $-$0.14 \\ \bottomrule
\end{longtable}
}
\end{landscape}
\restoregeometry

\begin{landscape}
\renewcommand\arraystretch{0.6}
\small{
\begin{longtable}{lr|rrrr|rrrr|rrrr}
    \caption{The experimental first ionization potentials of $N$-electron systems
            versus the negative LUMO energies of $(N-1)$-electron systems from different DFAs.
            Units are in eV.}
    \label{tab:IP_N-1} \\
    \toprule
             &       & \multicolumn{4}{c|}{DFAs}          & \multicolumn{4}{c|}{GSC1-DFAs}                & \multicolumn{4}{c}{GSC2-DFAs}                \\ \midrule
Mol          & Exp.  & LDA    & PBE    & BLYP   & B3LYP  & LDA      & PBE      & BLYP      & B3LYP      & LDA      & PBE      & BLYP      & B3LYP      \\ \midrule
    \endfirsthead
    \multicolumn{14}{l}{Table \ref{tab:IP_N-1} continued: IPs from $(N-1)$-systems.} \\ \toprule
             &       & \multicolumn{4}{c|}{DFAs}          & \multicolumn{4}{c|}{GSC1-DFAs}                & \multicolumn{4}{c}{GSC2-DFAs}                \\ \midrule
Mol          & Exp.  & LDA    & PBE    & BLYP   & B3LYP  & LDA      & PBE      & BLYP      & B3LYP      & LDA      & PBE      & BLYP      & B3LYP      \\ \midrule
    \endhead
    \midrule
    \endfoot
    \endlastfoot
         H4C & 14.40 & 18.72 & 18.44 & 18.43 & 17.57 & 14.54 & 14.33 & 14.32 & 14.35 & 13.99 & 13.96 & 13.84 & 14.11 \\
          H3N & 11.03 & 16.38 & 15.69 & 15.80 & 14.56 & 10.73 & 10.13 & 10.26 & 10.29 & 11.44 & 11.29 & 11.24 & 11.26 \\
           HO & 13.07 & 19.81 & 19.18 & 19.37 & 17.74 & 13.04 & 12.44 & 12.66 & 12.57 & 13.47 & 13.60 & 13.40 & 13.34 \\
          H2O & 12.74 & 19.13 & 18.31 & 18.39 & 16.82 & 12.53 & 11.81 & 11.91 & 11.83 & 13.10 & 13.02 & 12.81 & 12.78 \\
           HF & 16.20 & 23.92 & 22.97 & 23.06 & 21.16 & 16.09 & 15.24 & 15.35 & 15.19 & 16.63 & 16.51 & 16.32 & 16.33 \\
         H4Si & 12.84 &   NA & 15.77 & 15.72 & 15.28 &   NA & 12.97 & 12.92 & 13.06 &   NA & 12.09 & 12.00 & 12.42 \\
           HP & 10.18 &   NA & 14.27 & 14.07 & 13.29 &   NA & 10.57 & 10.39 & 10.40 &   NA & 10.40 & 10.06 & 10.23 \\
          H2P &  9.82 & 14.03 & 13.82 & 13.68 &   NA & 10.36 & 10.17 & 10.04 &   NA & 10.03 & 10.18 &  9.86 &   NA \\
          H3P & 10.61 & 14.53 & 14.14 & 14.22 & 13.47 & 10.92 & 10.56 & 10.64 & 10.63 & 10.73 & 10.59 & 10.65 & 10.76 \\
           HS & 10.41 &   NA & 14.50 & 14.61 & 13.72 &   NA & 10.46 & 10.58 & 10.57 &   NA & 10.57 & 10.35 & 10.47 \\
     H2S(2B1) & 10.48 & 14.92 & 14.42 & 14.44 & 13.57 & 10.92 & 10.44 & 10.48 & 10.46 & 10.60 & 10.51 & 10.31 & 10.44 \\
          HCl & 12.82 &   NA &   NA &   NA & 16.35 &   NA &   NA &   NA & 12.82 &   NA &   NA &   NA & 12.77 \\
         H2C2 & 11.51 & 16.21 & 15.62 & 15.59 & 14.57 & 12.03 & 11.46 & 11.45 & 11.32 & 11.70 & 11.46 & 11.25 & 11.31 \\
         H4C2 & 10.74 & 15.10 & 14.53 & 14.50 & 13.52 & 11.21 & 10.66 & 10.66 & 10.50 & 10.98 & 10.73 & 10.52 & 10.55 \\
           CO & 14.08 & 18.93 & 18.39 & 18.61 & 17.71 & 13.84 & 13.33 & 13.59 & 13.67 & 14.75 & 14.31 & 15.06 & 15.31 \\
      N2(2Sg) & 15.61 & 20.98 & 20.59 & 20.58 & 19.77 & 16.54 & 16.15 & 16.15 & 16.26 & 15.55 & 15.38 & 15.31 & 15.79 \\
           O2 & 12.49 &   NA & 18.52 & 18.47 & 17.53 &   NA & 13.67 & 13.61 & 13.70 &   NA & 12.50 & 12.47 & 13.03 \\
           P2 & 10.82 & 14.24 & 13.86 & 13.74 & 13.04 & 11.23 & 10.86 & 10.76 & 10.68 &  9.92 & 10.50 & 10.26 & 10.38 \\
           S2 &  9.56 &   NA & 13.25 & 13.05 & 12.55 &   NA & 10.50 & 10.31 & 10.38 &   NA &  9.51 &  9.31 &  9.69 \\
          Cl2 & 11.77 &   NA & 15.21 & 15.07 & 14.54 &   NA & 12.29 & 12.16 & 12.23 &   NA & 11.26 & 11.08 & 11.52 \\
          FCl & 12.95 &   NA & 17.11 & 17.05 & 16.34 &   NA & 13.44 & 13.38 & 13.38 &   NA & 12.52 & 12.37 & 12.80 \\
           CS & 11.51 & 15.42 & 14.93 & 15.05 & 14.20 & 11.06 & 10.62 & 10.76 & 10.79 & 11.87 & 11.91 & 11.89 & 12.07 \\
          BF3 & 16.18 & 19.33 & 18.91 & 18.83 & 18.57 & 16.40 & 15.98 & 15.91 & 16.24 & 14.73 & 14.42 & 14.35 & 15.21 \\
         BCl3 & 11.91 & 14.14 & 13.93 & 13.75 & 13.62 & 12.19 & 11.97 & 11.79 & 12.06 & 10.95 & 10.80 & 10.62 & 11.22 \\
          CO2 & 13.90 & 18.81 & 18.26 & 18.15 & 17.17 & 15.13 & 14.58 & 14.48 & 14.28 & 13.97 & 13.66 & 13.52 & 13.79 \\
          CF2 & 12.40 & 16.81 & 16.31 & 16.46 & 15.71 & 12.52 & 12.04 & 12.23 & 12.27 & 12.28 & 12.31 & 12.13 & 12.49 \\
          COS & 11.36 &   NA & 14.94 & 14.85 & 14.14 &   NA & 11.96 & 11.88 & 11.76 &   NA & 11.26 & 11.07 & 11.28 \\
          CS2 & 10.18 &   NA & 13.24 & 13.09 & 12.50 &   NA & 10.88 & 10.74 & 10.63 &   NA & 10.04 &  9.83 & 10.04 \\
          H2C & 10.40 & 15.82 & 15.42 & 15.18 & 14.08 & 10.70 & 10.36 & 10.16 & 10.19 & 10.72 & 10.87 & 10.49 & 10.58 \\
          H3C &  9.78 & 14.87 & 14.51 & 14.36 & 13.29 &  9.94 &  9.63 &  9.52 &  9.55 & 10.18 & 10.43 & 10.02 & 10.06 \\
         H5C2 &  8.60 & 12.52 & 12.32 & 12.19 & 11.39 &  8.58 &  8.36 &  8.22 &  8.25 &  8.63 &  8.85 &  8.58 &  8.77 \\
           CN & 14.22 & 19.72 & 19.11 & 19.12 & 17.90 & 14.97 & 14.39 & 14.43 & 14.23 & 14.53 & 14.33 & 14.12 & 14.13 \\
          HCO &  9.37 & 14.34 & 14.20 & 14.15 & 13.35 & 10.45 & 10.33 & 10.29 & 10.32 &  9.48 &  9.57 &  9.51 &  9.87 \\
        CH2OH &  8.18 & 12.68 & 12.45 & 12.34 & 11.51 &  8.66 &  8.44 &  8.34 &  8.35 &  8.25 &  8.38 &  8.21 &  8.50 \\
         CH3O & 10.94 & 15.27 & 15.02 & 15.00 & 14.03 & 10.81 & 10.46 & 10.45 & 10.36 & 10.63 & 10.65 & 10.57 & 10.89 \\
         H4CO & 11.17 & 15.39 & 15.09 & 15.04 & 14.13 & 11.40 & 10.98 & 10.93 & 10.74 & 10.88 & 10.74 & 10.67 & 10.91 \\
         H3CF & 13.47 &   NA & 17.28 & 17.22 & 16.48 &   NA & 13.51 & 13.45 & 13.45 &   NA & 12.68 & 12.62 & 13.02 \\
         H2CS &  9.47 & 13.34 & 12.94 & 12.91 & 12.13 &  9.90 &  9.47 &  9.44 &  9.35 &  9.44 &  9.37 &  9.17 &  9.34 \\
        CH2SH &  7.79 & 11.54 & 11.36 & 11.20 & 10.55 &  8.35 &  8.16 &  8.01 &  8.02 &  7.80 &  7.89 &  7.68 &  8.07 \\
        H3CCl & 11.49 &   NA & 15.11 & 15.04 & 14.33 &   NA & 11.61 & 11.54 & 11.44 &   NA & 11.20 & 11.05 & 11.30 \\
        H6C2O & 10.89 & 14.13 & 14.00 & 13.95 & 13.41 & 11.09 & 10.81 & 10.74 & 10.58 & 10.40 & 10.31 & 10.27 & 10.57 \\
        H4C2O & 10.38 & 14.43 & 14.07 & 14.01 & 13.07 & 10.55 & 10.11 & 10.04 &  9.81 & 10.23 & 10.07 &  9.98 & 10.20 \\
        H3COF & 11.68 & 15.55 & 15.28 & 15.24 & 14.55 & 11.87 & 11.50 & 11.46 & 11.35 & 11.18 & 11.02 & 10.98 & 11.45 \\
        H4C2S &  9.15 & 12.84 & 12.42 & 12.38 & 11.62 &  9.35 &  8.89 &  8.86 &  8.80 &  9.17 &  9.07 &  8.87 &  9.04 \\
         C2N2 & 13.59 & 17.25 & 16.80 & 16.65 & 15.98 & 14.51 & 14.06 & 13.92 & 13.81 & 13.33 & 13.02 & 12.84 & 13.11 \\
         H4B2 & 10.17 & 14.17 & 13.70 & 13.73 & 12.94 & 10.40 &  9.95 &  9.99 &  9.99 & 10.39 & 10.13 & 10.19 & 10.25 \\
           HN & 13.48 & 19.97 & 19.09 & 19.06 &   NA & 13.72 & 12.98 & 12.95 &   NA & 14.55 & 14.38 & 14.21 &   NA \\
          H2N & 12.12 & 17.51 & 16.95 & 17.23 & 15.98 & 11.84 & 11.32 & 11.63 & 11.64 & 12.71 & 12.72 & 13.01 & 12.80 \\
         H2N2 & 10.28 & 14.78 & 14.53 & 14.49 & 13.68 & 10.97 & 10.73 & 10.71 & 10.71 & 10.06 &  9.99 &  9.93 & 10.27 \\
         H3N2 &  8.34 & 13.14 & 12.92 & 12.81 & 11.95 &  9.08 &  8.87 &  8.77 &  8.76 &  8.46 &  8.46 &  8.36 &  8.78 \\
          HOF & 13.03 & 18.46 & 18.00 & 17.98 & 16.96 & 13.69 & 13.21 & 13.19 & 13.05 & 12.78 & 12.57 & 12.52 & 13.00 \\
         H2Si &  9.55 & 13.10 & 12.80 & 12.87 & 12.23 &  9.96 &  9.66 &  9.75 &  9.74 &  9.56 &  9.42 &  9.50 &  9.64 \\
         H3Si &  8.86 & 12.47 & 12.27 & 12.23 & 11.58 &  9.13 &  8.95 &  8.90 &  8.95 &  9.23 &  9.48 &  9.11 &  9.22 \\
        H2Si2 &  8.22 & 11.24 & 10.93 & 10.82 & 10.31 &  8.84 &  8.53 &  8.43 &  8.41 &  8.15 &  8.03 &  7.82 &  8.02 \\
        H4Si2 &  8.36 & 11.32 & 10.97 & 10.95 & 10.36 &  8.80 &  8.45 &  8.44 &  8.38 &  8.42 &  8.26 &  8.12 &  8.20 \\
        H5Si2 &  8.37 & 11.40 & 11.25 & 11.14 & 10.62 &  8.50 &  8.34 &  8.23 &  8.28 &  8.54 &  8.75 &  8.41 &  8.58 \\
        H6Si2 & 10.73 & 13.62 & 13.38 & 13.33 & 12.85 & 10.96 & 10.71 & 10.65 & 10.66 & 10.51 & 10.49 & 10.33 & 10.61 \\
         B2F4 & 13.30 & 16.20 & 15.87 & 15.83 & 15.67 & 13.74 & 13.41 & 13.38 & 13.56 & 12.44 & 12.19 & 12.14 & 12.83 \\
  H4C3(cyclo) & 10.04 & 13.64 & 13.33 & 13.23 & 12.49 & 10.65 & 10.30 & 10.20 & 10.06 &  9.85 &  9.70 &  9.55 &  9.73 \\
 H4C3(allene) & 10.31 & 13.84 & 13.50 & 13.39 & 12.66 & 10.91 & 10.55 & 10.44 & 10.31 & 10.20 & 10.00 &  9.85 & 10.05 \\
         H7C3 &  7.80 & 11.16 & 11.02 & 10.90 & 10.20 &  7.66 &  7.48 &  7.34 &  7.36 &  7.69 &  7.88 &  7.67 &  7.90 \\
         H4CS &  9.55 & 13.33 & 12.95 & 12.93 & 12.20 &  9.91 &  9.49 &  9.47 &  9.41 &  9.52 &  9.43 &  9.25 &  9.44 \\
        H4C4O &  9.09 & 12.60 & 12.15 & 12.05 & 11.34 &  9.84 &  9.41 &  9.30 &  9.18 &  9.19 &  8.92 &  8.75 &  8.89 \\
        H5C4N &  8.42 & 11.88 & 11.45 & 11.33 & 10.63 &  9.16 &  8.74 &  8.63 &  8.50 &  8.57 &  8.30 &  8.13 &  8.26 \\ \midrule
          MAE &       &  4.27 &  3.88 &  3.83 &  3.01 &  0.41 &  0.35 &  0.31 &  0.31 &  0.30 &  0.34 &  0.39 &  0.26 \\
          MSE &       &  4.27 &  3.88 &  3.83 &  3.01 &  0.34 &  0.04 &  0.00 & $-$0.02 & $-$0.01 & $-$0.10 & $-$0.24 & $-$0.02 \\ \bottomrule
\end{longtable}
}
\end{landscape}
\restoregeometry

\newgeometry{left=0.5in, right=0.5in, top=1in, bottom=1in}
\renewcommand\arraystretch{0.6}
\small{
\begin{longtable}{lrrrrr}
    \caption{The experimental first ionization potentials (in eV) of
            $N$-electron systems in comparison with results
            from $\Delta$-SCF method with different DFAs.
            $\Delta E_{\rm{DFA}} = E_{\rm{DFA}}(N-1) - E_{\rm{DFA}}(N)$.}
    \label{tab:IP_delta_scf} \\
    \toprule
Mol          & Exp.  & $\Delta E_{\rm{LDA}}$ & $\Delta E_{\rm{PBE}}$ & $\Delta E_{\rm{BLYP}}$  & $\Delta E_{\rm{B3LYP}}$ \\ \midrule
    \endfirsthead
    \multicolumn{6}{l}{Table \ref{tab:IP_delta_scf} continued: IPs from $\Delta$-SCF.} \\ \toprule
Mol          & Exp.  & $\Delta E_{\rm{LDA}}$ & $\Delta E_{\rm{PBE}}$ & $\Delta E_{\rm{BLYP}}$  & $\Delta E_{\rm{B3LYP}}$ \\ \midrule
    \endhead
    \midrule
    \endfoot
    \endlastfoot
          H4C & 14.40 & 14.04 & 13.92 & 13.87 & 14.14 \\
          H3N & 11.03 & 11.31 & 11.02 & 10.98 & 11.06 \\
           HO & 13.07 & 13.50 & 13.26 & 13.29 & 13.32 \\
          H2O & 12.74 & 13.15 & 12.76 & 12.73 & 12.77 \\
           HF & 16.20 & 16.72 & 16.27 & 16.26 & 16.27 \\
         H4Si & 12.84 &   NA & 12.12 & 12.05 & 12.45 \\
           HP & 10.18 &   NA & 10.23 &  9.97 & 10.16 \\
          H2P &  9.82 &  9.98 &  9.96 &  9.73 &   NA \\
          H3P & 10.61 & 10.64 & 10.47 & 10.42 & 10.58 \\
           HS & 10.41 &   NA & 10.37 & 10.32 & 10.46 \\
     H2S(2B1) & 10.48 & 10.61 & 10.37 & 10.27 & 10.42 \\
          HCl & 12.82 &   NA &   NA &   NA & 12.76 \\
         H2C2 & 11.51 & 11.74 & 11.40 & 11.28 & 11.35 \\
         H4C2 & 10.74 & 11.00 & 10.66 & 10.53 & 10.57 \\
           CO & 14.08 & 14.11 & 13.88 & 13.91 & 14.22 \\
      N2(2Sg) & 15.61 & 15.63 & 15.39 & 15.37 & 15.83 \\
           O2 & 12.49 &   NA & 12.62 & 12.59 & 13.11 \\
           P2 & 10.82 & 10.70 & 10.48 & 10.29 & 10.41 \\
           S2 &  9.56 &   NA &  9.55 &  9.35 &  9.71 \\
          Cl2 & 11.77 &   NA & 11.30 & 11.15 & 11.55 \\
          FCl & 12.95 &   NA & 12.51 & 12.42 & 12.82 \\
           CS & 11.51 & 11.51 & 11.32 & 11.27 & 11.52 \\
          BF3 & 16.18 & 14.80 & 14.46 & 14.40 & 15.24 \\
         BCl3 & 11.91 & 10.99 & 10.82 & 10.64 & 11.24 \\
          CO2 & 13.90 & 14.03 & 13.65 & 13.54 & 13.80 \\
          CF2 & 12.40 & 12.15 & 11.90 & 11.92 & 12.28 \\
          COS & 11.36 &   NA & 11.22 & 11.07 & 11.29 \\
          CS2 & 10.18 &   NA & 10.03 &  9.84 & 10.05 \\
          H2C & 10.40 & 10.71 & 10.57 & 10.32 & 10.45 \\
          H3C &  9.78 & 10.09 & 10.01 &  9.80 &  9.91 \\
         H5C2 &  8.60 &  8.54 &  8.53 &  8.36 &  8.58 \\
           CN & 14.22 & 14.57 & 14.24 & 14.15 & 14.17 \\
          HCO &  9.37 &  9.57 &  9.55 &  9.49 &  9.82 \\
        CH2OH &  8.18 &  8.27 &  8.20 &  8.08 &  8.36 \\
         CH3O & 10.94 & 10.64 & 10.55 & 10.52 & 10.82 \\
         H4CO & 11.17 & 10.89 & 10.71 & 10.65 & 10.93 \\
         H3CF & 13.47 &   NA & 12.67 & 12.63 & 13.05 \\
         H2CS &  9.47 &  9.45 &  9.25 &  9.15 &  9.33 \\
        CH2SH &  7.79 &  7.83 &  7.78 &  7.61 &  7.89 \\
        H3CCl & 11.49 &   NA & 11.13 & 11.01 & 11.29 \\
        H6C2O & 10.89 & 10.30 & 10.19 & 10.14 & 10.54 \\
        H4C2O & 10.38 & 10.24 & 10.02 &  9.96 & 10.20 \\
        H3COF & 11.68 & 11.16 & 10.97 & 10.94 & 11.43 \\
        H4C2S &  9.15 &  9.15 &  8.93 &  8.81 &  9.00 \\
         C2N2 & 13.59 & 13.37 & 13.04 & 12.87 & 13.14 \\
         H4B2 & 10.17 & 10.32 & 10.07 &  9.99 & 10.12 \\
           HN & 13.48 & 13.96 & 13.68 & 13.52 &   NA \\
          H2N & 12.12 & 12.38 & 12.22 & 12.26 & 12.31 \\
         H2N2 & 10.28 & 10.17 & 10.04 & 10.01 & 10.32 \\
         H3N2 &  8.34 &  8.57 &  8.49 &  8.39 &  8.71 \\
          HOF & 13.03 & 12.90 & 12.60 & 12.59 & 13.02 \\
         H2Si &  9.55 &  9.49 &  9.36 &  9.31 &  9.48 \\
         H3Si &  8.86 &  8.92 &  8.92 &  8.74 &  8.92 \\
        H2Si2 &  8.22 &  8.19 &  8.01 &  7.85 &  8.04 \\
        H4Si2 &  8.36 &  8.42 &  8.21 &  8.09 &  8.18 \\
        H5Si2 &  8.37 &  8.34 &  8.35 &  8.14 &  8.36 \\
        H6Si2 & 10.73 & 10.49 & 10.37 & 10.26 & 10.54 \\
         B2F4 & 13.30 & 12.43 & 12.17 & 12.12 & 12.81 \\
  H4C3(cyclo) & 10.04 &  9.90 &  9.70 &  9.58 &  9.75 \\
 H4C3(allene) & 10.31 & 10.25 & 10.02 &  9.88 & 10.07 \\
         H7C3 &  7.80 &  7.60 &  7.62 &  7.47 &  7.73 \\
         H4CS &  9.55 &  9.50 &  9.31 &  9.21 &  9.41 \\
        H4C4O &  9.09 &  9.23 &  8.91 &  8.77 &  8.91 \\
        H5C4N &  8.42 &  8.60 &  8.30 &  8.15 &  8.28 \\ \midrule
          MAE &       &  0.24 &  0.29 &  0.36 &  0.19 \\
          MSE &       & $-$0.03 & $-$0.23 & $-$0.33 & $-$0.09 \\ \bottomrule
\end{longtable}
}
\restoregeometry

\newgeometry{left=0.5in, right=0.5in, top=1in, bottom=1in}
\begin{landscape}
\subsection{The electron affinities (EAs)}
The same test set in Ref. \citenum{2020mei10277} was used in this work
to test the EAs. To make reliable comparison, molecules in the
test set with unbounded
$(N+1)$-electron systems are excluded in this work.
See Ref. \citenum{2020mei10277} for details
about the molecular geometries and the experimental energies.
6-311++G(3df, 3pd) is used as the basis set for all the calculations.

\renewcommand\arraystretch{0.6}
\small{
\begin{longtable}{lr|rrrr|rrrr|rrrr}
    \caption{The experimental first electron affinities of $N$-electron systems
            versus the negative LUMO energies of $N$-electron systems from
            different DFAs. Units are in eV.}
    \label{tab:EA_N} \\
    \toprule
             &       & \multicolumn{4}{c|}{DFAs}          & \multicolumn{4}{c|}{GSC1-DFAs}                & \multicolumn{4}{c}{GSC2-DFAs}                \\ \midrule
Mol          & Exp.  & LDA    & PBE    & BLYP   & B3LYP  & LDA      & PBE      & BLYP      & B3LYP      & LDA      & PBE      & BLYP      & B3LYP      \\ \midrule
    \endfirsthead
    \multicolumn{14}{l}{Table \ref{tab:EA_N} continued: EAs from $N$-systems.} \\ \toprule
             &       & \multicolumn{4}{c|}{DFAs}          & \multicolumn{4}{c|}{GSC1-DFAs}                & \multicolumn{4}{c}{GSC2-DFAs}                \\ \midrule
Mol          & Exp.  & LDA    & PBE    & BLYP   & B3LYP  & LDA      & PBE      & BLYP      & B3LYP      & LDA      & PBE      & BLYP      & B3LYP      \\ \midrule
    \endhead
    \midrule
    \endfoot
    \endlastfoot
     HC & 1.19 & 5.78 & 5.37 & 5.16 & 4.15 &  1.19 &  0.86 &  0.69 &  0.77 &  1.42 &  1.43 &  1.14 &  1.24 \\
    H2C & 0.83 & 3.90 & 3.32 & 3.60 &  NA & $-$0.25 & $-$0.79 & $-$0.47 &   NA &  0.47 &  0.76 &  0.62 &   NA \\
     HN & 0.33 & 4.86 & 4.32 & 4.50 & 3.30 & $-$0.11 & $-$0.61 & $-$0.38 & $-$0.31 &  0.44 &  0.52 &  0.31 & $-$0.31 \\
    H2N & 0.75 & 5.22 & 4.53 & 4.62 & 3.45 &  0.28 & $-$0.29 & $-$0.18 & $-$0.12 &  0.87 &  0.75 &  0.60 &  0.59 \\
     HO & 1.83 & 7.24 & 6.44 & 6.53 & 5.03 &  1.15 &  0.48 &  0.60 &  0.57 &  1.99 &  1.83 &  1.67 &  1.59 \\
    HSi & 1.26 & 4.44 &  NA &  NA & 3.44 &  1.52 &   NA &   NA &  1.22 &  1.33 &   NA &   NA &  1.21 \\
   H2Si & 1.07 & 4.25 & 4.05 & 3.91 & 3.34 &  1.36 &  1.19 &  1.07 &  1.14 &  1.23 &  1.31 &  0.98 &  1.14 \\
   H3Si & 0.93 & 3.98 & 3.53 & 3.59 & 3.01 &  1.05 &  0.63 &  0.71 &  0.78 &  1.09 &  1.03 &  1.17 &  1.25 \\
     HP & 1.02 & 4.41 & 3.97 & 4.08 & 3.39 &  1.18 &  0.75 &  0.90 &  0.95 &  1.09 &  1.10 &  0.94 &  1.06 \\
    H2P & 1.26 & 4.67 & 4.20 & 4.23 & 3.55 &  1.44 &  1.01 &  1.07 &  1.11 &  1.32 &  1.22 &  1.07 &  1.20 \\
     HS & 2.36 &  NA & 5.78 & 5.80 & 4.97 &   NA &  2.06 &  2.09 &  2.09 &   NA &  2.30 &  2.13 &  2.25 \\
     CN & 3.93 & 8.11 & 7.67 & 7.76 & 7.00 &  4.09 &  3.65 &  3.78 &  3.80 &  3.90 &  3.86 &  3.68 &  4.04 \\
     OP & 1.09 &  NA & 4.48 & 4.33 & 3.73 &   NA &  1.55 &  1.41 &  1.44 &   NA &  0.99 &  0.80 &  1.09 \\
     S2 & 1.64 &  NA & 4.62 & 4.52 & 3.97 &   NA &  2.03 &  1.94 &  1.94 &   NA &  1.36 &  1.22 &  1.49 \\
    Cl2 & 1.16 & 4.90 & 4.69 & 4.57 & 3.89 &  2.02 &  1.83 &  1.73 &  1.66 &  1.14 &  1.04 &  0.92 &  1.21 \\
     C2 & 3.24 &  NA &  NA & 7.10 & 6.24 &   NA &   NA &  3.46 &  3.39 &   NA &   NA &  3.34 &  3.44 \\
    C2O & 2.18 & 6.10 & 5.68 & 5.66 & 4.87 &  2.93 &  2.51 &  2.51 &  2.38 &  2.23 &  2.00 &  1.90 &  2.05 \\
    CNO & 3.55 & 7.81 & 7.28 & 7.20 & 6.32 &  4.38 &  3.87 &  3.81 &  3.65 &  3.60 &  3.30 &  3.17 &  3.35 \\
    NO2 & 1.53 & 5.56 & 5.17 & 5.18 & 4.46 &  2.07 &  1.70 &  1.73 &  1.73 &  1.20 &  0.96 &  0.96 &  1.34 \\
     O3 & 1.82 & 6.40 & 6.11 & 6.08 & 5.08 &  2.90 &  2.62 &  2.59 &  1.95 &  1.77 &  1.58 &  1.57 &  2.42 \\
     OF & 1.94 & 7.18 & 6.58 & 6.60 & 5.41 &  2.04 &  1.42 &  1.46 &  1.29 &  1.83 &  1.61 &  1.55 &  1.79 \\
    O2S & 1.06 & 4.88 & 4.66 & 4.57 & 4.03 &  2.03 &  1.81 &  1.73 &  1.76 &  0.96 &  0.86 &  0.77 &  1.17 \\
    OS2 & 1.88 & 5.14 & 4.97 & 4.83 & 4.43 &  2.75 &  2.58 &  2.45 &  2.53 &  1.75 &  1.67 &  1.54 &  1.96 \\
    HC2 & 2.95 & 6.89 & 6.37 & 6.50 & 5.61 &  2.35 &  1.99 &  2.27 &  2.29 &  3.40 &  2.91 &  3.25 &  3.43 \\
   H3C2 & 0.19 & 3.69 & 3.16 & 3.27 & 2.43 & $-$0.06 & $-$0.54 & $-$0.43 & $-$0.39 &  0.30 &  0.31 &  0.24 &  0.30 \\
   H2C3 & 1.70 & 5.35 & 5.08 & 4.89 & 4.24 &  2.58 &  2.31 &  2.15 &  2.08 &  1.89 &  1.79 &  1.59 &  1.81 \\
   H5C3 & 0.38 & 3.73 & 3.25 & 3.20 & 2.46 &  1.09 &  0.64 &  0.60 &  0.45 &  0.62 &  0.39 &  0.25 &  0.33 \\
    HCF & 0.24 & 4.54 & 3.97 & 3.98 & 3.06 &  0.43 & $-$0.10 & $-$0.07 & $-$0.05 &  0.32 &  0.78 &  0.55 &  0.64 \\
   CH3O & 1.43 & 5.84 & 5.34 & 5.32 & 4.12 &  1.21 &  0.62 &  0.62 &  0.42 &  1.52 &  1.33 &  1.22 &  1.23 \\
   H3CS & 1.88 & 3.06 &  NA &  NA &  NA &  0.04 &   NA &   NA &   NA & $-$0.18 &   NA &   NA &   NA \\
   H2CS & 0.31 & 3.84 & 3.65 & 3.50 & 2.88 &  0.98 &  0.79 &  0.66 &  0.67 &  0.34 &  0.30 &  0.15 &  0.42 \\
  CH2CN & 1.54 & 5.23 & 4.72 & 4.68 & 3.92 &  2.28 &  1.79 &  1.77 &  1.65 &  1.74 &  1.51 &  1.36 &  1.50 \\
  CH2NC & 0.99 & 4.68 & 4.15 & 4.11 & 3.31 &  1.73 &  1.24 &  1.22 &  1.06 &  1.29 &  1.07 &  0.91 &  0.97 \\
   HC2O & 2.28 & 6.13 & 5.63 & 5.58 & 4.81 &  2.98 &  2.51 &  2.48 &  2.36 &  2.35 &  2.07 &  1.95 &  2.11 \\
 CH2CHO & 1.71 & 5.47 & 4.96 & 4.92 & 4.08 &  2.44 &  1.97 &  1.94 &  1.76 &  1.96 &  1.68 &  1.56 &  1.63 \\
  CH3CO & 0.16 & 3.27 & 2.91 & 2.95 & 2.20 &  0.03 & $-$0.27 & $-$0.18 & $-$0.12 & $-$0.17 & $-$0.23 & $-$0.32 & $-$0.12 \\
  H5C2O & 1.57 & 5.75 & 5.23 & 5.20 & 4.05 &  1.29 &  0.69 &  0.68 &  0.47 &  1.77 &  1.53 &  1.41 &  1.40 \\
  H5C2S & 1.95 & 5.39 & 4.97 & 4.96 & 4.19 &  2.01 &  1.56 &  1.56 &  1.51 &  1.99 &  1.86 &  1.69 &  1.81 \\
    HLi & 0.31 &  NA &  NA & 1.72 &  NA &   NA &   NA & $-$0.37 &   NA &   NA &   NA &  4.67 &   NA \\
    HO2 & 0.62 & 5.31 & 4.82 & 4.82 & 3.71 &  0.76 &  0.27 &  0.27 &  0.12 &  0.43 &  0.20 &  0.16 &  0.37 \\ \midrule
    MAE &      & 3.82 & 3.45 & 3.40 & 2.62 &  0.54 &  0.53 &  0.45 &  0.38 &  0.21 &  0.16 &  0.34 &  0.17 \\
    MSE &      & 3.82 & 3.45 & 3.40 & 2.62 &  0.19 & $-$0.14 & $-$0.13 & $-$0.14 & $-$0.01 & $-$0.07 & $-$0.06 & $-$0.02 \\ \bottomrule
\end{longtable}
}
\end{landscape}
\restoregeometry

\newgeometry{left=0.5in, right=0.5in, top=1in, bottom=1in}
\begin{landscape}
\renewcommand\arraystretch{0.6}
\small{
\begin{longtable}{lr|rrrr|rrrr|rrrr}
    \caption{The experimental first electron affinities of $N$-electron systems
            versus the negative HOMO energies of $(N+1)$-electron systems from different DFAs.
            Units are in eV.}
    \label{tab:EA_N+1} \\
    \toprule
             &       & \multicolumn{4}{c|}{DFAs}          & \multicolumn{4}{c|}{GSC1-DFAs}                & \multicolumn{4}{c}{GSC2-DFAs}                \\ \midrule
Mol          & Exp.  & LDA    & PBE    & BLYP   & B3LYP  & LDA      & PBE      & BLYP      & B3LYP      & LDA      & PBE      & BLYP      & B3LYP      \\ \midrule
    \endfirsthead
    \multicolumn{14}{l}{Table \ref{tab:EA_N+1} continued: EAs from $(N+1)$-systems.} \\ \toprule
             &       & \multicolumn{4}{c|}{DFAs}          & \multicolumn{4}{c|}{GSC1-DFAs}                & \multicolumn{4}{c}{GSC2-DFAs}                \\ \midrule
Mol          & Exp.  & LDA    & PBE    & BLYP   & B3LYP  & LDA      & PBE      & BLYP      & B3LYP      & LDA      & PBE      & BLYP      & B3LYP      \\ \midrule
    \endhead
    \midrule
    \endfoot
    \endlastfoot
    HC & 1.19 & $-$1.95 & $-$2.06 & $-$2.24 & $-$1.26 &  1.74 &  1.63 &  1.43 & 1.73 &  1.18 &  1.05 &  0.84 &  1.05 \\
    H2C & 0.83 & $-$2.46 & $-$2.55 & $-$2.59 &   NA &  0.92 &  0.87 &  0.80 &  NA &  0.11 & $-$0.07 & $-$0.07 &   NA \\
     HN & 0.33 & $-$3.05 & $-$3.15 & $-$3.20 & $-$2.19 &  1.06 &  1.01 &  0.95 & 1.16 &  0.30 &  0.11 &  0.12 &  0.09 \\
    H2N & 0.75 & $-$2.65 & $-$2.83 & $-$2.90 & $-$1.84 &  1.57 &  1.39 &  1.32 & 1.58 &  0.72 &  0.42 &  0.38 &  0.36 \\
     HO & 1.83 & $-$2.35 & $-$2.55 & $-$2.59 & $-$1.26 &  3.01 &  2.82 &  2.78 & 3.08 &  1.82 &  1.50 &  1.47 &  1.40 \\
    HSi & 1.26 & $-$1.31 &   NA &   NA & $-$0.84 &  1.12 &   NA &   NA & 1.13 &  1.11 &   NA &   NA &  1.06 \\
   H2Si & 1.07 & $-$1.47 & $-$1.50 & $-$1.72 & $-$0.98 &  0.91 &  0.92 &  0.64 & 0.95 &  0.98 &  0.98 &  0.72 &  0.99 \\
   H3Si & 0.93 & $-$1.47 & $-$1.59 & $-$1.70 & $-$0.99 &  1.03 &  0.94 &  0.80 & 1.05 &  0.83 &  0.65 &  0.56 &  0.71 \\
     HP & 1.02 & $-$1.64 & $-$1.75 & $-$1.83 & $-$1.05 &  1.11 &  1.02 &  0.90 & 1.16 &  0.89 &  0.72 &  0.66 &  0.83 \\
    H2P & 1.26 & $-$1.48 & $-$1.60 & $-$1.72 & $-$0.91 &  1.31 &  1.19 &  1.04 & 1.33 &  1.11 &  0.91 &  0.80 &  0.98 \\
     HS & 2.36 &   NA & $-$1.02 & $-$1.13 & $-$0.16 &   NA &  2.36 &  2.24 & 2.56 &   NA &  2.02 &  1.91 &  2.07 \\
     CN & 3.93 &  0.14 & $-$0.00 & $-$0.02 &  1.26 &  3.83 &  3.66 &  3.65 & 4.28 &  3.66 &  3.46 &  3.44 &  3.82 \\
     OP & 1.09 &   NA & $-$1.98 & $-$2.13 & $-$1.22 &   NA &  0.55 &  0.38 & 0.86 &   NA &  0.71 &  0.53 &  0.94 \\
     S2 & 1.64 &   NA & $-$1.58 & $-$1.67 & $-$0.73 &   NA &  0.84 &  0.73 & 1.21 &   NA &  1.25 &  1.13 &  1.43 \\
    Cl2 & 1.16 & $-$1.83 & $-$1.92 & $-$2.03 & $-$1.14 &  0.58 &  0.48 &  0.35 & 0.84 &  0.94 &  0.84 &  0.73 &  1.12 \\
     C2 & 3.24 &   NA &   NA & $-$0.32 &  0.72 &   NA &   NA &  3.37 & 3.68 &   NA &   NA &  3.11 &  3.22 \\
    C2O & 2.18 & $-$1.18 & $-$1.34 & $-$1.43 & $-$0.42 &  1.82 &  1.66 &  1.57 & 2.00 &  2.12 &  1.88 &  1.79 &  1.98 \\
    CNO & 3.55 & $-$0.22 & $-$0.45 & $-$0.54 &  0.65 &  3.09 &  2.86 &  2.76 & 3.31 &  3.53 &  3.21 &  3.10 &  3.29 \\
    NO2 & 1.53 & $-$2.19 & $-$2.38 & $-$2.37 & $-$1.21 &  0.92 &  0.70 &  0.71 & 1.34 &  1.00 &  0.76 &  0.78 &  1.19 \\
     O3 & 1.82 & $-$2.29 & $-$2.42 & $-$2.42 & $-$0.79 &  1.07 &  0.93 &  0.93 & 1.90 &  1.72 &  1.53 &  1.53 &  2.13 \\
     OF & 1.94 & $-$2.72 & $-$2.92 & $-$2.91 & $-$1.40 &  2.12 &  1.92 &  1.93 & 2.54 &  1.68 &  1.37 &  1.37 &  1.60 \\
    O2S & 1.06 & $-$2.50 & $-$2.57 & $-$2.65 & $-$1.46 &  0.16 &  0.09 &  0.00 & 0.71 &  0.84 &  0.72 &  0.64 &  1.12 \\
    OS2 & 1.88 & $-$1.31 & $-$1.37 & $-$1.49 & $-$0.36 &  0.98 &  0.91 &  0.80 & 1.47 &  1.71 &  1.61 &  1.48 &  1.92 \\
    HC2 & 2.95 & $-$0.39 & $-$0.48 & $-$0.54 &  0.64 &  3.91 &  3.74 &  3.63 & 3.77 &  2.91 &  2.74 &  2.67 &  2.86 \\
   H3C2 & 0.19 & $-$2.44 & $-$2.53 & $-$2.56 & $-$1.81 &  0.70 &  0.62 &  0.58 & 0.82 & $-$0.47 & $-$0.76 & $-$0.76 & $-$0.34 \\
   H2C3 & 1.70 & $-$1.19 & $-$1.28 & $-$1.45 & $-$0.47 &  1.38 &  1.31 &  1.12 & 1.60 &  1.81 &  1.68 &  1.50 &  1.75 \\
   H5C3 & 0.38 & $-$2.10 & $-$2.27 & $-$2.37 & $-$1.58 &  0.30 &  0.12 &  0.00 & 0.33 &  0.53 &  0.27 &  0.16 &  0.26 \\
    HCF & 0.24 & $-$2.90 & $-$2.97 & $-$3.08 & $-$2.12 &  0.30 &  0.24 &  0.12 & 0.57 &  0.01 & $-$0.09 & $-$0.18 &  0.20 \\
   CH3O & 1.43 & $-$2.23 & $-$2.37 & $-$2.40 & $-$1.24 &  2.27 &  2.17 &  2.16 & 2.41 &  1.13 &  0.85 &  0.75 &  0.97 \\
   H3CS & 1.88 & $-$2.80 &   NA &   NA &   NA & $-$0.15 &   NA &   NA &  NA & $-$0.30 &   NA &   NA &   NA \\
   H2CS & 0.31 & $-$2.56 & $-$2.62 & $-$2.74 & $-$1.83 & $-$0.08 & $-$0.13 & $-$0.27 & 0.19 &  0.21 &  0.14 &  0.00 &  0.34 \\
  CH2CN & 1.54 & $-$1.38 & $-$1.57 & $-$1.66 & $-$0.72 &  1.35 &  1.17 &  1.05 & 1.48 &  1.62 &  1.35 &  1.24 &  1.41 \\
  CH2NC & 0.99 & $-$1.68 & $-$1.86 & $-$1.95 & $-$1.13 &  0.99 &  0.80 &  0.68 & 1.00 &  1.16 &  0.88 &  0.77 &  0.86 \\
   HC2O & 2.28 & $-$1.00 & $-$1.20 & $-$1.30 & $-$0.28 &  1.97 &  1.78 &  1.67 & 2.12 &  2.24 &  1.96 &  1.84 &  2.04 \\
 CH2CHO & 1.71 & $-$1.19 & $-$1.39 & $-$1.48 & $-$0.58 &  1.65 &  1.46 &  1.34 & 1.70 &  1.84 &  1.55 &  1.45 &  1.56 \\
  CH3CO & 0.16 & $-$1.67 & $-$1.78 & $-$1.85 & $-$1.28 &  0.54 &  0.41 &  0.35 & 0.59 & $-$0.17 & $-$0.56 & $-$0.40 & $-$2.15 \\
  H5C2O & 1.57 & $-$1.41 & $-$1.53 & $-$1.57 & $-$0.66 &  3.10 &  3.03 &  3.03 & 2.79 &  1.31 &  1.05 &  0.96 &  0.79 \\
  H5C2S & 1.95 & $-$1.02 & $-$1.14 & $-$1.24 & $-$0.34 &  2.17 &  2.07 &  1.97 & 2.26 &  1.76 &  1.55 &  1.42 &  1.63 \\
    HLi & 0.31 &   NA &   NA & $-$1.11 &   NA &   NA &   NA &  0.14 &  NA &   NA &   NA & $-$0.14 &   NA \\
    HO2 & 0.62 & $-$3.62 & $-$3.38 & $-$3.46 & $-$2.44 &  0.66 &  0.22 &  0.32 & 1.02 &  0.24 & $-$4.73 & $-$3.80 &  0.23 \\ \midrule
    MAE &      &  3.23 &  3.31 &  3.35 &  2.39 &  0.48 &  0.47 &  0.50 & 0.37 &  0.25 &  0.50 &  0.54 &  0.28 \\
    MSE &      & $-$3.23 & $-$3.31 & $-$3.35 & $-$2.39 &  0.00 & $-$0.10 & $-$0.18 & 0.20 & $-$0.21 & $-$0.50 & $-$0.54 & $-$0.25 \\ \bottomrule
\end{longtable}
}
\end{landscape}
\restoregeometry

\newgeometry{left=0.5in, right=0.5in, top=1in, bottom=1in}
\renewcommand\arraystretch{0.6}
\small{
\begin{longtable}{lrrrrr}
    \caption{The experimental first electron affinities (in eV) of
            $N$-electron systems in comparison with results
            from $\Delta$-SCF method with different DFAs.
            $\Delta E_{\rm{DFA}} = E_{\rm{DFA}}(N) - E_{\rm{DFA}}(N+1)$.}
    \label{tab:EA_delta_scf} \\
    \toprule
Mol          & Exp.  & $\Delta E_{\rm{LDA}}$ & $\Delta E_{\rm{PBE}}$ & $\Delta E_{\rm{BLYP}}$  & $\Delta E_{\rm{B3LYP}}$ \\ \midrule
    \endfirsthead
    \multicolumn{6}{l}{Table \ref{tab:EA_delta_scf} continued: EAs from $\Delta$-SCF.} \\ \toprule
Mol          & Exp.  & $\Delta E_{\rm{LDA}}$ & $\Delta E_{\rm{PBE}}$ & $\Delta E_{\rm{BLYP}}$  & $\Delta E_{\rm{B3LYP}}$ \\ \midrule
    \endhead
    \midrule
    \endfoot
    \endlastfoot
     HC & 1.19 & 1.70 &  1.50 &  1.29 & 1.34 \\
    H2C & 0.83 & 0.55 &  0.32 &  0.37 &  NA \\
     HN & 0.33 & 0.71 &  0.46 &  0.49 & 0.43 \\
    H2N & 0.75 & 1.12 &  0.75 &  0.72 & 0.69 \\
     HO & 1.83 & 2.26 &  1.84 &  1.83 & 1.75 \\
    HSi & 1.26 & 1.44 &   NA &   NA & 1.24 \\
   H2Si & 1.07 & 1.29 &  1.23 &  1.01 & 1.14 \\
   H3Si & 0.93 & 1.15 &  0.92 &  0.86 & 0.95 \\
     HP & 1.02 & 1.25 &  1.03 &  1.00 & 1.08 \\
    H2P & 1.26 & 1.47 &  1.22 &  1.14 & 1.23 \\
     HS & 2.36 &  NA &  2.31 &  2.23 & 2.32 \\
     CN & 3.93 & 4.01 &  3.76 &  3.76 & 4.06 \\
     OP & 1.09 &  NA &  1.12 &  0.96 & 1.18 \\
     S2 & 1.64 &  NA &  1.45 &  1.34 & 1.57 \\
    Cl2 & 1.16 & 1.37 &  1.24 &  1.12 & 1.30 \\
     C2 & 3.24 &  NA &   NA &  3.45 & 3.56 \\
    C2O & 2.18 & 2.37 &  2.09 &  2.03 & 2.16 \\
    CNO & 3.55 & 3.72 &  3.36 &  3.26 & 3.43 \\
    NO2 & 1.53 & 1.49 &  1.22 &  1.23 & 1.51 \\
     O3 & 1.82 & 1.96 &  1.75 &  1.74 & 2.23 \\
     OF & 1.94 & 2.07 &  1.71 &  1.72 & 1.90 \\
    O2S & 1.06 & 1.10 &  0.96 &  0.88 & 1.24 \\
    OS2 & 1.88 & 1.85 &  1.74 &  1.62 & 2.00 \\
    HC2 & 2.95 & 3.20 &  2.95 &  2.93 & 3.09 \\
   H3C2 & 0.19 & 0.43 &  0.19 &  0.18 & 0.22 \\
   H2C3 & 1.70 & 2.00 &  1.84 &  1.66 & 1.85 \\
   H5C3 & 0.38 & 0.73 &  0.43 &  0.34 & 0.39 \\
    HCF & 0.24 & 0.59 &  0.43 &  0.33 & 0.45 \\
   CH3O & 1.43 & 1.67 &  1.37 &  1.33 & 1.34 \\
   H3CS & 1.88 & 0.00 &   NA &   NA &  NA \\
   H2CS & 0.31 & 0.52 &  0.42 &  0.28 & 0.48 \\
  CH2CN & 1.54 & 1.85 &  1.53 &  1.44 & 1.55 \\
  CH2NC & 0.99 & 1.41 &  1.09 &  1.00 & 1.03 \\
   HC2O & 2.28 & 2.47 &  2.15 &  2.06 & 2.20 \\
 CH2CHO & 1.71 & 2.06 &  1.73 &  1.65 & 1.70 \\
  CH3CO & 0.16 & 0.44 &  0.23 &  0.22 & 0.20 \\
  H5C2O & 1.57 & 1.95 &  1.64 &  1.58 & 1.54 \\
  H5C2S & 1.95 & 2.09 &  1.85 &  1.76 & 1.86 \\
    HLi & 0.31 &  NA &   NA &  0.34 &  NA \\
    HO2 & 0.62 & 0.68 &  0.39 &  0.38 & 0.52 \\ \midrule
    MAE &      & 0.29 &  0.11 &  0.12 & 0.09 \\
    MSE &      & 0.16 & $-$0.03 & $-$0.09 & 0.05 \\ \bottomrule
\end{longtable}
}
\restoregeometry

\newgeometry{left=0.5in, right=0.5in, top=1in, bottom=1in}
\begin{landscape}
\subsection{Quasihole energiges}
The same test set in Ref. \citenum{2020yangc} was used in this work
to test the quasihole energies. See Ref. \citenum{2020yangc} for details
about the molecular geometries and the experimental energies.
aug-cc-pVTZ is used as the basis set for all the calculations.

\renewcommand\arraystretch{0.6}
\small{
\begin{longtable}{lll|rrrr|rrrr|rrrr}
    \caption{The experimental quasihole energies of $N$-electron systems
            versus the negative orbital energies of occupied orbitals
            of $N$-electron systems from different DFAs. Units are in eV.}
    \label{tab:quasihole} \\
    \toprule
              & &       & \multicolumn{4}{c|}{DFAs}          & \multicolumn{4}{c|}{GSC1-DFAs}                & \multicolumn{4}{c}{GSC2-DFAs}                \\ \midrule
Mol           & Symmetry & Exp.   & LDA   & PBE   & BLYP  & B3LYP & LDA   & PBE   & BLYP  & B3LYP & LDA   & PBE   & BLYP  & B3LYP \\ \midrule
    \endfirsthead
    \multicolumn{15}{l}{Table \ref{tab:quasihole} continued: quasihole energies.} \\ \toprule
              & &       & \multicolumn{4}{c|}{DFAs}          & \multicolumn{4}{c|}{GSC1-DFAs}                & \multicolumn{4}{c}{GSC2-DFAs}                \\ \midrule
Mol           & Symmetry & Exp.   & LDA   & PBE   & BLYP  & B3LYP & LDA   & PBE   & BLYP  & B3LYP & LDA   & PBE   & BLYP  & B3LYP \\ \midrule
    \endhead
    \midrule
    \endfoot
    \endlastfoot
      cyanogen &  4SGG & 22.80 & $-$18.95 & 18.85 & 18.72 & 20.69 & 24.43 & 24.29 & 24.22 & 25.08 & 23.74 & 23.55 & 23.42 & 23.95 \\
      cyanogen &  1PIU & 15.60 & $-$11.60 & 11.35 & 11.18 & 12.59 & 14.61 & 14.36 & 14.18 & 14.97 & 15.42 & 15.10 & 14.93 & 15.43 \\
      cyanogen &  4SGU & 14.86 & $-$10.44 & 10.39 & 10.28 & 12.02 & 13.24 & 13.19 & 13.08 & 14.27 & 14.18 & 14.09 & 13.97 & 14.78 \\
      cyanogen &  5SGG & 14.49 & $-$10.17 & 10.07 &  9.99 & 11.70 & 12.97 & 12.87 & 12.79 & 13.95 & 13.91 & 13.78 & 13.68 & 14.46 \\
      cyanogen &  1PIG & 13.51 & $-$9.59 &  9.35 &  9.18 & 10.38 & 12.29 & 12.05 & 11.88 & 12.54 & 13.36 & 13.04 & 12.87 & 13.15 \\
            CO &   4SG & 19.72 & $-$14.20 & 14.14 & 14.00 & 15.91 & 20.25 & 20.09 & 19.99 & 20.73 & 19.89 & 19.70 & 19.54 & 19.85 \\
            CO &   1PI & 16.91 & $-$12.13 & 11.87 & 11.73 & 13.20 & 17.45 & 17.19 & 17.05 & 17.48 & 17.55 & 17.16 & 17.00 & 17.02 \\
            CO &   5SG & 14.01 & $-$9.11 &  9.03 &  9.00 & 10.52 & 13.96 & 13.86 & 13.85 & 14.44 & 14.01 & 13.87 & 13.84 & 14.18 \\
     acetylene &  2SGG & 23.50 & $-$18.53 & 18.62 & 18.45 & 20.68 & 23.73 & 23.81 & 23.66 & 24.87 & 23.44 & 23.50 & 23.31 & 24.17 \\
     acetylene &  2SGU & 18.70 & $-$13.99 & 14.05 & 13.94 & 15.63 & 17.44 & 17.51 & 17.40 & 18.41 & 18.38 & 18.43 & 18.30 & 19.00 \\
     acetylene &  3SGG & 16.70 & $-$12.24 & 12.20 & 12.17 & 13.74 & 15.82 & 15.79 & 15.75 & 16.61 & 16.63 & 16.56 & 16.50 & 17.01 \\
     acetylene &  1PIU & 11.49 & $-$7.38 &  7.20 &  7.05 &  8.20 & 11.33 & 11.16 & 10.99 & 11.36 & 11.61 & 11.34 & 11.18 & 11.28 \\
         water &   1B2 & 18.55 & $-$13.26 & 13.19 & 13.13 & 14.71 & 18.76 & 18.69 & 18.62 & 19.11 & 18.95 & 18.80 & 18.70 & 18.82 \\
         water &   3A1 & 14.74 & $-$9.37 &  9.30 &  9.24 & 10.85 & 15.40 & 15.32 & 15.25 & 15.66 & 14.90 & 14.71 & 14.63 & 14.72 \\
         water &   1B1 & 12.62 & $-$7.39 &  7.24 &  7.20 &  8.82 & 13.39 & 13.27 & 13.23 & 13.67 & 12.82 & 12.55 & 12.49 & 12.59 \\
      ethylene &   2AG & 23.60 & $-$18.74 & 18.82 & 18.64 & 20.85 & 23.21 & 23.30 & 23.11 & 24.44 & 23.17 & 23.24 & 23.04 & 24.04 \\
      ethylene &  2B1U & 19.10 & $-$14.18 & 14.25 & 14.12 & 15.91 & 17.37 & 17.44 & 17.31 & 18.46 & 18.19 & 18.25 & 18.12 & 18.96 \\
      ethylene &  1B2U & 16.00 & $-$11.54 & 11.48 & 11.38 & 12.91 & 14.78 & 14.71 & 14.59 & 15.48 & 15.50 & 15.42 & 15.31 & 15.90 \\
      ethylene &   3AG & 14.80 & $-$10.29 & 10.21 & 10.15 & 11.63 & 13.97 & 13.89 & 13.84 & 14.58 & 14.41 & 14.28 & 14.20 & 14.60 \\
      ethylene &  1B3G & 12.80 & $-$8.48 &  8.51 &  8.48 &  9.81 & 11.31 & 11.34 & 11.30 & 12.08 & 12.33 & 12.34 & 12.29 & 12.75 \\
      ethylene &  1B3U & 10.68 & $-$6.96 &  6.78 &  6.61 &  7.67 & 10.65 & 10.47 & 10.29 & 10.62 & 10.90 & 10.62 & 10.46 & 10.53 \\
       ammonia &    1E & 16.00 & $-$11.37 & 11.34 & 11.27 & 12.74 & 16.25 & 16.21 & 16.14 & 16.64 & 16.44 & 16.35 & 16.25 & 16.42 \\
       ammonia &   3A1 & 10.80 & $-$6.19 &  6.09 &  6.02 &  7.41 & 11.33 & 11.25 & 11.18 & 11.55 & 10.91 & 10.72 & 10.63 & 10.76 \\
  acetonitrile &   5A1 & 24.90 & $-$19.36 & 19.39 & 19.22 & 21.36 & 23.55 & 23.58 & 23.42 & 24.73 & 23.66 & 23.68 & 23.50 & 24.47 \\
  acetonitrile &   6A1 & 17.40 & $-$12.88 & 12.81 & 12.69 & 14.27 & 16.82 & 16.72 & 16.62 & 17.39 & 17.04 & 16.90 & 16.77 & 17.23 \\
  acetonitrile &    1E & 15.70 & $-$11.59 & 11.51 & 11.40 & 12.86 & 15.38 & 15.34 & 15.26 & 15.94 & 15.74 & 15.63 & 15.53 & 15.93 \\
  acetonitrile &   7A1 & 13.17 & $-$8.57 &  8.52 &  8.44 & 10.09 & 14.18 & 14.07 & 14.03 & 14.54 & 13.20 & 13.04 & 12.95 & 13.14 \\
  acetonitrile &    2E & 12.46 & $-$8.30 &  8.12 &  7.99 &  9.24 & 12.04 & 11.92 & 11.80 & 12.28 & 12.42 & 12.18 & 12.05 & 12.22 \\
 fluoromethane &   4A1 & 23.40 & $-$17.25 & 17.30 & 17.16 & 19.24 & 21.19 & 21.29 & 21.15 & 22.43 & 21.83 & 21.89 & 21.75 & 22.69 \\
 fluoromethane &   5A1 & 17.00 & $-$12.13 & 11.94 & 11.82 & 13.41 & 16.81 & 16.68 & 16.59 & 17.20 & 17.03 & 16.75 & 16.62 & 16.82 \\
 fluoromethane &    1E & 17.00 & $-$11.75 & 11.62 & 11.55 & 13.26 & 15.48 & 15.34 & 15.26 & 16.28 & 16.22 & 16.04 & 15.97 & 16.60 \\
 fluoromethane &    2E & 13.10 & $-$8.17 &  8.07 &  8.03 &  9.67 & 12.32 & 12.32 & 12.33 & 12.85 & 12.84 & 12.71 & 12.67 & 13.00 \\
       benzene &  1A2U & 12.25 & $-$9.25 &  9.03 &   NA &   NA & 11.72 & 11.51 &   NA &   NA & 12.30 & 12.04 &   NA &   NA \\
       benzene &  1E1G &  9.24 & $-$6.50 &  6.29 &   NA &   NA &  8.90 &  8.70 &   NA &   NA &  9.53 &  9.25 &   NA &   NA \\
         furan &   3B2 & 23.00 & $-$18.47 & 18.44 & 18.22 & 20.36 & 21.60 & 21.54 & 21.31 & 22.81 & 22.12 & 22.06 & 21.84 & 23.01 \\
         furan &   4B2 & 19.70 & $-$14.87 & 14.76 & 14.58 & 16.38 & 17.52 & 17.39 & 17.20 & 18.48 & 18.42 & 18.28 & 18.09 & 19.01 \\
         furan &   6A1 & 18.80 & $-$13.96 & 13.95 & 13.80 & 15.59 & 16.48 & 16.44 & 16.29 & 17.56 & 17.46 & 17.42 & 17.27 & 18.23 \\
         furan &   7A1 & 17.50 & $-$13.40 & 13.27 & 13.11 & 14.84 & 15.88 & 15.72 & 15.55 & 16.75 & 16.81 & 16.65 & 16.48 & 17.40 \\
         furan &   1B1 & 15.60 & $-$11.20 & 10.96 & 10.76 & 12.37 & 15.05 & 14.80 & 14.61 & 15.50 & 14.99 & 14.67 & 14.49 & 14.98 \\
         furan &   5B2 & 15.25 & $-$10.93 & 10.86 & 10.75 & 12.23 & 13.29 & 13.24 & 13.16 & 14.17 & 14.35 & 14.26 & 14.15 & 14.79 \\
         furan &   6B2 & 14.40 & $-$9.98 &  9.89 &  9.84 & 11.32 & 12.55 & 12.45 & 12.41 & 13.37 & 13.44 & 13.33 & 13.27 & 13.89 \\
         furan &   8A1 & 13.80 & $-$9.72 &  9.59 &  9.49 & 10.94 & 12.60 & 12.48 & 12.41 & 13.21 & 13.24 & 13.07 & 12.97 & 13.49 \\
         furan &   9A1 & 13.00 & $-$9.02 &  8.86 &  8.74 & 10.27 & 12.46 & 12.32 & 12.24 & 12.84 & 12.81 & 12.60 & 12.47 & 12.88 \\
         furan &   2B1 & 10.40 & $-$7.02 &  6.80 &  6.63 &  7.78 &  9.86 &  9.65 &  9.47 & 10.05 & 10.47 & 10.17 & 10.01 & 10.26 \\
         furan &   1A2 &  9.00 & $-$5.84 &  5.62 &  5.46 &  6.45 &  8.54 &  8.32 &  8.16 &  8.61 &  9.18 &  8.89 &  8.73 &  8.88 \\
         HCOOH &   6A' & 22.00 & $-$16.15 & 16.07 & 15.94 & 17.79 & 20.45 & 20.21 & 20.08 & 20.97 & 20.83 & 20.62 & 20.47 & 21.08 \\
         HCOOH &   7A' & 17.80 & $-$14.63 & 14.51 & 14.42 & 16.21 & 18.57 & 18.40 & 18.31 & 19.23 & 19.12 & 18.94 & 18.84 & 19.49 \\
         HCOOH &   8A' & 17.10 & $-$11.45 & 11.22 & 11.11 & 12.88 & 15.00 & 15.74 & 15.70 & 16.52 & 15.66 & 15.68 & 15.56 & 16.01 \\
         HCOOH &   1A" & 15.80 & $-$11.35 & 11.20 & 11.07 & 12.63 & 15.91 & 14.75 & 14.61 & 15.48 & 15.90 & 15.34 & 15.21 & 15.72 \\
         HCOOH &   9A' & 14.80 & $-$9.98 &  9.86 &  9.78 & 11.41 & 13.64 & 13.53 & 13.51 & 14.36 & 14.26 & 14.08 & 14.00 & 14.50 \\
         HCOOH &   2A" & 12.60 & $-$8.44 &  8.22 &  8.12 &  9.50 & 11.98 & 11.76 & 11.66 & 12.31 & 12.65 & 12.33 & 12.21 & 12.44 \\
         HCOOH &  10A' & 11.50 & $-$7.05 &  6.89 &  6.84 &  8.31 & 11.44 & 11.32 & 11.30 & 11.84 & 11.37 & 11.12 & 11.04 & 11.26 \\ \midrule
           MAE &       &       &   4.51 &  4.61 &  4.79 &  3.18 &  0.94 &  1.00 &  1.08 &  0.63 &  0.49 &  0.54 &  0.66 &  0.32 \\
           MSE &       &       &   4.51 &  4.61 &  4.79 &  3.18 &  0.66 &  0.77 &  0.88 &  0.07 &  0.27 &  0.43 &  0.57 &  0.09 \\ \bottomrule
\end{longtable}
}
\end{landscape}
\restoregeometry

\newgeometry{left=0.5in, right=0.5in, top=1in, bottom=1in}
\begin{landscape}
\subsection{Low-lying excitation energies}
Low-lying excitation energies were calculated from QE-DFT method
\cite{2019mei673,2019mei2545}.
The same test set (with excluding the long-chained molecule octatetraene,
$\ce{C8H10}$) and the references data
in Ref. \citenum{2019mei673} were used in this work.
Notation S1 or S2 refer to the first or second singlet excitation,
and T1 or T2 refer to the first or second triplet excitation.
This test set includes 15 organic molecules with small or
moderate sizes. There are 15 cases for S1, 10 case for
S2, 12 case for T1, and 7 cases for T2 with available
experimental references. cc-pVTZ is used as the basis set for
all the calculations.

\renewcommand\arraystretch{0.6}
\small{
\begin{longtable}{l|rr|rrrr|rrrr|rrrr}
    \caption{The low-lying excitation (in eV) from QE-DFT method
        with different DFAs.}
    \label{tab:exci} \\
    \toprule
      &      &       & \multicolumn{4}{c|}{DFAs}          & \multicolumn{4}{c|}{GSC1-DFAs}                & \multicolumn{4}{c}{GSC2-DFAs}                \\ \midrule
Mol   & Symmetry  & Ref.  & LDA    & PBE    & BLYP   & B3LYP  & LDA      & PBE      & BLYP      & B3LYP      & LDA      & PBE      & BLYP      & B3LYP      \\ \midrule
    \endfirsthead
    \multicolumn{15}{l}{Table \ref{tab:exci} continued: low-lying excitation energies.} \\ \toprule
      &      &       & \multicolumn{4}{c|}{DFAs}          & \multicolumn{4}{c|}{GSC1-DFAs}                & \multicolumn{4}{c}{GSC2-DFAs}                \\ \midrule
Mol   & Symmetry  & Ref.  & LDA    & PBE    & BLYP   & B3LYP  & LDA      & PBE      & BLYP      & B3LYP      & LDA      & PBE      & BLYP      & B3LYP      \\ \midrule
    \endhead
    \midrule
    \endfoot
    \endlastfoot
          ethene &  $^1B1u$ & 7.80 &  7.27 &  7.72 &  7.48 &  7.79 &  6.53 &  6.88 &  6.64 &  7.07 &  6.54 &  6.66 &  6.59 &  7.20 \\
          ethene &  $^3B1u$ & 4.50 &  4.90 &  4.41 &  4.61 &  4.43 &  4.41 &  3.95 &  4.16 &  4.06 &  4.90 &  4.61 &  4.61 &  4.42 \\
           furan &   $^1A2$ & 6.03 &  6.93 &  7.01 &  6.94 &  7.33 &  6.61 &  6.62 &  6.57 &  6.97 &  6.67 &  6.70 &  6.62 &  7.08 \\
           furan &   $^1B2$ & 6.32 &  5.63 &  5.88 &  5.76 &  6.16 &  5.29 &  5.54 &  5.42 &  5.88 &  5.38 &  5.50 &  5.43 &  5.89 \\
           furan &   $^3A2$ & 5.99 &  6.00 &  5.75 &  5.79 &  5.87 &  5.86 &  5.63 &  5.67 &  5.75 &  5.96 &  5.79 &  5.77 &  5.84 \\
           furan &   $^3B2$ & 4.17 &  4.33 &  4.02 &  4.07 &  3.98 &  4.04 &  3.75 &  3.79 &  3.75 &  4.35 &  4.16 &  4.14 &  4.05 \\
    benzoquinone &  $^1B1g$ & 2.78 &  1.93 &  1.88 &  1.95 &  2.34 &  1.90 &  1.83 &  1.88 &  1.78 &  1.87 &  1.81 &  1.88 &  2.49 \\
    benzoquinone &  $^1B3u$ & 5.60 &  4.77 &  4.78 &  4.84 &  5.47 &  5.00 &  5.00 &  5.04 &  5.09 &  4.64 &  4.60 &  4.67 &  5.44 \\
    benzoquinone &  $^3B1g$ & 2.51 &  1.59 &  1.60 &  1.68 &  2.08 &  1.59 &  1.58 &  1.64 &  1.55 &  1.54 &  1.55 &  1.62 &  2.06 \\
    benzoquinone &  $^3B3u$ & 5.38 &  4.65 &  4.68 &  4.75 &  5.38 &  4.89 &  4.90 &  4.95 &  5.01 &  4.51 &  4.50 &  4.57 &  5.33 \\
 cyclopentadiene &   $^1A2$ & 5.65 &  6.81 &  6.92 &  6.84 &  7.25 &  6.49 &  6.53 &  6.47 &  6.90 &  6.54 &  6.58 &  6.50 &  7.01 \\
 cyclopentadiene &   $^1B2$ & 5.55 &  4.67 &  4.86 &  4.75 &  5.08 &  4.23 &  4.42 &  4.31 &  4.73 &  4.37 &  4.47 &  4.40 &  4.85 \\
 cyclopentadiene &   $^3A2$ & 5.61 &  5.87 &  5.64 &  5.68 &  5.81 &  5.70 &  5.49 &  5.53 &  5.66 &  5.80 &  5.63 &  5.61 &  5.76 \\
 cyclopentadiene &   $^3B2$ & 3.25 &  3.46 &  3.16 &  3.21 &  3.13 &  3.06 &  2.77 &  2.82 &  2.83 &  3.37 &  3.16 &  3.16 &  3.12 \\
       butadiene &   $^1Bg$ & 6.26 &  6.96 &  7.03 &  6.94 &  7.28 &  6.83 &  6.81 &  6.74 &  7.07 &  6.67 &  6.68 &  6.59 &  7.00 \\
       butadiene &   $^1Bu$ & 6.18 &  4.72 &  4.99 &  4.87 &  5.42 &  4.48 &  4.71 &  4.60 &  5.15 &  4.45 &  4.54 &  4.51 &  5.18 \\
       butadiene &   $^3Bg$ & 6.22 &  6.11 &  5.91 &  5.93 &  6.14 &  6.12 &  5.92 &  5.95 &  6.10 &  6.05 &  5.91 &  5.87 &  6.08 \\
       butadiene &   $^3Bu$ & 3.20 &  3.40 &  3.14 &  3.22 &  3.20 &  3.24 &  2.98 &  3.06 &  3.08 &  3.40 &  3.21 &  3.23 &  3.22 \\
      hexatriene &   $^1Au$ & 5.71 &  5.64 &  5.71 &  5.64 &  6.16 &  5.45 &  5.48 &  5.42 &  5.93 &  5.48 &  5.52 &  5.45 &  6.06 \\
      hexatriene &   $^1Bu$ & 5.10 &  3.55 &  3.73 &  3.66 &  4.24 &  3.43 &  3.58 &  3.51 &  4.04 &  3.43 &  3.52 &  3.49 &  4.14 \\
      hexatriene &   $^3Au$ & 5.68 &  4.97 &  4.84 &  4.85 &  5.15 &  4.86 &  4.72 &  4.73 &  5.03 &  4.90 &  4.78 &  4.76 &  5.10 \\
      hexatriene &   $^3Bu$ & 2.40 &  2.63 &  2.44 &  2.49 &  2.52 &  2.58 &  2.40 &  2.45 &  2.49 &  2.63 &  2.48 &  2.50 &  2.52 \\
    cyclopropene &   $^1B2$ & 7.06 &  5.99 &  6.44 &  6.32 &  6.74 &  6.13 &  6.48 &  6.32 &  6.62 &  5.72 &  5.90 &  5.88 &  6.44 \\
    cyclopropene &   $^3B2$ & 4.34 &  4.38 &  4.14 &  4.23 &  4.20 &  4.73 &  4.47 &  4.54 &  4.38 &  4.28 &  4.05 &  4.10 &  4.01 \\
   norbornadiene &   $^1A2$ & 5.34 &  4.58 &  4.72 &  4.67 &  4.94 &  4.24 &  4.35 &  4.29 &  4.62 &  4.33 &  4.42 &  4.38 &  4.75 \\
   norbornadiene &   $^3A2$ & 3.72 &  3.74 &  3.60 &  3.63 &  3.73 &  3.42 &  3.27 &  3.29 &  3.44 &  3.58 &  3.48 &  3.49 &  3.65 \\
       tetrazine &   $^1Au$ & 3.48 &   NA &  3.06 &  3.11 &  3.58 &   NA &  3.16 &  3.21 &  3.65 &   NA &  2.87 &  2.93 &  3.45 \\
       tetrazine &  $^1B3u$ & 2.24 &   NA &  1.70 &  1.75 &  2.15 &   NA &  1.81 &  1.86 &  2.24 &   NA &  1.67 &  1.72 &  2.10 \\
       tetrazine &   $^3Au$ & 3.52 &   NA &  2.90 &  2.96 &  3.44 &   NA &  3.02 &  3.08 &  3.52 &   NA &  2.72 &  2.79 &  3.30 \\
       tetrazine &  $^3B3u$ & 1.89 &   NA &  1.25 &  1.30 &  1.62 &   NA &  1.36 &  1.41 &  1.70 &   NA &  1.27 &  1.31 &  1.60 \\
    formaldehyde &   $^1A2$ & 3.88 &  3.89 &  3.58 &  3.64 &  3.42 &  3.54 &  3.14 &  3.16 &  2.92 &  3.82 &  3.59 &  3.59 &  3.53 \\
    formaldehyde &   $^3A2$ & 3.50 &  3.38 &  3.32 &  3.38 &  3.17 &  3.13 &  2.96 &  2.98 &  2.76 &  3.29 &  3.17 &  3.22 &  3.15 \\
         acetone &   $^1A2$ & 4.40 &  4.51 &  4.17 &  4.21 &  4.03 &  4.10 &  3.70 &  3.75 &  3.50 &  4.37 &  4.02 &  4.03 &  3.97 \\
         acetone &   $^1B2$ & 5.92 &  8.78 &  8.53 &  8.39 &  7.99 &  6.72 &  6.34 &  6.18 &  6.08 &  6.51 &  6.30 &  6.84 &  7.05 \\
         acetone &   $^3A2$ & 4.05 &  4.09 &  3.93 &  3.96 &  3.78 &  3.71 &  3.47 &  3.52 &  3.28 &  3.96 &  3.73 &  3.75 &  3.69 \\
         acetone &   $^3B2$ & 5.87 &  8.69 &  8.42 &  8.30 &  7.91 &  6.63 &  6.24 &  6.10 &  6.01 &  6.46 &  6.28 &  6.77 &  6.99 \\
        pyridine &   $^1A2$ & 5.11 &   NA &  5.20 &  5.35 &  5.23 &   NA &  3.87 &  4.48 &  3.87 &   NA &  5.02 &  5.57 &  5.21 \\
        pyridine &   $^1B1$ & 4.59 &   NA &  4.75 &  4.81 &  4.73 &   NA &  3.48 &  4.01 &  3.42 &   NA &  4.62 &  4.67 &  4.75 \\
        pyridine &   $^3A2$ & 5.28 &   NA &  5.17 &  5.01 &  5.23 &   NA &  3.83 &  4.15 &  3.87 &   NA &  4.98 &  4.52 &  5.20 \\
        pyridine &   $^3B1$ & 4.25 &   NA &  4.38 &  4.38 &  4.31 &   NA &  3.10 &  3.58 &  2.99 &   NA &  4.25 &  4.32 &  4.32 \\
      pyridazine &   $^1A2$ & 4.32 &  4.24 &  4.20 &  4.24 &  4.25 &  3.18 &  3.14 &  3.16 &  3.34 &  3.88 &  3.85 &  3.88 &  4.14 \\
      pyridazine &   $^1B1$ & 3.78 &  3.63 &  3.55 &  3.59 &  3.63 &  2.58 &  2.52 &  2.54 &  2.74 &  3.31 &  3.28 &  3.29 &  3.52 \\
        pyrizine &   $^1Au$ & 4.81 &  4.32 &  4.31 &  4.37 &  4.75 &  4.19 &  4.18 &  4.23 &  4.62 &  4.15 &  4.12 &  4.18 &  4.61 \\
        pyrizine &  $^1B3u$ & 3.95 &  3.51 &  3.48 &  3.52 &  3.90 &  3.48 &  3.45 &  3.48 &  3.85 &  3.44 &  3.39 &  3.44 &  3.82 \\
      pyrimidine &   $^1A2$ & 4.91 &  4.39 &  4.31 &  4.36 &  4.59 &  4.07 &  3.97 &  4.01 &  4.24 &  4.21 &  4.13 &  4.17 &  4.52 \\
      pyrimidine &   $^1B1$ & 4.55 &  4.01 &  3.91 &  3.96 &  4.17 &  3.66 &  3.54 &  3.58 &  3.81 &  3.86 &  3.77 &  3.81 &  4.12 \\ \midrule
        MAE (S1) &          &      &  0.70 &  0.57 &  0.60 &  0.34 &  1.01 &  0.94 &  0.94 &  0.74 &  0.91 &  0.83 &  0.84 &  0.47 \\
        MSE (S1) &          &      & $-$0.68 & $-$0.54 & $-$0.57 & $-$0.32 & $-$1.01 & $-$0.94 & $-$0.94 & $-$0.74 & $-$0.91 & $-$0.82 & $-$0.83 & $-$0.45 \\ \midrule
        MAE (S2) &          &      &  0.85 &  0.74 &  0.70 &  0.66 &  0.69 &  0.69 &  0.59 &  0.65 &  0.61 &  0.57 &  0.61 &  0.52 \\
        MSE (S2) &          &      &  0.41 &  0.30 &  0.30 &  0.55 & $-$0.07 & $-$0.24 & $-$0.21 & $-$0.00 & $-$0.05 & $-$0.13 & $-$0.04 &  0.34 \\ \midrule
        MAE (T1) &          &      &  0.23 &  0.23 &  0.19 &  0.17 &  0.30 &  0.50 &  0.42 &  0.48 &  0.26 &  0.25 &  0.24 &  0.20 \\
        MSE (T1) &          &      &  0.03 & $-$0.20 & $-$0.14 & $-$0.14 & $-$0.17 & $-$0.48 & $-$0.38 & $-$0.46 & $-$0.03 & $-$0.22 & $-$0.19 & $-$0.17 \\ \midrule
        MAE (T2) &          &      &  0.77 &  0.68 &  0.66 &  0.39 &  0.40 &  0.57 &  0.48 &  0.37 &  0.44 &  0.48 &  0.59 &  0.31 \\
        MSE (T2) &          &      &  0.25 & $-$0.03 & $-$0.04 &  0.17 & $-$0.12 & $-$0.47 & $-$0.43 & $-$0.33 & $-$0.18 & $-$0.37 & $-$0.36 &  0.01 \\ \midrule
     MAE (TOTAL) &          &      &  0.62 &  0.54 &  0.53 &  0.38 &  0.65 &  0.70 &  0.64 &  0.59 &  0.59 &  0.55 &  0.58 &  0.38 \\
     MSE (TOTAL) &          &      & $-$0.09 & $-$0.16 & $-$0.16 &  0.02 & $-$0.43 & $-$0.57 & $-$0.53 & $-$0.42 & $-$0.36 & $-$0.42 & $-$0.39 & $-$0.11 \\ \bottomrule
\end{longtable}
}
\end{landscape}
\restoregeometry

\newgeometry{left=0.5in, right=0.5in, top=0.6in, bottom=0.6in}
\subsection{Photoemission spectra}
10 organic molecules with
small or moderate sizes were selected from Ref. \citenum{2019mei673}
for testing. See Ref. \citenum{2019mei673} for details about the
molecular geometries.
Experimental spectra were used as the references.
The GSC and GW calculations were associated with the
PBE functional. cc-pVTZ was used as the basis set for all the
calculations. All the calculated spectra were obtained
from the Gaussian expansion of calculated quasiparticle energies
with a standard deviation of 0.2 eV.

\restoregeometry

\subsection{Size dependence calculations}
In Table~\ref{tab:size}, molecules are obtained from
the IP test set, polyacene (noted as "Ph\_n")
and polyacetylene (noted as "CH\_n") used in Ref. \citenum{2020mei10277},
and the quasiparticle test set used in Ref.~\citenum{2019mei673}.
The geometries and the reference
numbers are directly taken from the two references.
cc-pVTZ basis set is used for polyacene.
cc-pVDZ basis set is used for polyacene.
6-311++G(3df, 3pd) basis set is used for the IP test set.
cc-pVTZ basis set is used for the quasiparticle test.

\renewcommand\arraystretch{0.58}
\small{
\begin{longtable}{lrrrrrr}
    \caption{The experimental first ionization potentials of $N$-electron systems
            versus the negative HOMO energies of $N$-electron systems from different DFAs.
            Units are in eV.}
    \label{tab:size} \\
\toprule
Name                             & atoms & Ref   & BLYP  & GSC1-BLYP & GSC2-BLYP & LOSC-BLYP \\ \midrule
\endfirsthead
\multicolumn{7}{l}{Table \ref{tab:size} continued.}\\ \toprule
Name                             & atoms & Ref   & BLYP  & GSC1-BLYP & GSC2-BLYP & LOSC-BLYP \\ \midrule
\endhead
\midrule
\endfoot
\endlastfoot
Ph\_1                            & 12    & 9.24  & 6.09  & 8.49     & 9.06      & 8.49      \\
Ph\_2                            & 18    & 8.11  & 5.26  & 6.88     & 7.68      & 8.18      \\
Ph\_3                            & 24    & 7.47  & 4.76  & 6.07     & 6.87      & 7.42      \\
Ph\_4                            & 30    & 6.97  & 4.44  & 5.54     & 6.33      & 7.03      \\
Ph\_5                            & 36    & 6.63  & 4.22  & 5.17     & 5.94      & 6.46      \\
Ph\_6                            & 42    & 6.36  & 4.06  & 4.89     & 5.65      & 6.33      \\
CH\_1                            & 6     & 10.48 & 6.34  & 10.11    & 10.43     & 10.11     \\
CH\_2                            & 10    & 9.18  & 5.46  & 7.86     & 8.61      & 9.12      \\
CH\_3                            & 14    & 8.18  & 5.01  & 6.80     & 7.67      & 8.47      \\
CH\_4                            & 18    & 7.69  & 4.74  & 6.15     & 7.07      & 7.70      \\
CH\_5                            & 22    & 7.33  & 4.56  & 5.71     & 6.65      & 7.45      \\
CH\_6                            & 26    & 7.04  & 4.43  & 5.39     & 6.33      & 7.25      \\
CH\_7                            & 30    & 6.85  & 4.33  & 5.15     & 6.08      & 7.16      \\
CH\_8                            & 34    & 6.66  & 4.25  & 4.95     & 5.88      & 6.97      \\
CH\_9                            & 38    & 6.55  & 4.19  & 4.80     & 5.71      & 6.87      \\
CH\_10                           & 42    & 6.41  & 4.14  & 4.67     & 5.57      & 6.64      \\
Anthracene                       & 24    & 7.40  & 4.74  & 6.04     & 6.83      & 7.40      \\
Benzothiadiazole                 & 13    & 9.00  & 5.93  & 7.77     & 8.52      & 8.31      \\
Benzothiazole                    & 14    & 8.80  & 5.77  & 7.48     & 8.28      & 8.53      \\
C60                              & 60    & 7.60  & 5.53  & 6.42     & 7.08      & 7.66      \\
Fluorene                         & 23    & 7.90  & 5.21  & 6.74     & 7.41      & 7.59      \\
H2P                              & 38    & 6.90  & 4.78  & 5.79     & 6.53      & 6.84      \\
H2PC                             & 58    & 6.40  & 4.73  & 5.41     & 6.13      & 6.60      \\
H2TPP                            & 78    & 6.40  & 4.57  & 5.53     & 6.06      & 6.62      \\
Pentacene                        & 36    & 6.60  & 4.20  & 5.14     & 5.91      & 6.36      \\
PTCDA                            & 38    & 8.20  & 5.93  & 6.78     & 7.64      & 8.63      \\
Thiadiazole                      & 7     & 10.10 & 6.73  & 9.25     & 9.94      & 9.27      \\
thiophene                        & 9     & 8.85  & 5.59  & 8.18     & 8.62      & 8.19      \\
Benzoquinone                     & 12    & 10.03 & 6.17  & 8.08     & 9.10      & 10.92     \\
Cl4-isobenzofuranedione  & 15    & 10.80 & 6.88  & 8.37     & 9.15      & 9.91      \\
Dichlone                         & 18    & 9.59  & 6.39  & 8.27     & 8.93      & 11.11     \\
F4-benzoquinone          & 12    & 10.83 & 7.12  & 8.98     & 9.95      & 9.75      \\
Maleicanhydride                  & 9     & 11.09 & 6.93  & 9.23     & 10.23     & 11.82     \\
Nitrobenzene                     & 14    & 9.93  & 6.59  & 10.16    & 10.26     & 11.91     \\
Phenazine                        & 22    & 8.38  & 5.54  & 6.77     & 7.67      & 7.90      \\
Phthalimide                      & 16    & 9.84  & 6.15  & 8.38     & 9.07      & 10.93     \\
TCNE                             & 10    & 11.78 & 8.38  & 9.87     & 10.94     & 12.41     \\
Benzonitrile                     & 13    & 9.75  & 6.61  & 8.52     & 9.31      & 9.48      \\
Cl4-benzoquinone          & 12    & 9.82  & 6.76  & 8.14     & 9.05      & 9.19      \\
F4-benzenedicarbonitrile & 14    & 10.65 & 7.22  & 9.12     & 9.87      & 9.20      \\
Fumaronitrile                    & 8     & 11.23 & 7.62  & 9.57     & 10.60     & 11.13     \\
mDCBN                            & 14    & 10.40 & 7.15  & 8.75     & 9.64      & 9.98      \\
NDCA                             & 21    & 8.98  & 6.23  & 7.65     & 8.48      & 8.98      \\
Nitrobenzonitrile                & 15    & 10.40 & 7.08  & 10.64    & 10.67     & 12.39     \\
Phthalicanhydride                & 15    & 10.18 & 6.70  & 8.99     & 9.70      & 11.57     \\
TCNQ                             & 20    & 9.61  & 6.82  & 7.83     & 8.82      & 9.02      \\
Acridine                         & 23    & 7.99  & 5.18  & 6.42     & 7.29      & 7.66      \\
Azulene                          & 18    & 7.43  & 4.73  & 6.53     & 7.15      & 7.24      \\
Naphthalenedione                 & 18    & 9.54  & 5.98  & 7.87     & 8.64      & 10.68     \\
cyanogen                         & 4     & 13.51 & 9.18  & 11.88    & 12.87     & 13.51     \\
CO                               & 2     & 14.01 & 9.00  & 13.85    & 13.84     & 13.95     \\
acetylene                        & 4     & 11.49 & 7.05  & 10.99    & 11.18     & 11.01     \\
water                            & 3     & 12.62 & 7.20  & 13.23    & 12.49     & 13.24     \\
ethylene                         & 6     & 10.68 & 6.61  & 10.29    & 10.46     & 10.32     \\
ammonia                          & 4     & 10.80 & 6.02  & 11.18    & 10.63     & 11.24     \\
acetonitrile                     & 6     & 12.46 & 7.99  & 11.80    & 12.05     & 12.08     \\
fluoromethane                    & 5     & 13.10 & 8.03  & 12.33    & 12.42     & 12.51     \\
furan                            & 9     & 9.00  & 5.46  & 8.16     & 8.73      & 8.20      \\
HCOOH                            & 5     & 11.50 & 6.84  & 11.30    & 11.05     & 11.64     \\
H4C                              & 5     & 14.40 & 9.37  & 13.68    & 13.97     & 13.75     \\
H3N                              & 4     & 11.03 & 6.17  & 11.34    & 10.81     & 11.40     \\
HO                               & 2     & 13.07 & 7.35  & 13.53    & 13.01     & 13.54     \\
H2O                              & 3     & 12.74 & 7.18  & 13.21    & 12.51     & 13.24     \\
HF                               & 2     & 16.20 & 9.59  & 16.79    & 15.98     & 16.80     \\
H4Si                             & 5     & 12.84 & 8.43  & 11.96    & 12.33     & 11.99     \\
HP                               & 2     & 10.18 & 5.89  & 9.30     & 9.85      & 9.33      \\
H2P                              & 3     & 9.82  & 5.78  & 9.16     & 9.65      & 9.21      \\
H3P                              & 4     & 10.61 & 6.59  & 10.02    & 10.36     & 10.06     \\
HS                               & 2     & 10.41 & 6.09  & 9.85     & 10.18     & 9.87      \\
H2S(2B1)                         & 3     & 10.48 & 6.15  & 9.87     & 10.16     & 9.91      \\
HCl                              & 2     & 12.82 & 7.90  & 12.15    & 12.47     & 12.16     \\
H2C2                             & 4     & 11.51 & 7.04  & 10.97    & 11.16     & 10.98     \\
H4C2                             & 6     & 10.74 & 6.61  & 10.28    & 10.45     & 10.31     \\
CO                               & 2     & 14.08 & 9.03  & 13.85    & 13.86     & 13.95     \\
N2(2Sg)                          & 2     & 15.61 & 10.26 & 14.49    & 15.21     & 14.51     \\
O2                               & 2     & 12.49 & 6.88  & 11.50    & 12.38     & 11.50     \\
P2                               & 2     & 10.82 & 6.88  & 9.74     & 10.22     & 9.67      \\
S2                               & 2     & 9.56  & 5.72  & 8.35     & 9.27      & 8.37      \\
Cl2                              & 2     & 11.77 & 7.29  & 10.11    & 11.07     & 10.12     \\
FCl                              & 2     & 12.95 & 7.87  & 11.42    & 12.32     & 11.53     \\
CS                               & 2     & 11.51 & 7.34  & 11.51    & 11.26     & 11.69     \\
BF3                              & 4     & 16.18 & 10.03 & 12.92    & 14.34     & 16.52     \\
BCl3                             & 4     & 11.91 & 7.57  & 9.50     & 10.61     & 11.64     \\
CO2                              & 3     & 13.90 & 8.99  & 12.56    & 13.47     & 12.56     \\
CF2                              & 3     & 12.40 & 7.35  & 11.52    & 11.83     & 11.59     \\
COS                              & 3     & 11.36 & 7.32  & 10.26    & 11.02     & 10.42     \\
CS2                              & 3     & 10.18 & 6.62  & 8.91     & 9.80      & 8.93      \\
H2C                              & 3     & 10.40 & 5.49  & 9.99     & 10.12     & 10.03     \\
H3C                              & 4     & 9.78  & 5.21  & 9.58     & 9.67      & 9.65      \\
H5C2                             & 7     & 8.60  & 4.43  & 8.26     & 8.35      & 8.41      \\
CN                               & 2     & 14.22 & 9.25  & 13.74    & 14.02     & 13.74     \\
HCO                              & 3     & 9.37  & 4.94  & 8.47     & 9.32      & 8.54      \\
CH2OH                            & 5     & 8.18  & 3.85  & 7.54     & 7.95      & 7.63      \\
CH3O                             & 5     & 10.94 & 6.03  & 10.76    & 10.49     & 10.94     \\
H4CO                             & 6     & 11.17 & 6.26  & 10.80    & 10.64     & 11.01     \\
H3CF                             & 5     & 13.47 & 8.04  & 12.20    & 12.64     & 12.34     \\
H2CS                             & 4     & 9.47  & 5.40  & 8.77     & 9.08      & 8.86      \\
CH2SH                            & 5     & 7.79  & 4.03  & 7.01     & 7.53      & 7.17      \\
H3CCl                            & 5     & 11.49 & 6.97  & 10.69    & 11.01     & 10.87     \\
H6C2O                            & 9     & 10.89 & 6.17  & 10.45    & 10.28     & 10.86     \\
H4C2O                            & 7     & 10.38 & 5.92  & 10.06    & 9.89      & 10.30     \\
H3COF                            & 6     & 11.68 & 6.62  & 10.75    & 10.95     & 10.92     \\
H4C2S                            & 7     & 9.15  & 5.23  & 8.67     & 8.77      & 8.75      \\
C2N2                             & 4     & 13.59 & 9.15  & 11.83    & 12.82     & 13.45     \\
H4B2                             & 6     & 10.17 & 6.22  & 9.82     & 9.94      & 9.86      \\
HN                               & 2     & 13.48 & 7.73  & 13.36    & 13.33     & 13.37     \\
H2N                              & 3     & 12.12 & 7.22  & 12.48    & 12.10     & 12.53     \\
H2N2                             & 4     & 10.28 & 5.64  & 9.21     & 9.85      & 9.25      \\
H3N2                             & 5     & 8.34  & 4.07  & 7.85     & 8.24      & 7.90      \\
HOF                              & 3     & 13.03 & 7.31  & 11.94    & 12.43     & 12.04     \\
H2Si                             & 3     & 9.55  & 5.76  & 8.72     & 9.24      & 8.77      \\
H3Si                             & 4     & 8.86  & 5.12  & 8.26     & 8.71      & 8.34      \\
H2Si2                            & 4     & 8.22  & 4.92  & 7.21     & 7.77      & 7.38      \\
H4Si2                            & 6     & 8.36  & 5.24  & 7.63     & 8.05      & 8.30      \\
H5Si2                            & 7     & 8.37  & 5.01  & 7.85     & 8.16      & 8.07      \\
H6Si2                            & 8     & 10.73 & 7.16  & 9.88     & 10.26     & 9.99      \\
B2F4                             & 6     & 13.30 & 8.39  & 11.11    & 12.15     & 11.64     \\
H4C3(cyclo)                      & 7     & 10.04 & 5.98  & 8.93     & 9.51      & 9.01      \\
H4C3(allene)                     & 7     & 10.31 & 6.42  & 9.73     & 9.94      & 9.89      \\
H7C3                             & 10    & 7.80  & 3.94  & 7.45     & 7.50      & 7.66      \\
H4CS                             & 6     & 9.55  & 5.48  & 8.92     & 9.16      & 9.04      \\
H4C4O                            & 9     & 9.09  & 5.53  & 8.21     & 8.73      & 8.26      \\
H5C4N                            & 10    & 8.42  & 5.01  & 7.64     & 8.10      & 7.88 \\ \bottomrule
\end{longtable}
\clearpage

\bibliography{qm4d,SI,my,GSC,DFT,add}